\documentclass[12p,a4paper]{article}
\usepackage{amssymb}
\usepackage{graphicx}
\usepackage{scrextend}   
\usepackage[utf8]{inputenc}
\usepackage{url}
\usepackage{lipsum}
\usepackage[all]{xy}
\usepackage{xcolor}
\usepackage{authblk}
\usepackage{amsmath}
\usepackage{bbm}

\usepackage{blindtext}  
\addtokomafont{labelinglabel}{\sffamily}

\newcommand{\vect}[1]{\boldsymbol{#1}}

\urldef{\mailsa}\path|{marina.martinez, jesus.malo}@uv.es|

\newcommand{\red}[1]{\textcolor[rgb]{0,0,0}{#1}}
\newcommand{\azul}[1]{\textcolor[rgb]{0,0,1}{#1}}

\title{Derivatives and Inverse of a Linear-Nonlinear\\ Multi-Layer Spatial Vision Model}
\date{\vspace{-5ex}}
\author[1]{$\,\,\,\,\,\,$ B. Galan}
\author[1,2]{M. Martinez-Garcia}
\author[3]{P. Cyriac}
\author[3]{T. Batard}
\author[3]{$\,\,\,\,\,\,\,\,\,\,\,$ M. Bertalmio}
\author[1]{ J. Malo\vspace{-0.3cm}}
\affil[1]{\vspace{-0.00cm}\small{Image Processing Lab. Parc Científic, Universitat de València, Spain}}
\affil[2]{Instituto de Neurociencias, CSIC, Spain}
\affil[3]{\vspace{-0.00cm}Dept. Tecnol. Inf. Comunic., Universitat Pompeu Fabra, Spain}

\begin{document}

\maketitle

\begin{abstract}
Linear-nonlinear transforms are interesting in vision science because
they are key in modeling a number of perceptual experiences such as color, motion or spatial texture.
Here we first show that a number of issues in vision may
be addressed through an analytic expression of the Jacobian of these linear-nonlinear transforms. The particular model analyzed afterwards (an extension of \cite{Malo15}) is illustrative because it consists of a cascade of standard
linear-nonlinear modules. Each module roughly corresponds to a known psychophysical mechanism:
(1) linear spectral integration and nonlinear brightness-from-luminance computation,
(2) linear pooling of local brightness and nonlinear normalization for local contrast computation,
(3) linear frequency selectivity and nonlinear normalization for spatial contrast masking, and
(4) linear wavelet-like decomposition and nonlinear normalization for frequency-dependent masking. Beyond being the appropriate technical report with the missing details in \cite{Malo15}, the interest of the presented analytic results and numerical methods transcend the particular model because of the ubiquity of the linear-nonlinear structure.\\
\azul{Part of this material was presented at MODVIS 2016 (see slides of the conference talk in the appendix at the end of this document).}
\end{abstract}

In Section \ref{general} we describe the general architecture using a unified notation in which every
module is composed by isomorphic linear and nonlinear transforms and we introduce a convenient
matrix-vector form for the \emph{divisive normalization}.
Section \ref{implications_of_jacobian} gives explicit Jacobian-related expressions for the considered architecture in a number of relevant issues in vision science.
These include (1) discrimination, (2) subjective distortions, (3) adaptive receptive fields and features,
(4) stimulus design through the inverse, through sensor-selective features, or through Maximum Differentiation, and finally (5) redundancy reduction effects of perceptual transforms also depend on the Jacobian.
Even though generic invertibility conditions are studied in Section \ref{inv_and_jacobian},
Section \ref{inverse} shows a more specific (analytic) expression for the inverse of
\emph{divisive normalization}. Then, Sections \ref{forward} and \ref{inverse_example} consider the details of the four layers in our illustrative model,
both in the forward and inverse directions respectively.
We explicitly list the derivatives of every module, and analyze the structure of the corresponding matrices. We find different analytical and numerical problems in each specific module.
Solutions are proposed for all of them.
Finally, Section \ref{section_MAD} describes an illustrative application of
the above theory to generate maximally/minimally visible distortions with the
provided toolbox (code described in Section \ref{toolbox}).

\section{Linear-nonlinear multi-layer models}
\label{general}

\subsection{The input image.}
An \emph{image} in the retina, $\vect{x}^0(\vect{p},\lambda)$, is a function describing the spectral irradiance in each spatial location, $\vect{p}$, and wavelength, $\lambda$.
Assuming a dense enough sampling\footnote{\label{sampling}Note on sampling. In the case of images intended for human observers, the required sampling frequencies are limited by (1) the spatial cut-off frequency of the Contrast Sensitivity Functions (CSFs) \cite{Mullen85}, and (2) the smoothness of the spectral sensitivities (both achromatic $V_{\lambda}$, and chromatic, $T_{c \lambda}$, where $c = 1,2,3$) \cite{Stiles82,Fairchild13}. Therefore, the spatial dimensions may be sampled at about 80 cycles/degree (cpd) and
the spectral dimension at about 0.1 samples/nm \cite{Wandell95}.
With a dense enough sampling, the specific sampling pattern has no major relevance since the continuous signal can always be obtained from the discrete signal \cite{Tekalp95}. Here we will assume cartesian sampling in every dimension.}, the continuous input image can be represented
by a discrete hyperspectral array with no information loss. The hyperspectral array consists of $b$ matrices of size $h \times w$, where the $l$-th matrix represents the discrete spatial distribution of the energy of $l$-th discrete wavelength ($l = 1,\ldots, b$).

Using an appropriate rearrangement\footnote{\label{arrangement}The particular scanning pattern has no major relevance as long as it can be inverted back to the original spatio-spectral domain. Here we will use the \emph{last-dimension-first} convention used in the \texttt{Matlab} functions \texttt{im2col.m} and \texttt{col2im.m}. The \texttt{BasicVideoTools} toolbox \cite{BasicVideoTools} has convenient generalizations of these vectorization functions to be applied in spatio-spectral (or spatio-temporal) arrays (namely \texttt{im2colcube.m} and \texttt{col2imcube.m}).
The selected rearrangement pattern has no fundamental effect, but it has to be taken into account to make sense of the structure of the matrices acting on the input vector.} the discrete input image can be thought as a \emph{vector} in a $d_0$-dimensional space,
\vspace{-0.4cm}
\begin{equation}
   \vect{x}^0(\vect{p},\lambda) \,\,\,\,\,\, \longrightarrow \,\,\,\,\,\, \vect{x}^0 = \left(\begin{array}{c}
                                             x^0_1 \\
                                             x^0_2 \\
                                             \vdots \\
                                             x^0_k \\
                                             \vdots \\
                                             x^0_{d_0} \\
                                         \end{array}
                                         \right)
  \label{the_image}
\end{equation}
i.e. the \emph{image}, $\vect{x}^0$, is a \emph{column vector}, $\vect{x}^0 \in \mathbb{R}^{d_0 \times 1}$, where $d_0 = h \times w \times b$.

Note that with the considered sampling\footref{sampling} the dimension of the stimulus is \emph{huge}
even for moderate image sizes (small angular field and spectral range).

\subsection{The (visual) system.}
The visual system may be thought as an operator, $S$, transforming the input $d_0$-dimensional vectors (stimuli) into $d_n$-dimensional output vectors (or sets of $d_n$ responses),
\vspace{-0.25cm}
\begin{equation}
  \xymatrixcolsep{2pc}
  \xymatrix{ \vect{x}^0  \,\,\,\, \ar@/^1pc/[r]^{\scalebox{1.1}{\emph{S}}} & \,\,\,\, \vect{x}^n
  }
  \label{global_response}
\end{equation}
where, $\vect{x}^n \in \mathbb{R}^{d_n \times 1}$. This is equivalent to considering $d_n$ separate sensors (or mechanisms) acting on the stimulus, $\vect{x}^0$, leading to the corresponding individual responses, $x^n_k$, where $k = 1, 2, \ldots d_n$.
In this view, the $k$-th sensor would be responsible for the $k$-th dimension of the response vector, $\vect{x}^n$.
The number of separate sensors analyzing the signal may not be the same as the input dimension, so in general $d_n \neq d_0$.
\subsection{Modular architecture.}
\subsubsection*{Multi-layer structure.} The global response described above may be decomposed as a set of elementary operations,
or a cascade of modules (stages or layers), $S^{(i)}$, where $i = 1,2,\cdots,n$,
\vspace{-0.2cm}
\begin{equation}
  \xymatrixcolsep{2pc}
  \xymatrix{ \vect{x}^0 \ar@/_1pc/[r]_{S^{(1)}} \ar@/^2pc/[rrrrrr]^{\scalebox{1.1}{\emph{S}}} & \vect{x}^1  \ar@/_1pc/[r]_{S^{(2)}}  & \vect{x}^2 \cdots \!\!\!\!\!\!\!\!\!\!\!\!\!\! & \vect{x}^{i-1} \ar@/_1pc/[r]_{S^{(i)}}  & \vect{x}^i  \cdots \!\!\!\!\!\!\!\!\!\!\!\!\!\! &  \vect{x}^{n-1} \ar@/_1pc/[r]_{S^{(n)}}  & \vect{x}^n
  }
  \label{modular}
\end{equation}
\vspace{-0.0cm}
The intermediate representations of the signal along this response path may have different dimension, i.e. $\vect{x}^i \in \mathbb{R}^{d_i \times 1}$, because the number of mechanisms in stage $S^{(i)}$ may be different from the number of mechanisms in $S^{(i-1)}$.
\subsubsection*{Linear-nonlinear modules.}
Each layer in the above deep network architecture performs a linear-nonlinear operation:
\vspace{-0.35cm}
\begin{equation}
  \xymatrixcolsep{2pc}
  \xymatrix{ \cdots \vect{x}^{i-1} \ar@/_2pc/[rr]_{S^{(i)}} \ar[r]^{\,\, L^{(i)}} & \vect{y}^i \ar[r]^{N^{(i)}} & \vect{x}^i \cdots}
  \label{module}
  \vspace{-0.1cm}
\end{equation}
The linear operation is represented by a matrix $L^{(i)} \in \mathbb{R}^{d_i \times d_{i-1}}$. The number of rows in $L^{(i)}$ corresponds to the number of \emph{linear} sensors in layer $S^{(i)}$. This number of mechanisms determines the dimension of the linear output, $\vect{y}^i \in \mathbb{R}^{d_i \times 1}$, where the subscript $L$ stands for \emph{linear},
\begin{equation}
       \vect{y}^i = L^{(i)} \cdot \vect{x}^{i-1}
       \label{linear}
\end{equation}
In the nonlinear operation, $N^{(i)}$, each response of the linear output undergoes a saturation. Phenomena such as \emph{masking} or \emph{lateral inhibition} imply that the saturation of $y^i_{k}$ should depend on the neighbors $y^i_{k'}$ with $k' \neq k$. This saturation is usually formalized using \emph{divisive normalization} \cite{Carandini12}.
This adaptive saturation is a canonical neural operation and it is at the core of models for color \cite{Fairchild13}, motion \cite{Heeger99}, and spatial texture vision \cite{Watson97}.
\subsubsection*{Divisive normalization in matrix-vector form.}
The classical expressions of the divisive normalization saturation \cite{Carandini12} make extensive use of element-wise operations combined with matrix-on-vector operations\footnote{\label{classical_div_norm} Example of classical \emph{explicit element-wise} expression of divisive normalization,
\begin{equation}
       x^i_k =  N^{(i)}(\vect{y}^i)_k = \textrm{sign}(y^i_{k}) \,\, \frac{|y^i_{k}|^{\gamma^i}}{b^i_k + \sum_{k'} H^{(i)}_{k k'} |y^i_{ k'}|^{\gamma^i}} = \textrm{sign}(y^i_{k}) \,\, \frac{|y^i_{k}|^{\gamma^i}}{\mathcal{D}^{(i)}(|\vect{y}^i)|)_k}
       \nonumber
\end{equation}
This expression combines conventional \emph{matrix-on-vector} operations (such as the effect of the activity of the neighbor mechanisms, $k'$, in the $k$-th mechanism through the matrix $H^{(i)}$)
with a number of \emph{element-wise} operations: the division of each coefficient of a vector by the corresponding coefficient of a \emph{denominator vector}, $\mathcal{D}_i$,
the element-wise absolute value (or rectification), the element-wise exponentiation, the element-wise computation of sign, and its preservation in the response through an element-wise product.
}.
Such combination makes differentiation and inversion from the explicit expression cumbersome. This could be alleviated by a matrix-vector expression where the individual coefficients, $k$, are not explicitly present. Incidentally, this matrix expression will imply more efficient code in matrix-oriented environments
such as \texttt{Matlab}.

In order to get such matrix-vector form, it is convenient to remind the equivalence between the element-wise (or Haddamard) product and the operation with diagonal matrices \cite{Minka00}.
Given two vectors $a$ and $b$, its Haddamard product is:
\begin{equation}
      a \odot \, b = D_a \cdot b = D_b \cdot a
      \nonumber
      \label{haddamard}
\end{equation}
where $D_a$ is the diagonal matrix with vector $a$ in the diagonal.
Using this, and the separation in \emph{sign} and \emph{absolute value}, we can re-write the classical expression of divisive normalization in matrix-vector form, as:
%
%

\begin{equation}
 \begin{aligned}
 \vect{x}^i &= D_{\textrm{sign}(\vect{y}^i)} \cdot \mathcal{N}(\vect{y}^i); \\
 \mathcal{N}(\vect{y}^i) & = \frac{K^{(i)} \cdot a}{b^i + H^{(i)} \cdot a} = \frac{K^{(i)} \cdot a}{\mathcal{D}^{(i)}( a )} = D_{\frac{1}{\mathcal{D}^{(i)}(a)}} \cdot K^{(i)} \cdot a\\
  \textrm{where} \,\,\,\, & a = |\vect{y}^i|^{\gamma^i}
            \end{aligned}
        \label{divisive_norm}
\end{equation}

%
where the \emph{normalization}, $\mathcal{N}(\vect{y}^i) $,
is a Haddamard quotient: division of each element in the vector $K^{(i)} \cdot a$ by the corresponding element in the \emph{denominator vector} $\mathcal{D}^{(i)}(a)$. This denominator is responsible for the adaptive saturation, which is key in \emph{masking} and \emph{adaptation} phenomena. Here the normalization only acts on the absolute values, but the matrix $D_{\textrm{sign}(\vect{y}^i)}$ acting on the vector $\mathcal{N}^{(i)}(a)$ preserves the sign of each $y^i_{k}$. Note that all the operations in $\mathcal{N}^{(i)}(a)$ are element-wise except the matrix-on-vector, $H^{(i)} \cdot a$, within the denominator vector $\mathcal{D}^{(1)}(a)$. This matrix-on-vector operation is important because the $k$-th row of $H^{(i)}$ describes how the activities $a_{k'}$ saturate (or mask) the response of the $k$-th nonlinear response. We will see below that Eq. \ref{divisive_norm} is extremely useful to avoid cumbersome individual element-wise partial derivatives.

Of course variations exist (\emph{Note!}): \red{Naka-Rushton, different excitation/inhibition exponents, different regularized versions (e.g. see \cite{Carandini12})}, but the important concept is \emph{saturation} and \emph{interaction} through $H$. This leads to a neighbor-dependent adaptive saturation. Simplified dimension-wise saturation (logarithm, exponents $<1$,...) neglect
the interactions. \red{(\emph{Note!}) Figure and connections with deep-networks.} Throughout the document the text (\emph{Note!}) means that the corresponding paragraph requires additional elaboration.

%
%

\section{The Jacobian matrix: implications in vision}
\label{implications_of_jacobian}

\subsection{Local-linear approximation}
The response function, $S$, can be seen as a nonlinear change of coordinates,
and its properties depend on how the output depends on the input, i.e. its properties depend
on the matrix of derivatives (or Jacobian) \cite{Dubrovin82,Spivak},
\vspace{-0.3cm}
\begin{eqnarray}
      \nabla S(\vect{x}^0_A) &=&
      \Bigg[\frac{\partial \vect{x}^n(\vect{x}^0_A)}{\partial x^0_1}, \cdots, \frac{\partial \vect{x}^n(\vect{x}^0_A)}{\partial x^0_j}, \cdots ,\frac{\partial \vect{x}^n(\vect{x}^0_A)}{\partial x^0_{d_0}} \Bigg] \\[0.5cm]
          &=&
      \begin{bmatrix}
      \dfrac{\partial x^n(\vect{x}^0_A)_1}{\partial x_1^0} & \cdots & \dfrac{\partial x^n(\vect{x}^0_A)_1}{\partial x_j^0} & \cdots & \dfrac{\partial x^n(\vect{x}^0_A)_1}{\partial x_{d_0}^0}\\
      \vdots& \ddots& \vdots &\ddots &\vdots\\
      \dfrac{\partial x^n(\vect{x}^0_A)_k}{\partial x_1^0} & \cdots &\dfrac{\partial x^n(\vect{x}^0_A)_k}{\partial x_j^0}  & \cdots &  \dfrac{\partial x^n(\vect{x}^0_A)_k}{\partial x_{d_0}^0}\\
       \vdots& \ddots& \vdots &\ddots &\vdots\\
       \dfrac{\partial x^n(\vect{x}^0_A)_{d_n}}{\partial x_1^0} & \cdots &\dfrac{\partial x^n(\vect{x}^0_A)_{d_n}}{\partial x_j^0}  & \cdots &  \dfrac{\partial x^n(\vect{x}^0_A)_{d_n}}{\partial x_{d_0}^0}
      \end{bmatrix}
      \nonumber
\end{eqnarray}

Note that this Jacobian may depend on the input, $\vect{x}^0$, because the \emph{slope} of the response
(the behavior of the system) may be different in different points of the stimulus space. Note also the matrix size:
$\nabla S \in \mathbb{R}^{d_n \times d_0}$

The global nonlinear behavior of the system can be linearly approximated in a neighborhood of some stimulus, $\vect{x}^0_A$,
using the Jacobian:
\begin{eqnarray}
      \nonumber S(\vect{x}^0_A + \Delta \vect{x}^0) & \approx & S(\vect{x}^0_A) + \nabla S(\vect{x}^0_A) \cdot \Delta \vect{x}^0 \\[2mm]
                 \Delta \vect{x}^n & \approx & \nabla S(\vect{x}^0_A) \cdot \Delta \vect{x}^0
      \label{linear_approx}
\end{eqnarray}
i.e. variations of the response linearly depend on variations of the input for small distortions $\Delta \vect{x}^0$.
\subsection{Adaptive receptive fields and Jacobian.}
If the whole system is \emph{linear}, it can be simply represented by a single matrix, so $S$ becomes $\mathcal{S}$ where  $\mathcal{S} \in \mathbb{R}^{d_n \times d_0}$, and it holds $\vect{x}^n = \mathcal{S} \cdot \vect{x}^0$, and $\nabla \mathcal{S} (\vect{x}^0) = \mathcal{S}, \,\,\, \forall \vect{x}^0$ because the derivative of a linear function is the (point-independent) matrix itself.
In this linear case, each element of the response vector, $x^i_k$, is the scalar product of the $k$-th row of $S$ and the input: $$x^n_k = \sum_{j=1}^{d_0} \mathcal{S}_{k j} \, x^0_j$$
the $k$-th row of $\mathcal{S}$ can be defined as the \emph{receptive field} of the $k$-th linear sensor of the system.

For nonlinear systems the conventional receptive field concept can be extended using
the Jacobian. Following the dot-product definition, Eq. \ref{linear_approx} implies that the vectors extracted from the rows of the Jacobian may be referred to as \emph{\bf{local receptive fields}} in a neighborhood of $\vect{x}^0_A$:
\begin{equation}
       \vect{w}^{(n)}_k(\vect{x}^0_A)
        = \Bigg [\dfrac{\partial x_k^n}{\partial x_1^0},  \cdots ,\dfrac{\partial x_k^n}{\partial x_j^0},  \cdots , \dfrac{\partial x_k^n}{\partial x_{d_0}^0} \Bigg]^T
       \label{recept_field}
\end{equation}
where we used the superscript $n$ because the $k$-th receptive field, the vector $\vect{w}^{(n)}_k$, approximates the variations of the $k$-th response at the last stage $n$,
\begin{equation}
       \Delta \vect{x}^{n}_k = {\vect{w}^{(n)}_k}^\top \cdot \Delta \vect{x}^0 
       \label{resp_recept_field}
\end{equation}
in other words, $\nabla \mathcal{S}$ may be thought as consisting of a set of local receptive fields:
\begin{equation}
   \nabla \mathcal{S}(\vect{x}^0_A) = \left[\begin{array}{c}
                                            ( \;\;\;\;\;\;\;\;\;\;\;\; \vect{w}^{(n)}_1(\vect{x}^0_A)^\top \;\;\;\;\;\;\;\;\;\;\;\;)\\
                                             \vdots \\
                                             ( \;\;\;\;\;\;\;\;\;\;\;\; \vect{w}^{(n)}_k(\vect{x}^0_A)^\top \;\;\;\;\;\;\;\;\;\;\;\;)\\
                                             \vdots \\
                                            ( \;\;\;\;\;\;\;\;\;\;\;\; \vect{w}^{(n)}_{d_n}(\vect{x}^0_A)^\top \;\;\;\;\;\;\;\;\;\;\;\;)\\
                                         \end{array}
                                         \right]
  \label{nabla_S_sensors}
\end{equation}
If $S$ does not preserve the dimensionality, $\nabla \mathcal{S}$ is not square, but rectangular, being either \emph{fat} (when reducing the dimensionality, $d_n < d_0$) or \emph{tall}, when $d_n > d_0$.

Note that these receptive fields, $\vect{w}^{(n)}_k$, live in the input (image) domain: the dot-product in Eq. \ref{resp_recept_field} is between an image $\Delta \vect{x}^0$, and an image-like vector $\vect{w}^{(n)}_k$. The elements of the $k$-th receptive field, $\vect{w}^{(n)}_{k \, l}$, indicate how the irradiance in the spatio-spectral location $l$ affects the response of the $k$-th sensor.

In this context, these receptive fields are called \emph{local} because they provide a valid linear approximation in a small neighborhood of $\vect{x}^0_A$ (according to Eq. \ref{linear_approx}).
Here (in an image space of dimension $d_0$) \emph{local} does not necessarily means having a \emph{small spatial size}\footnote{\label{local_vs_local}The concept of \emph{neighborhood} or \emph{locality} in the image space of dimension $d_0$ has not to be confused with spatial proximity in a two-dimensional image. Later on we will also use \emph{local} in this more intuitive fashion when referring to spatially small receptive fields or effects with small spatial extent. Note that spatially close \emph{pixels} may not be close dimensions (or close directions in the image space) depending on the rearrangement pattern used in Eq. \ref{the_image} (see discussion in footnotefootnote\footref{arrangement}).}.
On the contrary, here \emph{local} means \emph{background-dependent} or \emph{adaptive}.
The orientation and length of these receptive fields in the image space of dimension $d_0$ (i.e. their shape, energy, contrast or gain) may change depending on the background image.
\subsection{Chain-rule: \\ Network Jacobian in terms of elementary Jacobians}
The Jacobian of the composition of functions of several variables
(e.g. the multi-layer architecture we have here), can be decomposed as the product of the individual Jacobian matrices. For example:
\begin{equation}
     \nabla_{\!\!x} \, f \Big(g\big(h(x)\big)\Big) = \nabla_{\!\!g} \, f \cdot \nabla_{\!\!h} \, g \cdot \nabla_{\!\!x} \, h
     \nonumber
\end{equation}
where the derivation variable is explicitly indicated as a subindex at each derivative symbol.
Whenever it is possible we will omit the derivation variable for simplicity.
Note that in this \emph{matrix chain-rule} the order is important for obvious reasons.
\begin{equation}
      \nabla S = \nabla S^{(n)} \cdot \nabla S^{(n-1)} \cdot \ldots \cdot \nabla S^{(i)} \cdot \ldots \cdot \nabla S^{(2)} \cdot \nabla S^{(1)} = \prod_{i=n}^{1} \nabla S^{(i)}
      \label{chain_rule}
\end{equation}
where $\nabla S^{(i)} = \partial \vect{x}^{i}/\partial \vect{x}^{i-1}$. Similarly to $\nabla S$, in general $\nabla S^{(i)}$ is point-dependent and rectangular. The general considerations made on the following sections for $\nabla S$ will also apply for each $\nabla S^{(i)}$.
\subsection{Elementary Jacobian matrix.}
Here we explicitly show the convenience of the matrix form in Eqs. \ref{linear} and \ref{divisive_norm} to compute the Jacobian matrix.
The derivatives of Eq. \ref{divisive_norm} are illustrative because the procedure can be extended to variations of the normalization (as shown in Section \ref{forward}).

First, we have:
\begin{equation}
      \nabla S^{(i)}(\vect{x}^{i-1}) = \nabla N^{(i)}(\vect{y}^i) \cdot L^{(i)}
      \label{elementary_jacobian}
\end{equation}
where $L^{(i)}$ may be rectangular (it determines the number of mechanisms, $d_i$), i.e. $L^{(i)} \in \mathbb{R}^{d_i \times d_{i-1}}$,
but point-independent. Note that $\nabla N^{(i)}$ is point-dependent and square (it describes the nonlinear interaction between the $d_i$ linear mechanisms), i.e. $\nabla N^{(i)} \in \mathbb{R}^{d_i \times d_i}$.


Then, derivation of Eq. \ref{divisive_norm}, using the diagonal matrix notation for the Haddamard products/quotients, $D_{(a)} = \texttt{diag(a)}$, leads to$^5$, 
\begin{equation}
      \hspace{-1.4cm}
      \nabla N^{(i)}(\vect{y}^i) =  D_{\textrm{sign}(\vect{y}^i)} \cdot
      \left[
      K^{(i)} \cdot D_{\left(\frac{\gamma^i |\vect{y}^i|^{\gamma^i -1}}{\mathcal{D}^{(i)}(a)} \right)} - D_{\left(\frac{K^{(i)} \cdot |\vect{y}^i|^{\gamma^i}}{{\mathcal{D}^{(i)}(a)}^2}\right)} \cdot H^{(i)} \cdot D_{\left(\gamma^i |\vect{y}^i|^{\gamma^i -1}\right)}
      \right] \cdot D_{\textrm{sign}(\vect{y}^i)}
      \label{deriv_DN}
\end{equation}

 Eq. \ref{deriv_DN} shows that the Jacobian, $\nabla N^{(i)}$, depends on the subtraction of two matrices,
 where the first one is \emph{diagonal} if the linear gain $K^{(i)}$ is scalar or point-wise,
 and the second one depends on $H^{(i)}$, the matrix describing the interaction between the intermediate linear responses.
 Note that the role of the interaction is \emph{subtractive}, i.e. it reduces the slope (for positive $H^{(i)}$).
In situations where there is no interaction between the different coefficients of $\vect{y}^i$, $H^{(i)}_{kl}=0 \,\,\, \forall k\neq l$, the resulting $\nabla N^{(i)}$ is point-dependent, but diagonal.

Eq. \ref{deriv_DN} also shows that the sign of the linear coefficients has to be considered \emph{twice} (through the multiplication by the diagonal matrices at the left and right). This detail (which, for instance, is relevant in MAD gradient descent), was not properly addressed in previous reports of the Jacobian (e.g. in \cite{Malo06,Laparra10a,Malo10}) because they were focused on properties which are independent of the sign (diagonal nature, effect on the metric, and determinant respectively).
\subsection{Inverse and Jacobian.}\label{inv_and_jacobian}
The possibility to obtain the input stimulus from the output response (the invertibility of $S$) depends on the Jacobian, $\nabla S$. This is easy to see in the dimension preserving case ($d_n = d_0$).

By definition, the inverse, $S^{-1}$, is the integral of the derivative:
\begin{equation}
\vect{x}^0 \,\,\,=\,\,\, S^{-1}(\vect{x}^n) \,\,\,=\,\,\, \vect{x}^0_A + \int_{\vect{x}^n_A}^{\vect{x}^n} \nabla S^{-1}(\vect{x}^n_*) \cdot dx^{n}_*
\nonumber
\end{equation}
where we assume we know the stimulus-response correspondence at some point denoted by the subscript $A$, i.e.
for certain $\vect{x}^n_*= \vect{x}^n_A$ we know the inverse $\vect{x}^0_*= \vect{x}^0_A$. We select this point as the origin of the integral. This point which may be as trivial as zero-response for zero-stimulation.
\footnote{\label{proof_deriv_DN} \noindent $^5$ A note on the derivative of divisive normalization. Explicitly considering the derivation variable and the separation in sign and absolute value, we have:
\begin{alignat}{2}
      \nonumber
      \nabla_{\!\!x} N^{(i)} & = \nabla_{\!\!x} \textrm{sign}(\vect{y}^i) \cdot D_{\mathcal{N}^{(i)}(a)} + D_{\textrm{sign}(\vect{y}^i)} \cdot \nabla_{\!\!x} \mathcal{N}^{(i)}(a) &   \\ \nonumber
        & \quad \;\;  \text{where,} \\ \nonumber
        & \quad \;\; \nabla_{\!\!x} \textrm{sign}(\vect{y}^i) \cdot D_{\mathcal{N}^{(i)}(a)} = 0, \,\,\, \text{since sign($\vect{x}^i_{L \,\, k}$) is constant  $\forall \vect{x}^i_{L \,\, k} \neq 0$} \\  \nonumber
        & \quad \;\; \nabla_{\!\!x} \mathcal{N}^{(i)}(a) = \nabla_{\!\!a} \mathcal{N}^{(i)}(a) \cdot \nabla_{\!\!x} a & \\   \nonumber
        & \quad \;\;\;\; \quad \;\;\;\; \quad \;\;\;\;\;\;\;\; \text{where,} \\  \nonumber
        & \quad \;\;\;\; \quad \;\;\;\; \quad \;\;\;\;\;\;\;\; \nabla_{\!\!a} \mathcal{N}^{(i)}(a) = \nabla_{\!\!a} K^{(i)} \!\!\cdot\! a \,\, \cdot D_{\left(\frac{1}{\mathcal{D}^{(i)}(a)} \right)} + D_{K^{(i)} \cdot a} \cdot \nabla_{\!\!a} \, \frac{1}{\mathcal{D}^{(i)}(a)} \\  \nonumber
        & \quad \;\;\;\; \quad \;\;\;\; \quad \;\;\;\;\;\;\;\; \qquad \qquad \text{where,} \\  \nonumber
        & \quad \;\;\;\; \quad \;\;\;\; \quad \;\;\;\;\;\;\;\; \qquad \qquad \nabla_{\!\!a} K^{(i)} \!\!\cdot\! a = K^{(i)} \\  \nonumber
        & \quad \;\;\;\; \quad \;\;\;\; \quad \;\;\;\;\;\;\;\; \qquad \qquad \nabla_{\!\!a} \, \frac{1}{\mathcal{D}^{(i)}(a)} = -D_{\frac{1}{{\mathcal{D}^{(i)}(a)}^2}} \cdot \nabla_{\!\!a} \mathcal{D}^{(i)}(a) = -D_{\frac{1}{{\mathcal{D}^{(i)}(a)}^2}} \cdot H^{(i)}  \\  \nonumber
        & \quad \;\;\;\; \quad \;\;\;\; \quad \;\;\;\;\;\;\;\; \qquad \qquad \text{therefore,} \\  \nonumber
        & \quad \;\;\;\; \quad \;\;\;\; \quad \;\;\;\;\;\; \qquad \qquad = K^{(i)} \cdot D_{\left(\frac{1}{\mathcal{D}^{(i)}(a)} \right)} - D_{K^{(i)} \cdot a} \cdot D_{\frac{1}{{\mathcal{D}^{(i)}(a)}^2}} \cdot H^{(i)} \\  \nonumber
        & \quad \;\;\;\; \quad \;\;\;\; \quad \;\;\;\;\;\; \nabla_{\!\!x} a = \nabla_{\!\!|x|} a \cdot \nabla_{\!\!x} |x| \\  \nonumber
        & \quad \;\;\;\; \quad \;\;\;\; \quad \;\;\;\;\;\;\;\; \qquad \qquad \text{where,} \\  \nonumber
        & \quad \;\;\;\; \quad \;\;\;\; \quad \;\;\;\;\;\;\;\; \qquad \qquad \nabla_{\!\!|x|} a = D_{\gamma^i |\vect{y}^i|^{\gamma^i -1}} \\  \nonumber
       & \quad \;\;\;\; \quad \;\;\;\; \quad \;\;\;\;\;\;\;\; \qquad \qquad \nabla_{\!\!x} |x| = D_{\textrm{sign}(\vect{y}^i)}, \,\, \text{since the slope of $|x|$ is sign($x$)} \\  \nonumber
        & \quad \;\;\;\; \quad \;\;\;\; \quad \;\;\;\;\;\;\;\; \qquad \qquad \text{therefore,} \\  \nonumber
        & \quad \;\;\;\; \quad \;\;\;\; \quad \;\;\;\;\;\; \qquad \qquad = D_{\gamma^i |\vect{y}^i|^{\gamma^i -1}} \cdot D_{\textrm{sign}(\vect{y}^i)} \\  \nonumber
        & \quad \;\;\;\; \quad \;\;\;\; \quad \; \text{therefore,} \\  \nonumber
        & \quad \;\;\;\; \quad \;\;\;\; \quad \; = \left[ K^{(i)} \cdot D_{\left(\frac{1}{\mathcal{D}^{(i)}(a)} \right)} - D_{K^{(i)} \cdot a} \cdot D_{\frac{1}{{\mathcal{D}^{(i)}(a)}^2}} \cdot H^{(i)} \right] \cdot D_{\gamma^i |\vect{y}^i|^{\gamma^i -1}} \cdot D_{\textrm{sign}(\vect{y}^i)} \\  \nonumber
        & \text{therefore, putting all the pieces together, we have Eq. \ref{deriv_DN}}.
\end{alignat}
}

If we don't know the analytic inverse ($S^{-1}$ and $\nabla S^{-1}$ unknowns), when $d_n = d_0$ Jacobian matrices are square (and eventuallyinvertible) so we can use this relation between Jacobians \cite{Spivak}: $\nabla S^{-1}(\vect{x}^n_*) = \nabla S(\vect{x}^0_*)^{-1}$, where $\vect{x}^0_* = S^{-1}(\vect{x}^n_*)$.

Therefore, if $\nabla S$ is known, we may compute the unknown $S^{-1}$ by solving this differential equation (or \emph{initial value problem}), $d\vect{x}^0_* = \nabla S(\vect{x}^0_*)^{-1} \cdot d\vect{x}^n_*$
\begin{equation}
\vect{x}^0 \,\,\,=\,\,\, S^{-1}(\vect{x}^n) \,\,\,=\,\,\, \vect{x}^0_A + \int_{\vect{x}^n_A}^{\vect{x}^n} \nabla S(\vect{x}^0_*)^{-1} \cdot dx^{n}_*
\label{inverse_through_integral}
\end{equation}
where the \emph{initial value} of $\vect{x}^0_*$ is known at an arbitrary origin (or \emph{initial conditions}, $\vect{x}^n_A$), and $\vect{x}^0_*$ is updated along the integration path in this way: for a new point in the path ($\vect{x}^n_{*'}= \vect{x}^n_* + d\vect{x}^n_{*'}$), we take,
\begin{equation}
        \vect{x}^0_{*'}= \vect{x}^0_* + \nabla S(\vect{x}^0_*)^{-1} \cdot d\vect{x}^n_{*'}
        \nonumber
\end{equation}
The theoretical consequence of Eq. \ref{inverse_through_integral} is that a vision model is invertible
if the Jacobian is nonsingular $\forall \,\, \vect{x}^0$.
The chain-rule decomposition implies that this condition applies to every stage: every elementary Jacobian, $\nabla S^{(i)}$, has to be nonsingular.
When dealing with a specific vision model (e.g. the one in Section \ref{forward}), invertibility implies certain restrictions in the parameters (at every stage of the model).

Another theoretical advantage of this Jacobian matrix view of invertibility is that information loss reduces to matrix algebra: the determination of the null subspaces of matrices $\nabla S^{(i)}$ (directions with small singular value).

Taking into account that $\nabla N^{(i)}$ are square and $L^{(i)}$ may be rectangular,
the above \emph{inverse-through-integration} approach may always be used for the nonlinear stages
(where matrices in the integral are square) and the inverse of the linear stages could be addressed using the pseudoinverse when the dimension is not preserved.

In practice, this \emph{inverse-through-integration} idea can be applied to actual vision models through techniques such as Runge-Kutta integration \cite{Numerical92},
as shown in Section \ref{inverse}.

\subsection{Optimal features: stimuli that isolate the response of a sensor. Definitions from inverse and Jacobian.}
Using carefully crafted stimuli to isolate the response of a specific mechanism is a common practice in visual psychophysics \cite{isolation}. Isolating the response means generating a particular image that stimulates a single sensor (e.g. the $k$-th sensor at the $n$-th stage) and elicits no response in the other sensors, i.e. $x^n_k \neq 0$ while $x^n_{k'}=0 \;\; \forall k'\neq k$. This is (proportional to) a Dirac delta response at the $n$-th layer centered at coefficient $k$: $\delta^n_{k-k'}$. Given the general nonlinearity of the system, this \emph{sensor-isolating stimulus} may depend on the background image, i.e. different backgrounds may require different stimuli to elicit isolated variation of response at the $k$-th sensor. As a result, it is better to use an incremental definition and take a specific background, $\vect{x}^0_A$, as a reference. If we refer to this incremental sensor-isolating stimulus as $\vect{f}^{(n)}_k(\vect{x}^0_A)$, its definition implies,
\begin{equation}
  \xymatrixcolsep{4pc}
  \xymatrix{ \vect{x}^0_A + \vect{f}^{(n)}_k(\vect{x}^0_A)  \,\,\,\, \ar@/^1pc/[r]^{  \scalebox{1.1}{\emph{S}}  }  & \,\,\,\, \vect{x}^n_A + \delta^n_{k-k'} \ar@/^1pc/[l]^{ \scalebox{1.1}{\emph{S}}^{-1} }
  }
  \label{isolatioN^{(i)}nv}
\end{equation}
\noindent namely, given a certain background image $\vect{x}^0_A$, the stimulus $\vect{f}^{(n)}_k(\vect{x}^0_A)$ is the distortion or increment that has to be applied to $\vect{x}^0_A$ such that at the $n$-th stage the only sensor that modifies its response is the $k$-th sensor.

According to this definition, the inverse analyzed above can be used to compute these stimuli for any background and the appropriate delta:
\begin{equation}
      \vect{f}^{(n)}_k(\vect{x}^0_A) =  \left[S^{-1}\left(S(\vect{x}^0_A) + \delta^n_{k-k'}\right)\right] - \vect{x}^0_A
      \label{feature_inverse}
\end{equation}
This expression cannot be further simplified because of the nonlinearity of $S$.
However, in the local linear approximation, Eq. \ref{linear_approx}, the variation in the response is mediated by the Jacobian. As a result,
\begin{equation}
      \vect{f}^{(n)}_k(\vect{x}^0_A) \approx \nabla S(\vect{x}^0_A)^+ \cdot \delta^n_{k-k'}
      \label{feature_inverse_jacobian}
\end{equation}
The Dirac delta selects the $k$-th column in $\nabla S(\vect{x}^0_A)^+ \in \mathbb{R}^{d_0 \times d_n}$. Note that these columns, $\vect{f}^{(n)}_k$, are vectors that live in the input (image) domain.

Similarly to the view of the Jacobian as a set of linear receptive fields, Eq. \ref{nabla_S_sensors}, the (pseudo)inverse of the Jacobian may be seen as composed by a set of \emph{sensor-isolating stimuli}:
\begin{equation}
   \hspace{-1.5cm}
   \nabla S(\vect{x}^0_A)^{+} = \left[\begin{array}{cccccc}

                                    \left(\begin{array}{c}
                                           \\

                                           \\
                                           \\
                                           \\
                                           \\
                                           \!\!\!\vect{f}^{(n)}_1(\vect{x}^0_A) \!\!\! \\
                                           \\
                                           \\
                                           \\

                                           \\
                                           \\
                                    \end{array}
                                    \right)
                                           &

                                    \left(\begin{array}{c}
                                           \\
                                           \\
                                           \\

                                           \\
                                           \\
                                           \!\!\! \vect{f}^{(n)}_2(\vect{x}^0_A) \!\!\! \\

                                           \\
                                           \\
                                           \\
                                           \\
                                           \\

                                    \end{array}
                                    \right)

                                           & \; \cdots \;&

                                    \left(\begin{array}{c}
                                           \\
                                           \\
                                           \\
                                           \\
                                           \\

                                           \!\!\! \vect{f}^{(n)}_k(\vect{x}^0_A) \!\!\! \\
                                           \\
                                           \\
                                           \\
                                           \\
                                           \\

                                    \end{array}
                                    \right)

                                           & \; \cdots \;&

                                    \left(\begin{array}{c}
                                           \\
                                           \\
                                           \\

                                           \\
                                           \\
                                           \!\!\! \vect{f}^{(n)}_{d_n}(\vect{x}^0_A) \!\!\! \\
                                           \\
                                           \\
                                           \\
                                           \\
                                           \\

                                    \end{array}
                                    \right)

                                         \end{array}
                                         \right]
  \label{nabla_S_features}
\end{equation}
The $k$-th sensor of the system is distinctly stimulated by $\vect{f}^{(n)}_k$. Therefore, these \emph{sensor-isolating stimuli} can also be referred to as \emph{optimal features}: each separate sensor is distinctly tuned to (or distinctly stimulated by) certain
image feature. The response of the sensor with receptive field, $\vect{w}^n_{k'}$, acting on the feature, $\vect{f}^{(n)}_k$, is either 0 or 1:
\begin{equation}
      \delta^n_{k-k'} = {\vect{w}^{(n)}_{k'}}^\top \cdot\vect{f}^{(n)}_k
      \label{filter_on_feature}
\end{equation}
given that $\nabla S \cdot \nabla S^+ = \mathbbm{1}$.

Since $\Delta \vect{x}^0 = \nabla S(\vect{x}^0)^{+} \cdot \Delta \vect{x}^n$ in a neighborhood of $\vect{x}^0_A$, the set of features $\big\{\vect{f}^{(n)}_k\big\}_{k=1}^{d_n}$
is a basis of the image space in a neighborhood of $\vect{x}^0_A$ because every $\Delta \vect{x}^0$ can be expressed as a linear combination of
sensor-isolating features, the vectors $\vect{f}^{(n)}_k$, each one weighted by the corresponding scalar coefficient, $\Delta x^n_k$:
\begin{equation}
      \Delta \vect{x}^0 = \sum_{k=1}^{d_n} \vect{f}^{(n)}_k \; \Delta x^n_k
\end{equation}
Therefore, Eqs. \ref{nabla_S_sensors} and \ref{nabla_S_features} can be seen as \emph{change-of-basis} matrices.
The \emph{features} are the basis vectors used to synthesize any image, and the \emph{receptive fields} are the
analysis functions used to obtain the coefficients in the transformed
representation\footnote{This \emph{analysis}/\emph{synthesis} view of \emph{receptive fields} and \emph{features}
is consistent with the notation in the wavelet and filterbank literature \cite{Simoncelli90} or in the \emph{linear} Independent Component Analysis literature \cite{Hyvarinen01,Hyvarinen09}, and more generally in linear feature extraction. Filters and features are the same if the set of basis function is orthogonal (i.e. if $\nabla S^{-1} = \nabla S^\top$).
}.

\subsection{Adaptive receptive fields and features at $i$-th layer}
The receptive field definition in Eq. \ref{recept_field} applies to sensors at the last stage (the $n$-th stage).
Taking into account the modular structure, these sensors, $\vect{w}^{(n)}_k$, are affected by all the transforms along
the pathway, i.e. by $S^{(i)}$ with $i=1,2, \ldots, n$.

However, following physiology, one could think on equivalent sensors measured or defined at previous stages (think of sensors at LGN versus sensors at V1). We could consider sensors at the $i$-th layer: $\vect{w}^i_k$. In this case one would have $d_i$ of such sensors, i.e. $k = 1,2, \ldots, d_i$. These sensors would be affected only by stages $S^{(1)}, S^{(2)}, \ldots, S^{(i)}$, with $i<n$.
Similarly to the receptive fields for the whole system, $\vect{w}^i_k$ can be defined from the $k$-th row of the Jacobian matrix \emph{up to the} $i$-th \emph{stage}, i.e. from $\prod_{j=i}^1 \nabla S^{(j)} \in \mathbb{R}^{d_i \times d_0}$,
\begin{equation}
       \vect{w}^{(i)}_k(\vect{x}^0_A) = \left( \left( \prod_{j=i}^1 \nabla S^{(j)} \right)_{k \, 1}, \left( \prod_{j=i}^1 \nabla S^{(j)} \right)_{k \, 2}, \ldots, \left( \prod_{j=i}^1 \nabla S^{(j)} \right)_{k \, d_0} \right)^\top
       \label{recept_field_i}
\end{equation}
\noindent where we omitted the point-dependence of the Jacobian matrices $\nabla S^{(j)}(\vect{x}^0_A)$ for clarity.

Similarly to Eq. \ref{resp_recept_field}, these local receptive fields describe the variations of the response at the $i$-th layer for a
given background, $\vect{x}^0_A$:
\begin{equation}
     \Delta \vect{x}^i_k = \vect{w}^{(i)}_k(\vect{x}^0_A)^\top \cdot \Delta \vect{x}^0
     \nonumber
\end{equation}
Similarly to Eq. \ref{nabla_S_sensors}, the Jacobian up to the $i$-th stage be thought as composed by the above receptive fields.

Similarly to the application of the receptive field concept to different layers, the concept of sensor-isolating stimulus (or optimal feature) can also be applied to previous stages. One may consider $\vect{f}^i_k(\vect{x}^0_A)$: the distortion over $\vect{x}^0_A$
so that the only sensor that modifies its response at the $i$-th stage is the $k$-th sensor.
As in the case of the features for the $n$-th stage, one may consider two kinds of definitions for the features of the $i$-th stage:
a \emph{general} definition based on the inverse (similar to Eq. \ref{feature_inverse}), and a \emph{restricted} definition
based on the local-linear approximation (similar to Eq. \ref{feature_inverse_jacobian}).
In the general case $S$ and $S^{-1}$ have to be trivially substituted by the corresponding transforms \emph{up to the $i$-th layer}:
 \begin{equation}
      \hspace{-0.7cm}
      \vect{f}^{(i)}_k(\vect{x}^0_A) =  \left[ \left(S^{(i)} \circ S^{(i-1)} \circ \cdots \circ S^{(1)}\right)^{-1} \left( S^{(i)} \circ S^{(i-1)} \circ \cdots \circ S^{(1)} (\vect{x}^0_A) + \delta^i_{k-k'}\right)\right] - \vect{x}^0_A
      \label{feature_inverse_i}
\end{equation}
Following the above, the features $\vect{f}^{(i)}_k(\vect{x}^0_A)$ can be defined from the columns of $\left( \prod_{m = i}^1 \nabla S^{(j)}(x^m_A) \right)^{+}$.
In any case, similarly to Eq. \ref{filter_on_feature}, the response of the sensor with receptive field, $\vect{w}^i_{k'}$, acting on the feature, $\vect{f}^{(i)}_k$, is either 0 or 1:
\begin{equation}
      \delta^{(i)}_{k-k'} = {\vect{w}^{(i)}_{k'}}^\top \cdot \vect{f}^{(i)}_k
      \nonumber
\end{equation}

\subsection{Perceptual distance, discrimination regions and Jacobian.}

In this input-output setting of vision, perceptual decisions (e.g. discrimination between stimuli) will be made on the basis of the available information in the response space and not in the input space. This is consistent with (1) the psychophysical practice that relates incremental thresholds with the slope of the response under the assumption of Euclidean discrimination in the response domain \cite{Brainard05,Laparra12}, and (2) the formulation of subjective distortion metrics as Euclidean measures in the response domain \cite{Teo94a,Epifanio03,Malo06a,Laparra10a}.

\subsubsection*{Perceptual distance} Following this, the perceptual distance, $d_p$, between two images, $\vect{x}^0_A$ and $\vect{x}^0_B$, can be defined as the \emph{Euclidean distance in the response domain}\footnote{\label{summation}A note on summation. Euclidean distance not only implies equal relevance of all the dimensions in the response domain (which is a quite sensible assumption), but also implies a \emph{quadratic summation} of the distortions in each dimension.
Different summation exponents affect the relative relevance of distortions which are either localized in a small set of components or extended over multiple dimensions of the representation \cite{Osher92,total_variation_example}.
Localized distortions are amplified by high exponents ($>2$) while extended distortions are amplified by low exponents ($<2$).
Quadratic summation has been questioned (or modified to improve correlation with subjective distortion judgements) \cite{General_on_summation,Malo10,cita_Bovik_Wang}.
Nevertheless, for mathematical simplicity, we will assume quadratic summation because it simplifies (1) the expressions of the 2nd order metric in different spaces or layers, and (2) the process of extending
patch-wise results to big visual field images \cite{TechReportBarna}. On the positive side, note that keeping quadratic summation
means focusing on the point-dependence, size and orientation of the discrimination regions more than on
their convexity (which are more relevant to characterize sensitivity over the whole image space).}:
\begin{equation}
      d_p(\vect{x}^0_A, \vect{x}^0_B) = \left|\vect{x}^n_B - \vect{x}^n_A\right|_2 = \sqrt{\left(\vect{x}^n_B - \vect{x}^n_A\right)^\top \cdot \left(\vect{x}^n_B - \vect{x}^n_A\right)} = \sqrt{\Delta {\vect{x}^n}^\top \cdot \Delta {\vect{x}^n}}
      \label{distance}
\end{equation}

For nontrivial systems, an Euclidean distance in the response domain implies a \emph{quite} non-Euclidean measure
in the input image domain. One may imagine that, for nontrivial $S^{-1}$, the inverse of the points in the sphere
of radius $|\Delta \vect{x}^n|_2$ around the point $\vect{x}^n_A$ will no longer be a sphere (not even a convex region!) in
the input space. The size and orientation of these \emph{discrimination regions} determine the visibility of distortions $\Delta \vect{x}^0$
on top of certain background image, $\vect{x}^0_A$: different Euclidean lengths in the image space (different $|\Delta \vect{x}^0|_2$) will be
required in different directions to lead to the same perceptual distance $d_p$.

The variety of orientations and sizes of the well-known Brown-MacAdam color discrimination regions \cite{Brown49} is an intuitive (just 3-d) example of the above.

The local-linear approximation through the Jacobian allows a more specific formalization of the rich geometrical behavior suggested above. Under that approximation (which is standard in differential geometry \cite{Dubrovin82}), the general non-regular discrimination regions are approximated by \emph{ellipsoids}.

In the local-linear approximation\footnote{Remember it implies, $\Delta \vect{x}^n = \nabla S(\vect{x}^0_A) \cdot \Delta \vect{x}^0$, where $\Delta \vect{x}^0 = \vect{x}^0_B - \vect{x}^0_A$}, the (square of the) distance reduces to:
\begin{equation}
      d_p(\vect{x}^0_A, \vect{x}^0_A + \Delta \vect{x}^0)^2  = \Delta {\vect{x}^0}^\top \cdot \nabla S(\vect{x}^0_A)^\top \cdot \nabla S(\vect{x}^0_A) \cdot \Delta {\vect{x}^0}
      \label{distance_2nd}
\end{equation}

\subsubsection*{Perceptual metric} Following the \emph{metric matrix} concept in differential geometry\footnote{\label{note_jetric}Metric matrix. In differential geometry \cite{Dubrovin82}, the matrix that combines the distortions in the different dimensions is called \emph{metric matrix}, $M(x)$:
\begin{equation}
       d(x,x+\Delta x)^2 = \Delta x^\top \cdot M(x) \cdot \Delta x = \sum_{k \, l} \Delta x_k \, M(x)_{k l} \, \Delta x_l
       \nonumber
\end{equation}
where $M(x)$ is a symmetric definite-positive matrix.
Note that the surface of points at a fixed distance $\tau$ from a given point $x$, defined by the equation $\Delta x^\top \cdot M(x) \cdot \Delta x = \tau^2$,
is an \emph{ellipsoid}. The properties of this ellipsoid are determined by $M(x) = R \cdot \Lambda \cdot R^\top$, where $R^\top$ is a rotation that determines the orientation, and $\Lambda$ is a diagonal matrix with the squares of the widths of the ellipsoid (i.e. $\Lambda$ determines the volume).
}
the metric induced by the model $S$ in the input domain assuming an identity matrix (Euclidean metric) in the output is:
\begin{equation}
      M^{(0)}(\vect{x}^0_A) = \nabla S(\vect{x}^0_A)^\top \cdot \nabla S(\vect{x}^0_A)
      \label{metric_nabla}
\end{equation}
Therefore, according to\footref{note_jetric}, relevant magnitudes in psychophysics such as (1) the size of the discrimination ellipsoids (or \emph{Just Noticeable Differences}, JNDs), and (2) the orientation of the ellipsoids (which distortions are more or less visible) also depend on the Jacobian.
The eigenvalues of $\nabla S^\top \cdot \nabla S$ determine the size of the ellipsoid and its eigenvectors, the orientation.

This local-linear metric concept can be applied at any representation domain along the modular structure: one may talk about $M^{(0)}$, $M^{(1)}$, etc... In particular, the psychophysically meaningful assumption we are doing can be formulated as: $M^{(n)} = \mathbbm{1}$, i.e. \emph{the metric in the last domain is the identity}. In the multilayer context, it is important to know how the metric changes from one representation to the other. Again one finds a dependence with the Jacobian:
\begin{equation}
      M^{(i-1)}(\vect{x}^{i-1}) = \nabla S^{(i)}(\vect{x}^{i-1})^\top \cdot M^{(i)}(\vect{x}^{i}) \cdot \nabla S^{(i)}(\vect{x}^{i-1})
      \label{change_jetric}
\end{equation}
where $\vect{x}^i = S^{(i)}(\vect{x}^{i-1})$. Note that Eq. \ref{metric_nabla} is just a particular case of Eq. \ref{change_jetric} for the whole system and the assumption $M^{(n)} = \mathbbm{1}$.

\subsubsection*{Optimization of perceptual distance. Derivative for gradient descent}
Applications involving the optimization of subjective distortion will depend on the derivatives of the above distortion. This includes engineering applications, but also psychophysical techniques such as Maximum Differentiation (MAD) \cite{Wang08,Malo15,Malo16}, in which the stimuli are designed to maximize/minimize the perceived distortion. Given a reference image, $\vect{x}^0_A$, MAD looks for the best $\vect{x}^0_B$ by starting from a guess and following the direction of the gradient of the perceptual distance. In our setting (Eq. \ref{distance}) this gradient vector depends on the Jacobian matrix\footnote{\label{demo_grad_distanc}\red{(\emph{Note!}) On the derivative of the distance.}}:
\begin{equation}
  \frac{\partial d_p(\vect{x}^0_A, \vect{x}^0_B)}{\partial \vect{x}^0_B} = \,\, \frac{1}{d_p} \,\, \nabla S(\vect{x}^0_B)^\top \cdot (\vect{x}^n_B - \vect{x}^n_A)
  \label{grad_distance}
\end{equation}

\subsection{Redundancy reduction and Jacobian.}
Appropriate characterisation of the redundancy of the signal at the different stages of the model is relevant in interpretations of the visual function in terms of information theory \cite{Barlow61,Olshausen96,Schwartz01,Barlow01,Malo06b,Malo10,Laparra15,Martinez16}
because the information transmitted through a network depends on the reduction of the redundancy \cite{Linsker82,Parga92}.

Measures of redundancy include the correlation between components (or the covariance matrix, a restricted, 2nd order, description), and the multi-information (a general description) \cite{Cardoso03}. Interestingly, the way these magnitudes change across the model also depend on the Jacobian.

The covariance of the signal around a certain stimulus\footnote{\label{covariance}Covariance. $C^{(i)}(\vect{x}^i_A) = \mathcal{E}\left[ (\vect{x}^i-\vect{x}^i_A) \cdot (\vect{x}^i-\vect{x}^i_A)^\top \right]$ (where $\mathcal{E}\left[ \cdot \right]$ stands for expected value). 2nd order relations are described by the nondiagonal nature of $C$ \cite{Cardoso03}.} at the $i$-th layer, $C^{(i)}(\vect{x}^i_A)$, depends on the covariance at the previous stage, $C^{(i-1)}(\vect{x}^{i-1}_A)$, and on the transform $S^{(i)}$. In the local-linear approximation, the local covariance changes as:
\begin{equation}
       C^{(i)}(\vect{x}^i_A) = \nabla S^{(i)}(\vect{x}^{i-1}_A) \cdot C^{(i-1)}(\vect{x}^{i-1}_A) \cdot \nabla S^{(i)}(\vect{x}^{i-1}_A)^\top
       \label{covariance_change}
\end{equation}
Note that this change of $C$ is consistent with the change of the metric, Eq. \ref{change_jetric}, for covariance-based metrics such as Mahalanobis, $M(x) = C(x)^{-1}$ \cite{Mahalanobis36}.

The multi-information\footnote{\label{multi_info}Multi-information.
$I(\vect{x}^i) = \textsf{D}(p(\vect{x}^i)|\prod_k p_k(\vect{x}^i_k))$ where \textsf{D}
is the KL-divergence between the joint PDF, $p(\vect{x}^i)$, and the product of the marginal PDFs $\prod_k p_k(\vect{x}^i_k)$, thus providing a measure of distance from the condition of statistical independence.}, decreases in this way under dimension preserving transforms $S^{(i)}$ \cite{Studeny98}:
\begin{equation}
      \Delta I^{(i)} = I(\vect{x}^{i-1}) - I(\vect{x}^i) = \sum_{k=1}^{d_{i-1}} h(x^{i-1}_k) - \sum_{k=1}^{d_{i}} h(x^{i}_k) + \mathcal{E}[ log |\nabla S^{(i)}| ]
      \label{multi_info_change}
\end{equation}
where $h(\cdot)$ is the entropy of the corresponding univariate variable.
This expression is interesting because a multivariate (and hence hard-to-estimate) quantity such as multi-information is reduced to univariate estimations (entropy from univariate histograms) and the average
of the Jacobian.

\section{Analytic inverse}
\label{inverse}

In section \ref{inv_and_jacobian}, Eq. \ref{inverse_through_integral}, we proposed a general  \emph{inversion-through-integration} approach that links inverse and (non-singular) Jacobian. However, one may also address the reconstruction of the input by inverting each individual linear-nonlinear transform. If the expression of the nonlinearity is simple enough, one may find an analytic inverse which does not depend on numerical integration of the Jacobian.

In this section we focus on the inversion of the divisive normalization (the nonlinear part of the elementary modules) because the linear part can be addressed by standard matrix inversion (eventually including pseudoinverse and regularization) \cite{Nemerical92,Golub}.

Analytic inversion of standard divisive normalization, Eq. \ref{divisive_norm}, is straightforward
using the diagonal matrix notation for the Haddamard product \cite{Malo06a},
\begin{equation}
       \vect{y}^i = {N^{(i)}}^{-1}(\vect{x}^i) = D_{\text{sign}(\vect{x}^i)} \cdot \left[ \left( \mathbbm{1} - {K^{(i)}}^{-1} \cdot D_{|\vect{x}^i|} \cdot H^{(i)}  \right)^{-1} {K^{(i)}}^{-1} \cdot D_{b^i} \cdot |\vect{x}^i| \right]^{\frac{1}{\gamma^i}}
       \label{inv_DN}
\end{equation}
where $[v]^{\frac{1}{\gamma^i}}$ is element-wise exponentiation of elements of the vector $v$. See the details of the inversion below\footnote{\label{proof_inverse}
A note on the analytic inverse of divisive normalization. The element-wise quotient in
Eq. \ref{divisive_norm} implies that the denominator in,
\begin{equation}
      |\vect{x}^i| = \frac{K^{(i)} \cdot a}{b^i + H^{(i)} \cdot a} \nonumber
\end{equation}
can be written in the left hand side as a Haddamard product,
\begin{equation}
       (b^i + H^{(i)} \cdot a) \odot |\vect{x}^i| = K^{(i)} \cdot a \nonumber
\end{equation}
therefore, using the diagonal matrix notation:
\begin{gather*}
      D_{b^i} \cdot |\vect{x}^i| + D_{|\vect{x}^i|} \cdot H^{(i)} \cdot a = K^{(i)} \cdot a \\
      {K^{(i)}}^{-1} \cdot D_{b^i} \cdot |\vect{x}^i| = \left( \mathbbm{1} - {K^{(i)}}^{-1} \cdot  D_{|\vect{x}^i|} \cdot H^{(i)} \right) \cdot a \\
      a = \left( \mathbbm{1} - {K^{(i)}}^{-1} \cdot  D_{|\vect{x}^i|} \cdot H^{(i)} \right)^{-1} \cdot {K^{(i)}}^{-1} \cdot D_{b^i} \cdot |\vect{x}^i|
\end{gather*}
and taking into account that $a = |\vect{y}^i|^{\gamma^i}$, it leads to Eq. \ref{inv_DN}.
}.
Finally, the signal representation at the previous layer can be obtained from the linear responses
through the inverse of the linear transform: $\vect{x}^{i-1} = {L^{(i)}}^+ \cdot \vect{y}^i$.

As anticipated by the generic inverse-through-integration approach based on $\nabla S^{-1}$ (Eq. \ref{inverse_through_integral}), here Eq. \ref{inv_DN} shows more specifically that in this linear-nonlinear architecture, \emph{inversion} reduces to \emph{matrix inversion}.
While the linear filtering operations, $L^{(i)}$ and $K^{(i)}$, may be inverted without the need of an explicit matrix inversion through surrogate signal representations (deconvolution in the Fourier or Wavelet domains), the inverse of the form $\left( \mathbbm{1} - \varepsilon \right)^{-1}$ in Eq. \ref{inv_DN} may pose numerical problems because there is no way to avoid it (see section \ref{inverse_example}).

\section{A specific four-layer model (I): \\ Forward transform and Jacobian matrices}
\label{forward}

The model considered for this illustration was originally intended to provide a psychophysically meaningful alternative to the \emph{modular concept} in Structural Similarity measures (SSIM) \cite{Wang04}.
Its authors suggest a separate consideration of luminance, contrast and structure (which is a sensible approach), but the definition of such factors has no obvious perceptual meaning in SSIM.

The idea for a \emph{more perceptual} alternative in \cite{Malo15} was addressing one psychophysical factor at a time (i.e. modular), by using a cascade of linear-nonlinear transforms:
\begin{labeling}{Layer $S^{(4)}$}
\item [Layer $S^{(1)}$] linear spectral integration to compute luminance and opponent tristimulus channels, and nonlinear brightness/color response.
\item [Layer $S^{(2)}$] definition of local contrast by using linear filters and divisive normalization.
\item [Layer $S^{(3)}$] linear LGN-like contrast sensitivity filter and nonlinear local contrast masking in the spatial domain.
\item [Layer $S^{(4)}$] linear V1-like wavelet decomposition and nonlinear divisive normalization to account for
orientation and scale-dependent masking.
\end{labeling}
Here we extend previous results by considering two extra layers (1-st and 4-th layers were not considered in \cite{Malo15}).
On top of its interpretability, the modular structure simplifies the use of MAD to set the free parameters by determining only one layer at a time. As a result, this model substantially improves the performance of SSIM and related models in image quality assessment.

As shown below, all layers involve a different kind of saturation/divisive normalization. Layers 1 and 2 have no analytic inverse, but while the nonlinearity in the 1st layer poses no major problem (it is a point-wise operation), the 2nd layer has to be inverted by using its (analytical) Jacobian in a (numerical) Runge-Kutta integration. Layer 3 and 4 are analytically invertible. However, while layer 3 poses no major problem to naively apply the analytic inverse in current computers (with reasonable image sizes), the redundant nature of the linear wavelet transform in the 4th stage makes matrices huge, and one is forced to apply other iterative inversion methods based on series expansions.
\subsection{First layer. Brightness: \\ Linear luminance and nonlinear Weber-like saturation}
This first layer addresses (1) the energy integration at the photoreceptors, (2) the separate processing of achromatic and opponent chromatic information, and (3) the nonlinear relations between (a) luminance and brightness, and (b) opponent tristimulus values and colorfulness. \red{(\emph{Note!}) Figure on basic color stimuli and perceptual dimensions of color.} This layer assumes no interaction between spatial locations. It only addresses the spectral information. The spatial meaning of the coefficients is not changed at all.

\subsubsection{Linear transform:}
In this case, the matrix $L^{(1)}$ describes linear energy integration at photoreceptors: (1) linear spectral integration in each spatial location, and (2) a transform to an opponent color space.
\begin{eqnarray}
    \vect{y}^1 = A \cdot T^\lambda \cdot \vect{x}^0
    \label{spectraL^{(i)}ntegration}
\end{eqnarray}
where $T^\lambda$ contains the spectral sensitivities (or color matching functions) for each spatial location, (e.g. functions tuned to long, medium and short wavelengths, LMS), and the matrix $A$ performs the LMS-to-opponent color space transform. This is a pure \emph{linear tristimulus colorimetry} transform \cite{Fairchild13} whose structure is determined by the (arbitrary) rearrangement operation\footref{arrangement}

\red{(\emph{Note!}) See Figure on the sensitivities and matrix structure. Include a note on Von-Kries adaptation: adaptivity in matrix $A$ could be seen as divisive normalization -the classical gray world assumption reduces to dividing by the average color of spatial neighbors-.}

%
%

\subsubsection{Nonlinear transform:}
Here the saturation is a simple exponential function with $\gamma^1 < 1$ and no interaction between neighbor dimensions (i.e. $K^{(1)} = 0$ and $\mathcal{D}^{(1)} = 1$),
\begin{equation}
    \vect{x}^1 = \textrm{sign}(y^1) \odot |y^1|^{\gamma^1(\,|y^1|\,)}
    \label{stage1}
\end{equation}
where all operations (sign, rectification, exponentiation) are dimension-wise.
However, note the exponent is a function of the magnitude of the input tristimulus value. Specifically,
\begin{equation}
      \gamma^1(|y^1|)  =  \gamma_H - (\gamma_H-\gamma_L)\cdot \dfrac{\mu^m}{(\mu^m + |y^1|^m)}
      \nonumber
\end{equation}
The exponent has different values for low and high inputs, $\gamma_L$ and $\gamma_H$ respectively. The transition of $\gamma^1$ between $\gamma_L$ and $\gamma_H$ happens around the value $y^1 = \mu$.
This transition is smooth, and its sharpness is controlled by the exponent $m$. \red{(\emph{Note!}) Figure on the nonlinear response ($\gamma$ on inset).}

This peculiar expression for $\gamma^1$ has statistical grounds since the resulting nonlinearity approximately equalizes the PDF of luminance values in natural scenes \red{(\emph{Note!}) \cite{MumfordXX,BertalmioYY}}, which is a sensible goal in the information maximization context \cite{Laughlin83}, also applicable to the opponent color channels \cite{MacLeod03b,Laparra12}.

\subsubsection{Jacobian:}
The Jacobian of the first stage is the product of three matrices:
\begin{equation}
   \nabla S^{(1)}(\vect{x}^0) =
   D_{ \Bigg(    \gamma^1 |y^1|^{\gamma^1 - 1}  + \frac{\partial \gamma^1}{\partial |y^1|} |y^1|^{\gamma^1} \ln (|y^1|) \Bigg)}
   \cdot A \cdot T^\lambda
   \label{deriv_stage1}
\end{equation}
where $D_{(v)}$ is a diagonal matrix with vector $v$ in the diagonal, the dependence of $\gamma^1$ has been omitted for simplicity, all the operations in the diagonal are dimension-wise\footnote{ Notes on the derivative:
\begin{itemize}
    \item Note the similarity of this diagonal matrix with Eq. \ref{deriv_DN} taking no normalization $\mathcal{D}^{(1)} = 1$ (and hence $H^{(1)} = 0$), and $K^{(1)} = 1$. The only difference here is the \emph{correction} term accounting for the dependence of $\gamma$ on $x$.
    \item \red{(\emph{Note!}) on the use of logarithm to compute the derivative of $x^{\gamma(x)}$. }
    \item Similarly to Eq. \ref{deriv_DN}, the Jacobian of $|y^1|^{\gamma^1}$ is multiplied (left and right) by the diagonal matrices with the sign. Nevertheless, the sign matrices cancel out if the matrix in the center is diagonal, as is the case here.
\end{itemize}\label{notes_deriv1}
}

, and the derivative of $\gamma^1$ is:
\begin{equation}
      \dfrac{\partial \gamma^1(|y^1|)}{\partial |y^1|} = (\gamma_H-\gamma_L)\cdot \dfrac{m \, |y^1|^{ (m-1) }\cdot  \mu^m}{ (\mu^m + |y^1|^{m})^2 }
      \nonumber
\end{equation}

\red{(\emph{Note!}) on the strong dimensionality reduction at stage 1: $L^{(1)}$ and $\nabla S^{(1)}$ are \emph{strongly fat} rectangular matrices}. Assuming the input signal is a \emph{tristimulus image}, not a \emph{hyperspectral image} (i.e. neglecting the spectral integration stage $T^\lambda$), this layer preserves dimension. This means $A$ and $\nabla \mathcal{N}^1$ are square matrices $\in \mathbb{R}^{d_1 \times d_1}$, with $d_1 = h \times w \times 3$ for the three spectral channels.
\subsection{Second layer: local brightness and local contrast}
The second layer accounts for the different perceptual relevance of local brightness and local contrast \cite{Burt83}. \red{(\emph{Note!}) Figure local brightness and local contrast components}. Here \emph{local} does mean the intuitive \emph{small spatial neighborhood} (in subtended degrees). \emph{Local brightness} at a certain point is computed by pooling the brightness values in neighbor locations. The \emph{contrast} concept implies subtracting the mean and scaling by the mean. Therefore, \emph{local contrast} is computed at each spatial point by subtracting the local brightness and dividing by the local mean. Computations of the local mean can be done by convolving the signal with the appropriate Gaussian kernel whose width determines the extent of the neighborhood. Convolution is equivalent to the application of a circulant matrix made of these Gaussian receptive fields.

Given the strong spatial localization of these Gaussian receptive fields, the meaning of the responses, $x^2_k$, emerging from this layer is still spatial (for each chromatic channel).

\subsubsection*{Linear-nonlinear transform.} As a result, relative weighting of the local brightness and contrast can be done by adding these two terms:

\begin{align}
      \vect{x}^{2}& = S^{(2)}(\vect{x}^1)= \alpha_1 \; H^{(2)}_A \cdot \vect{x}^1+\alpha_2\, \mathcal{N}(\vect{x}^1)\\ \label{Astage2}
      &=\alpha_1 \; H^{(2)}_A \cdot \vect{x}^1 + \alpha_2 \; \dfrac{K^{(2)} \cdot \vect{x}^{1}}{b^{2}+ H^{(2)}_C \cdot \vect{x}^{1}} \nonumber
\end{align}
where, as though out the work, the superscripts refer to the 1st and 2nd stages (they are not exponents), the first term corresponds to the local brightness since $H^{(2)}_A \cdot \vect{x}^1$ is a smoothed version of $\vect{x}^1$, and the second term corresponds to the contrast: note the subtraction of the local mean, $H^{(2)}_B \cdot \vect{x}^1$, in the numerator and the division by the local mean, $H^{(2)}_C \cdot \vect{x}^1$, in the denominator (plus an additive term, the vector $b^{2}$, to avoid singularities).
$H^{(2)}_A$, $H^{(2)}_B$, and $H^{(2)}_C$, are convolution-like matrices with Gaussian receptive fields of spatial width $\sigma_A$,  $\sigma_B$, $\sigma_C$ (measured in degrees).
This stage preserves dimension: $d_2 = d_1$, and matrices are square.

\red{(\emph{Note!}) Comment on the applicability to color channels: absolute value and sign separation required in opponent chromatic channels.}

\subsubsection*{Jacobian}
Using Eq. \ref{deriv_DN} and the associated discussion, we can see:
\begin{equation}
      \nabla S^{(2)}(\vect{x}^1) = \alpha_1 \; H^{(2)}_A + \alpha_2 \; 
      \left( K^{(2)}\cdot D_{\left( \frac{1}{\mathcal{D}^{(2)}(\vect{x}^1)} \right)} -
      D_{\left( \frac{K^{(2)} \cdot \vect{x}^1}{(\mathcal{D}^{(2)}(\vect{x}^1))^2} \right)} \cdot H^{(2)}_C \right)
      \label{deriv_stage2}
\end{equation}

\subsection{Third layer. \\ Frequency sensitivity and local masking}

The third layer accounts for the \emph{frequency-dependent sensitivity} (through the Contrast Sensitivity Function, CSF \cite{Campbell68,Mullen85}) and the \emph{local masking} (in certain spatial location $p$, or certain coefficient $k$)
due to high energy (or high contrast) of the image around this location (or coefficient) \cite{MejorCitaLocal,Watson02}).
\red{(\emph{Note!}) Figure: illustration of frequency sensitivity and local masking.}

\subsubsection*{Linear-nonlinear transform.} The linear stage, $L^{(3)}$, is a circulant convolution matrix, $L_{\text{CSF}}$, made of center-surround receptive fields obtained from the inverse Fourier transform of the CSF. The non-linear stage is just the standard Eq. \ref{divisive_norm} with no extra gain in the numerator ($K^{(3)}=\mathbbm{1}$):
\begin{eqnarray}
      \vect{y}^{3} &=& L_{\text{CSF}} \cdot \vect{x}^2 \\ \nonumber
      \vect{x}^3  &=&N^{(3)}(\vect{y}^{3})= \text{sign}(\vect{y}^{3})\odot \mathcal{N}^{(3)}(\vect{y}^{3})  = \text{sign}(\vect{y}^{3})\odot \frac{|\vect{y}^{3}|^{\gamma^3}}{b^3 + H^{(3)} \cdot |\vect{y}^{3}|^{\gamma^3}}
      \label{stage3}
\end{eqnarray}
\red{(\emph{Note!}) Figure: CSF and corresponding receptive fields.}

Given the strong spatial localization of the LGN-like receptive fields derived from the CSFs \cite{UriegasXX}, the meaning of the responses, $x^3_k$, emerging from this layer is still spatial (for each chromatic channel). The interaction kernel, $H^{(3)}$, describes how the activity in \emph{spatial} neighbors
attenuates each nonlinear sensor. Therefore, $H^{(3)}$ is a convolution-like matrix with Gaussian receptive fields of \emph{spatial} width $\sigma^3$ (measured in degrees).

Image representations $\vect{x}^1$, $\vect{x}^2$, and $\vect{x}^3$, are \emph{retinotopic}: i.e. despite the transformations, responses can still be considered as \emph{intensities of certain feature at each spatial location}. The difference with a conventional image is that the \emph{features} are not trivial Dirac deltas (irradiance at each point), but something else. \red{(\emph{Note!}) Figure: show examples of features $f^{(1)}_k$ = deltas of \emph{luminance-adaptive} height, $f^{(2)}_k$ = center surround, $f^{(3)}_k$ = center surround. Possibility: make this comment and unified exercise after introducing the 4-th layer.}.

In our implementation, this stage preserves dimension: $d_3 = d_2$ and matrices are square.
Here, for mathematical convenience, we assume that information loss is exclusively described by the band limitation introduced by the CSF and not by undersampling.
However, note that more realistic implementation of the retina-LGN path would imply a \emph{fat} rectangular  matrix $L_{\text{CSF}}$
combining band limitation and undersampling \cite{Martinez+Neuron+2014,Martinez-Garcia16}.

\subsubsection*{Jacobian.} Straightforward application of Eqs. \ref{elementary_jacobian} and \ref{deriv_DN} leads to:
\begin{equation}
      \hspace{-1.4cm}
      \nabla S^{(3)}(\vect{x}^2) =  D_{\textrm{sign}(\vect{y}^{3})} \cdot
      \left[
      D_{\left(\frac{\gamma^3 |\vect{y}^{3}|^{\gamma^3 -1}}{\mathcal{D}^{(3)}(|\vect{y}^{3}|)} \right)} - D_{\left(\frac{|\vect{y}^{3}|^{\gamma^3}}{{\mathcal{D}^{(3)}(|\vect{y}^{3}|)}^2}\right)} \cdot H^{(3)} \cdot D_{\left(\gamma^3 |\vect{y}^{3}|^{\gamma^3 -1}\right)}
      \right] \cdot D_{\textrm{sign}(\vect{y}^{3})} \cdot L_{\text{CSF}}
      \label{deriv_stage3}
\end{equation}

\subsection{Fourth layer. \\ Wavelet filters and masking in the wavelet domain}

The fourth layer accounts for the existence of band-pass spatially-local oriented filters in V1 \cite{Marcelja80,Watson87b,Watson87a} and its mutual inhibitory interactions responsible for frequency-dependent masking \cite{Carandini94,Foley94,Watson97}).
\red{(\emph{Note!}) Figure: evidence of oriented filters by adaptation \cite{Legge80} (check citation), and illustration of oriented masking}.

\subsubsection*{Linear-nonlinear transform.}
The linear stage, $L^{(4)}$, is made of wavelet-like receptive fields,
that is why we call this matrix $W$.
Following the overcompleteness of the image representation at V1 \cite{Olshausen13},
the wavelet matrix is usually taken to be \emph{tall}, so $d_4>d_3$, consisting of multiple staked circulant convolution matrices, one per \emph{subband}.

The non-linear stage is just the standard Eq. \ref{divisive_norm} with no extra gain in the numerator ($K^{(4)}=\mathbbm{1}$):
\begin{eqnarray}
      \vect{y}^4 &=& W \cdot \vect{x}^3\\ \nonumber
      \vect{x}^4  &=& N^{(4)}(\vect{y}^4) = \text{sign}(\vect{y}^4)   \odot \mathcal{N}^{(4)}(\vect{y}^4)= \text{sign}(\vect{y}^4)
       \odot \frac{|\vect{y}^4|^{\gamma^4}}{b^4 + H^{(4)} \cdot |\vect{y}^4|^{\gamma^4}}
      \label{stage4}
\end{eqnarray}
The linear operation $W$ introduces frequency meaning in the coefficients in such a way that the $\vect{x}^4$ representation is no longer retinotopic (or more specifically, it is not 3 retinoptopic representations, one per chromatic channel), but a set of multiple retinotopic representations, one per spatial subband (and chromatic channel).

Interaction in $H^{(4)}$ is more complicated than in the interaction kernels of previous layers. While $H^{(2)}_C$ and $H^{(3)}$ are plain convolutional matrices with \emph{spatial} Gaussian receptive fields, in $H^{(4)}$ the neighborhood defined in the kernel has spatial meaning (within a subband), but also frequency and orientation meaning (between subbands).

\subsubsection*{Jacobian.} Straightforward application of Eqs. \ref{elementary_jacobian} and \ref{deriv_DN} leads to:
\begin{equation}
      \hspace{-1.4cm}
      \nabla S^{(4)}(\vect{x}^3) =  D_{\textrm{sign}(\vect{y}^4)} \cdot
      \left[
      D_{\left(\frac{\gamma^4 |\vect{y}^4|^{\gamma^4 -1}}{\mathcal{D}^{(4)}(|\vect{y}^4|)} \right)} - D_{\left(\frac{|\vect{y}^4|^{\gamma^4}}{{\mathcal{D}^{(4)}(|\vect{y}^4|)}^2}\right)} \cdot H^{(4)} \cdot D_{\left(\gamma^4 |\vect{y}^4|^{\gamma^4 -1}\right)}
      \right] \cdot D_{\textrm{sign}(\vect{y}^4)} \cdot W
      \label{deriv_stage4}
\end{equation}

\red{(\emph{Note!}) Figure: Matrix $W$ and examples of receptive fields.}

\red{(\emph{Note!}) Figure: show adaptive receptive fields / features for layers 1-4 in different backgrounds
(with the synthetic image).}
\section{A specific four-layer model (II): Inverse}
\label{inverse_example}
\subsection*{Inverse of 4-th layer: expansion method.}
Given the increment in dimension implied by the wavelet transform $W$, the analytical inverse in Eq. \ref{inv_DN} is not feasible because $(\mathbbm{1} - D_{|\vect{x}^4|} \cdot H^{(4)})$ is huge\footnote{\label{steerable} Sensible wavelet models (as for instance steerable wavelets \cite{Simoncelli92,Simoncelli95} with 4 scales and 5 orientations), have a $7 \times$ overcompleteness. This implies that working with images as small as $100\times100$ pixel would imply inverting matrices of size $75000 \times 75000$.}

The numerical integration of the Jacobian also implies inverting matrices of the same size. Nevertheless, the series expansion of the analytical inverse avoids the need of inverting such huge matrices. Using $(\mathbbm{1} - \varepsilon)^{-1} = \sum_{p=0}^{\infty} \varepsilon_p$ one may approximate the inverse through a fast iterative process that only involves matrix-on-vector operations (not even exponential of matrices) \cite{Malo06a}:
\begin{align}
      a_{(0)} & = D_{b^4} \cdot |\vect{x}^4| \\ \nonumber
      a_{(j)} & = D_{b^4} \cdot |\vect{x}^4| + D_{|\vect{x}^4|} \cdot H^{(4)} \cdot a_{(j-1)}
      \label{iterat_inv_4}
\end{align}
where the subindex $(j)$ indicates the iteration, and after convergence of $a_{(j)}$, the  representation at 3rd stage is computed through the inverse wavelet transform:
\begin{equation}
       \vect{x}^3= W^+ \cdot D_{\text{sign}(\vect{x}^4)} \cdot a_{(j)}^{\frac{1}{\gamma^4}}.
       \label{inv_wavel_4}
\end{equation}

\subsection*{Inverse of 3-rd layer: analytical inverse.}
Dimension at this layer coincides with the input spatial size so this gives rise to matrices of moderate size and the analytic inverse, Eq. \ref{inv_DN}, can be applied:
\begin{equation}
       \vect{x}^2 = L_{\text{CSF}}^{-1} \cdot D_{\text{sign}(\vect{x}^3)} \cdot \left[ \left( \mathbbm{1} -  D_{|\vect{x}^3|} \cdot H^{(3)}  \right)^{-1} \cdot D_{b^3} \cdot |\vect{x}^3| \right]^{\frac{1}{\gamma^3}}
       \label{inv_stage_3}
\end{equation}

\subsection*{Inverse of 2-nd layer: Runge Kutta integration.}
The 2nd-stage considered here, \azul{Eq. 35},
does not have an obvious analytic inverse. However, since we know the analytical Jacobian and the dimension at this stage is moderate, we can use the inverse-through-integration of ${\nabla S^{(2)}}^{-1}$ described in section \ref{inv_and_jacobian}. In particular we used 6-th order Runge-Kutta \cite{Press92} to solve the initial value problem.

\subsection*{Inverse of 1-st layer: iterative inverse.}
Coupling between the input luminance (or tristimulus value) and the saturation exponent may be avoided by making an initial guess of the exponent (for instance the average value between the two extremes), and then obtaining the first guess for the luminance assuming that approximate exponent. Then, the estimate of the exponent is recomputed from the new luminance estimate, and so on. At the $j$-th iteration,
\begin{eqnarray}
j &=& 0 \,\,
\begin{cases}
\,\, \gamma^1_{(0)} = \frac{1}{2}(\gamma_{L} + \gamma_{H}) \\[0.2cm] \nonumber
\,\, \vect{x}^0_{(0)} = {\vect{x}^1}^{\frac{1}{\gamma^1_{(0)}}}
\end{cases}
\\[0.2cm]
j &>& 0 \,\,
\begin{cases}
\,\, \gamma^1_{(j)} = \gamma^1(\vect{x}^0_{(j-1)}) \\[0.2cm]
\,\, \vect{x}^0_{(j)} = {\vect{x}^1}^{\frac{1}{\gamma^1_{(j)}}}
\end{cases}
\label{inv_stage1}
\end{eqnarray}
where $(j)$ indicates the iteration and $\gamma^1(x)$ is computed using Eq. \ref{stage1}.

\section{Application: MAximum Differentiation (MAD)}
\label{section_MAD}

As discussed in the appendix, one of the advantages of the vector notation explored in this work
is that it enables a geometric view of the neural image representation (\cite{Olshausen96,Hyvarinen09,Field16}).
Specifically, the notation consistent with this geometric view allows original experimental procedures to measure the parameters of the model such as Maximum Differentiation \cite{Wang08,Malo15}.

\begin{figure}[!t]
	\centering
    \small
    \setlength{\tabcolsep}{2pt}
    \begin{tabular}{ccc}
    \hspace{-1.7cm} \includegraphics[width=5cm,height=5cm]{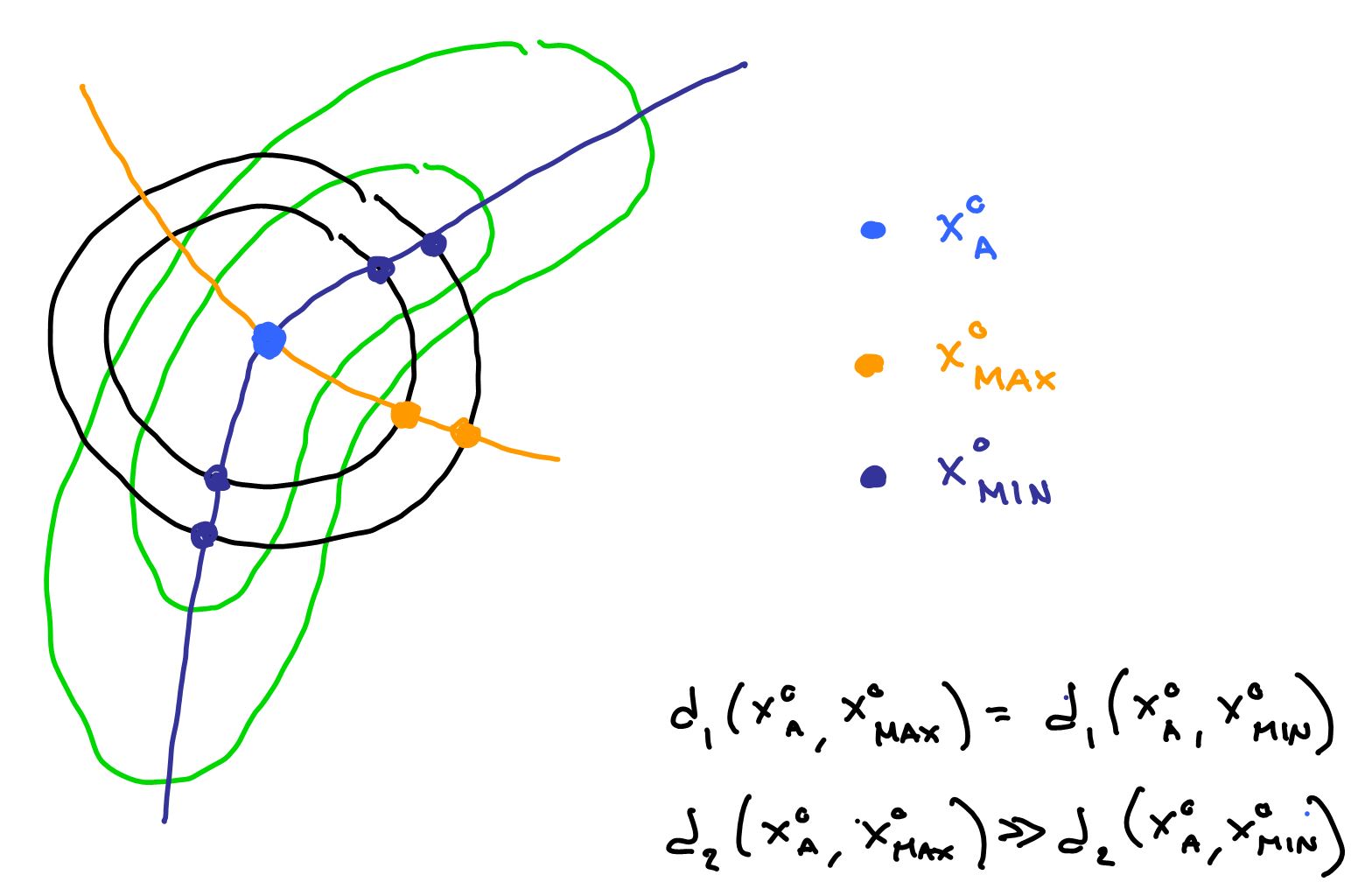} &
                    \includegraphics[width=6cm,height=5cm]{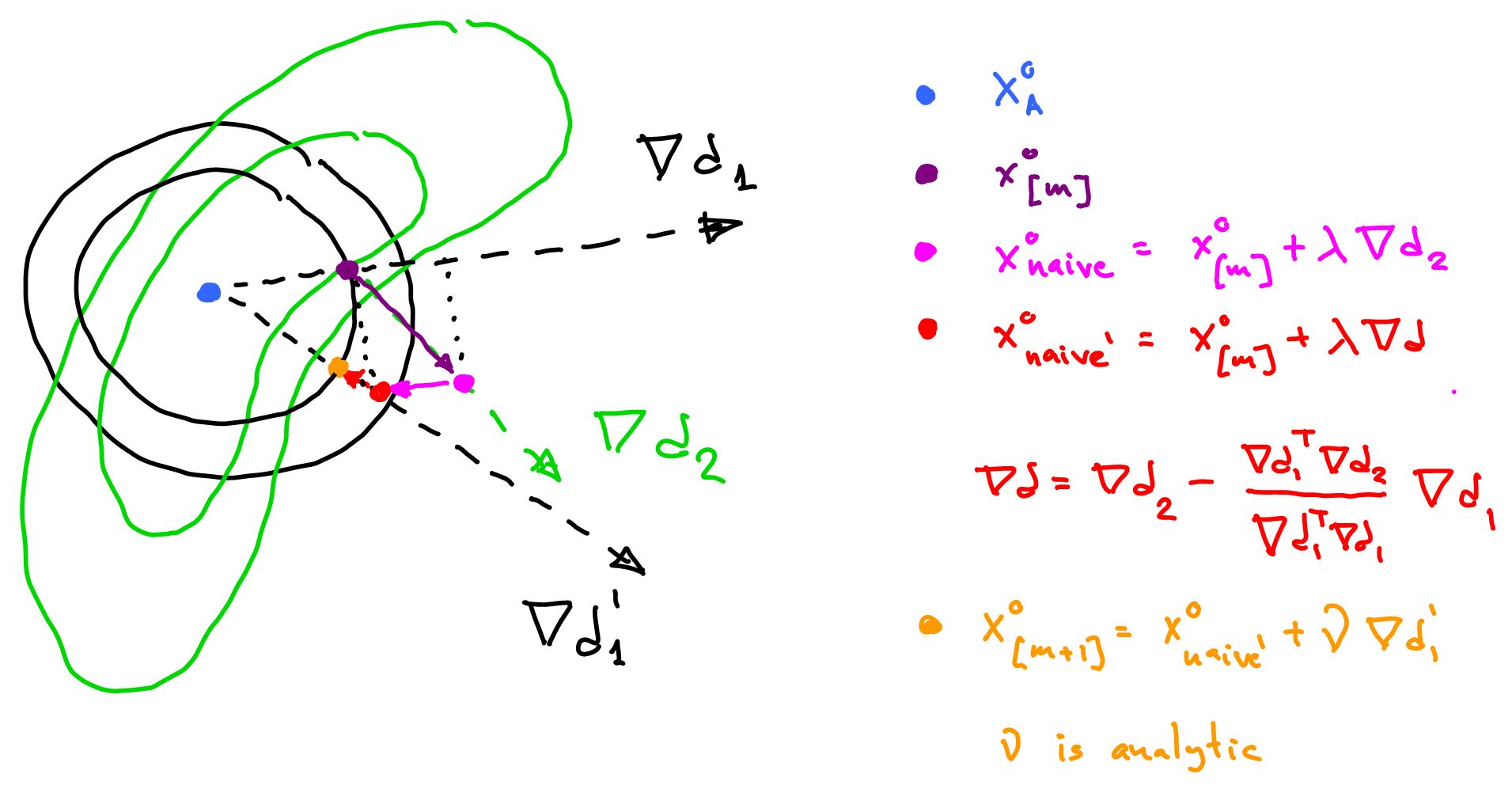} \hspace{0.0cm} &
    \hspace{0.0cm}  \includegraphics[width=2.8cm,height=4.5cm]{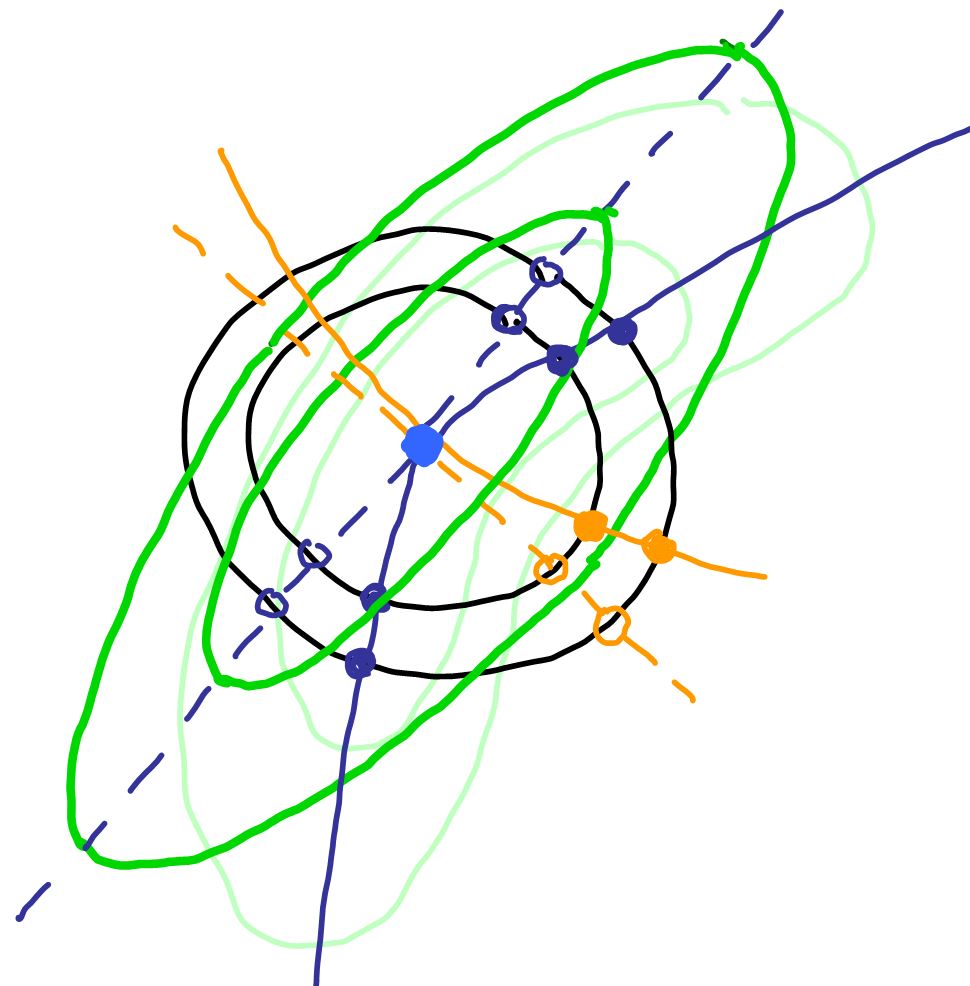} \\
    (a) & (b) & (c) \\
    \end{tabular}
    \vspace{-0.15cm}
	\caption{MAximum Differentiation technique (MAD). (a) The MAD concept: given an image metric coming from a vision model (discrimination regions in green) and a certain fixed Euclidean distance (spheres in black), look for the best and worst images according to the perceptual metric. The model makes perceptual sense if these images are subjectively quite different between them. (b) The MAD algorithm: start from a random point at the sphere and modify to increase or decrease the perceptual distance following the gradient. (c) Second order approximation: approximate the perceptual discrimination regions by ellipsoids (local linear approximation of the vision model. In this way the MAD images are given by the directions of the maximum and minimum eigenvalue of the 2nd order metric matrix.)}\label{MAD_technique}
    \vspace{-0.15cm}
\end{figure}

\begin{itemize}
  \item General (iterative) solution.

  \item Local-linear iterative solution.

  \item Local-linear analytic solution.
\end{itemize}

\section{Associated toolbox}
\label{toolbox}

The concepts associated to the notation and model proposed in this work can be further explored by using the Matlab code
available on-line\footnote{\texttt{http://isp.uv.es/docs/MODVIS16linearnonlinear.zip}}. The preliminary versions of this code were originally developed by J. Malo and E.P. Simoncelli to measure stages 2 and 3 using MAD \cite{Malo15}. Then coauthors from the UV and UPF extended the model to include stages 1 and 4.

\section{Appendix: Slides of the talk at MODVIS}

Below (after the references) you can find the slides of the presentation at MODVIS 2016 and the extension with the relations to deep networks presented at the Image Processing Lab, back in Valencia.

\subsubsection*{Acknowledgments.}
This work was partially funded by the MINECO projects CICYT TEC2013-50520-EXP
and CICYT BFU2014-59776-R.

\newpage

\begin{figure}[h]
	\centering
    \small
    \setlength{\tabcolsep}{2pt}
    \vspace{-2cm}
    \begin{tabular}{c}
    \hspace{-1.5cm} \includegraphics[width=14cm]{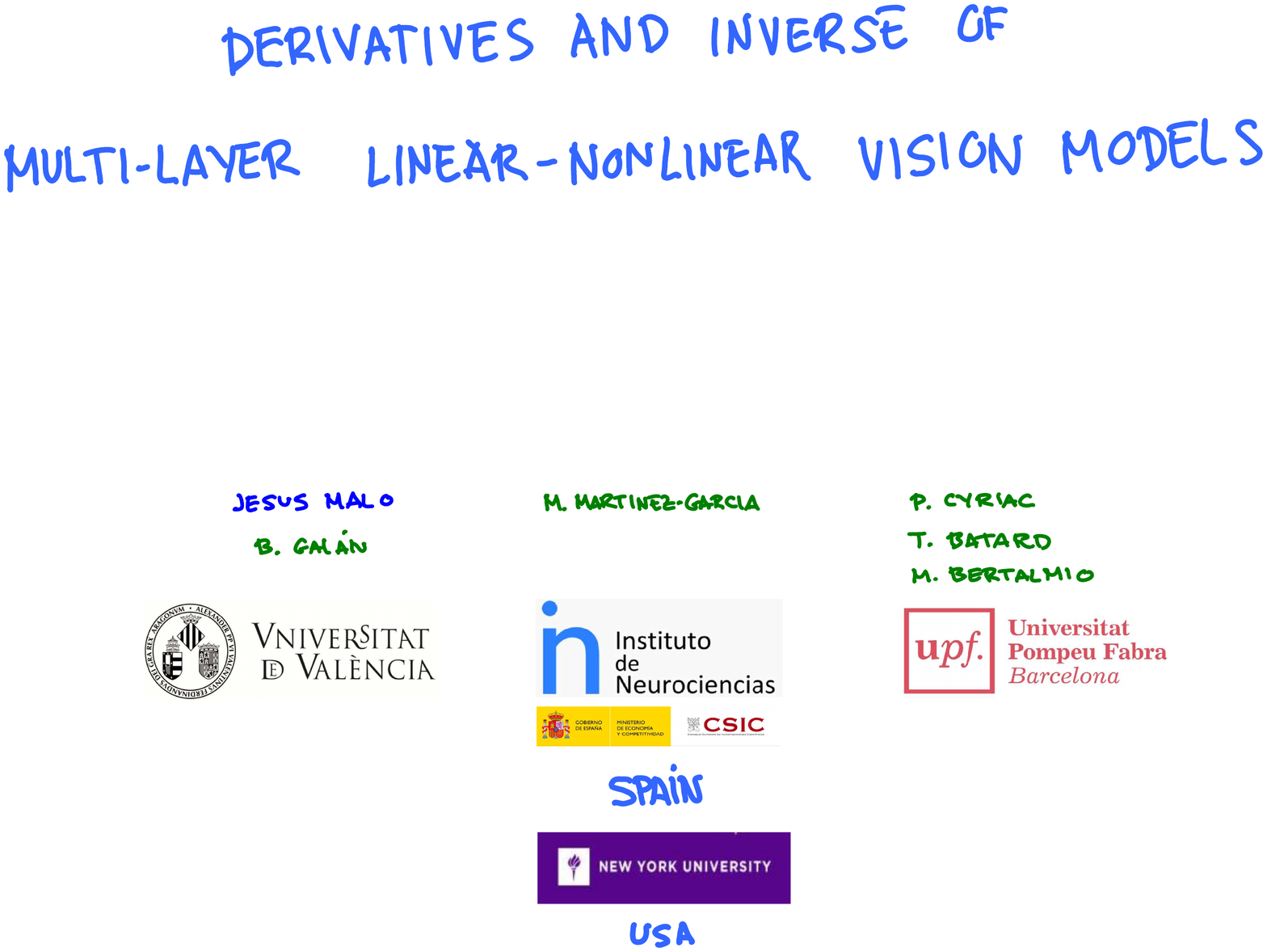} \\ [2cm]
    \hspace{-1.5cm} \includegraphics[width=14cm]{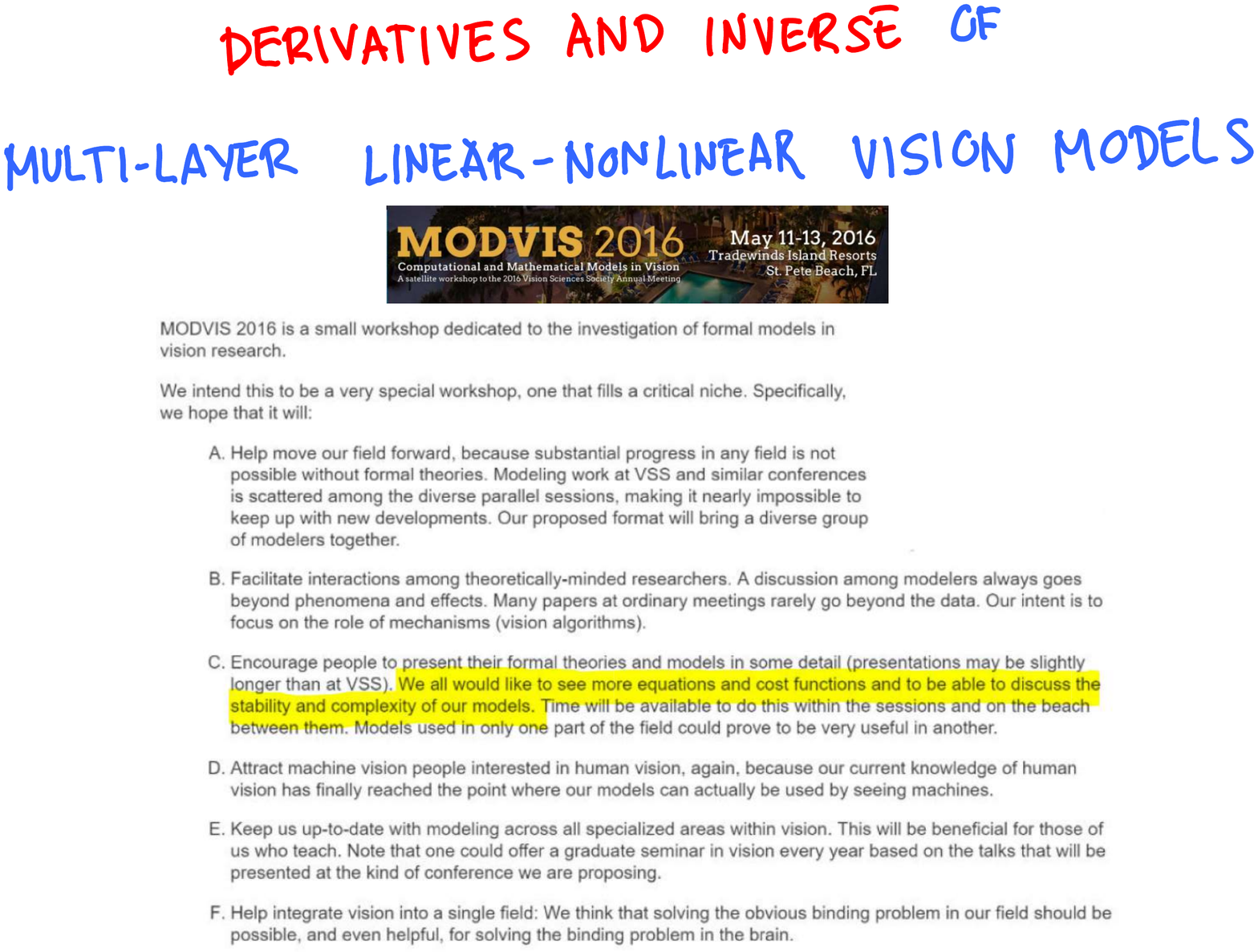}
    \end{tabular}
\end{figure}

\begin{figure}[h]
	\centering
    \small
    \setlength{\tabcolsep}{2pt}
    \vspace{-2cm}
    \begin{tabular}{c}
    \hspace{-1.5cm} \includegraphics[width=14cm]{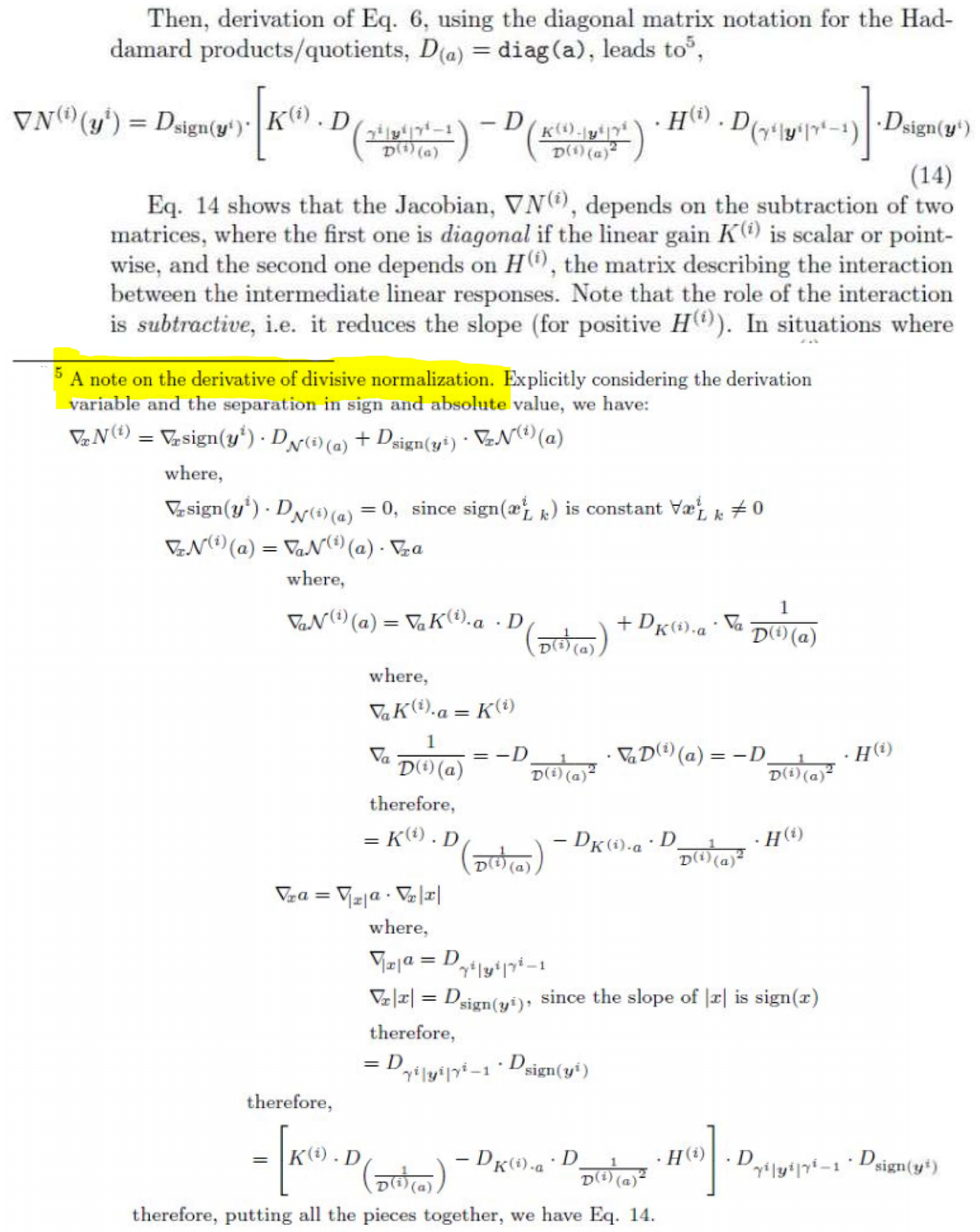} \\ [2cm]
    \hspace{-1.5cm} \includegraphics[width=14cm]{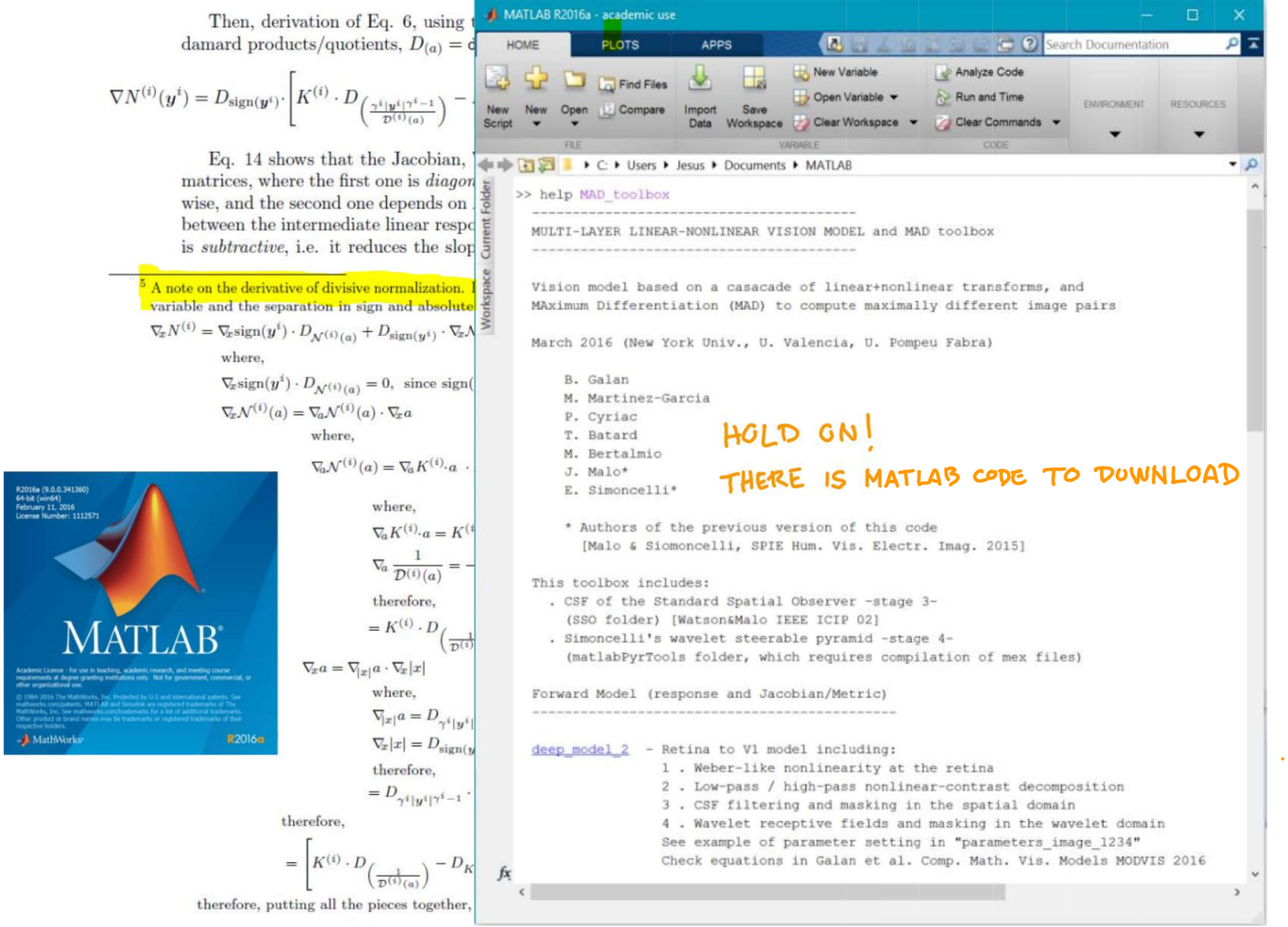}
    \end{tabular}
\end{figure}

\begin{figure}[h]
	\centering
    \small
    \setlength{\tabcolsep}{2pt}
    \vspace{-2cm}
    \begin{tabular}{c}
    \hspace{-1.5cm} \includegraphics[width=14cm]{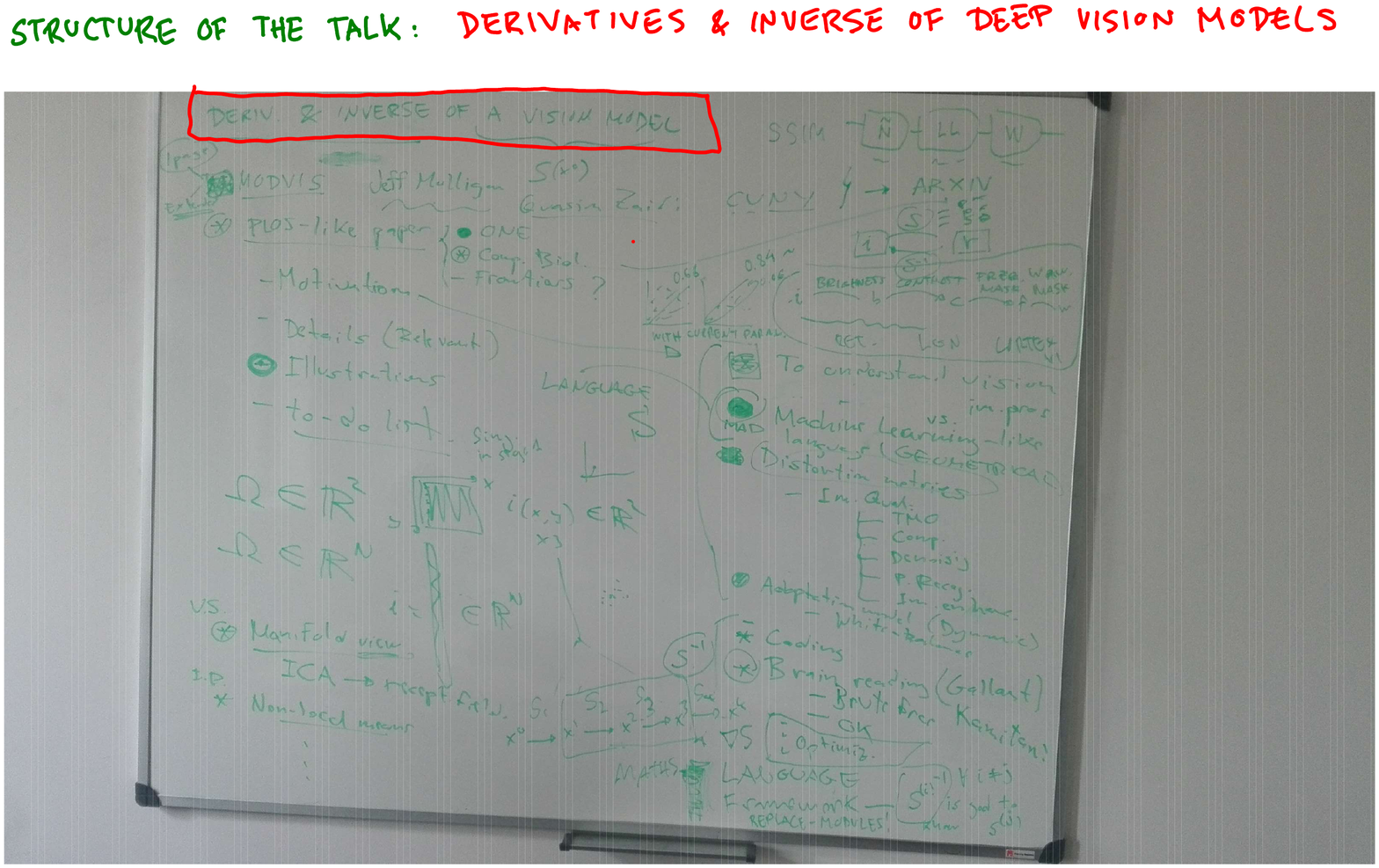} \\ [2cm]
    \hspace{-1.5cm} \includegraphics[width=14cm]{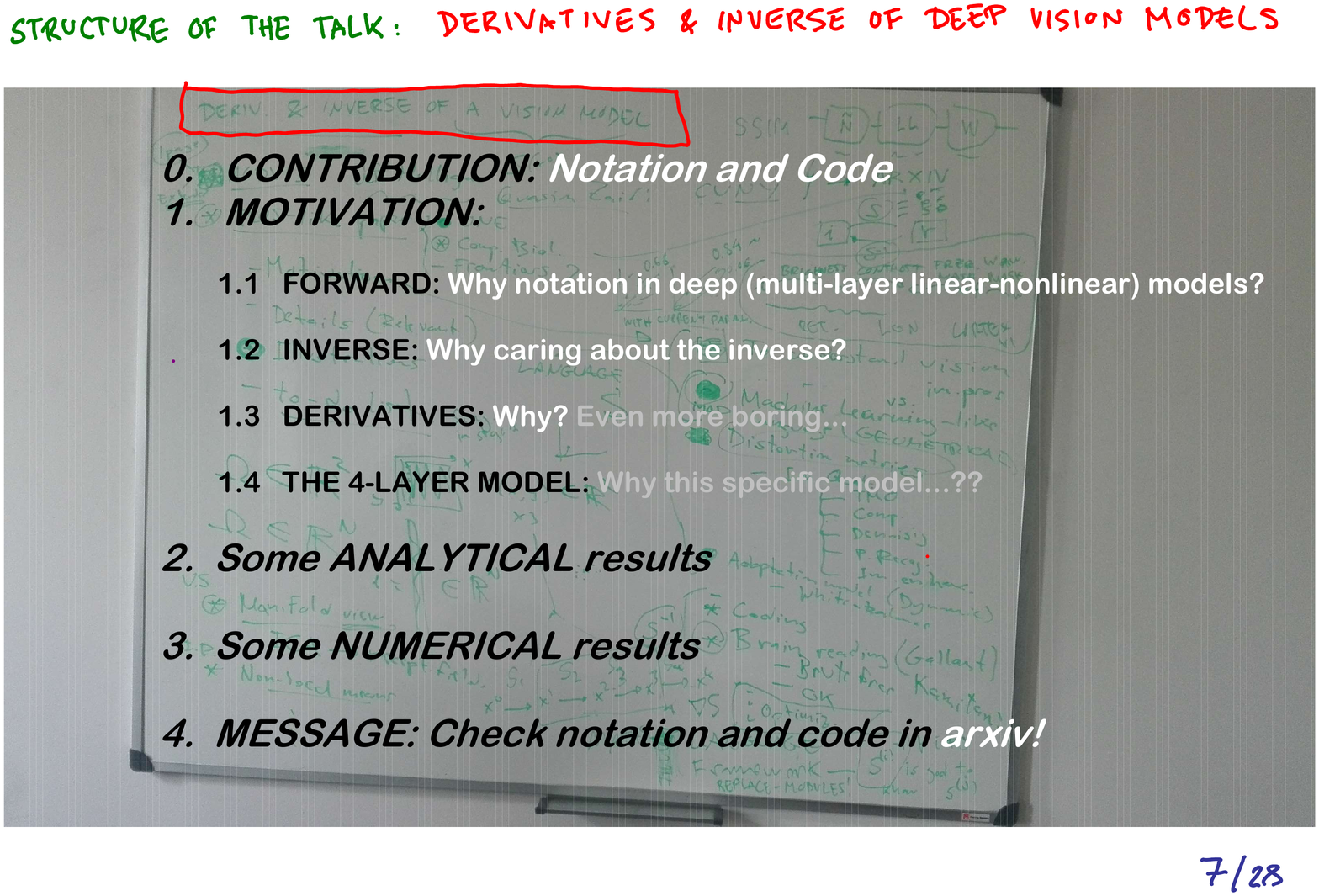}
    \end{tabular}
\end{figure}

\begin{figure}[h]
	\centering
    \small
    \setlength{\tabcolsep}{2pt}
    \vspace{-2cm}
    \begin{tabular}{c}
    \hspace{-1.5cm} \includegraphics[width=14cm]{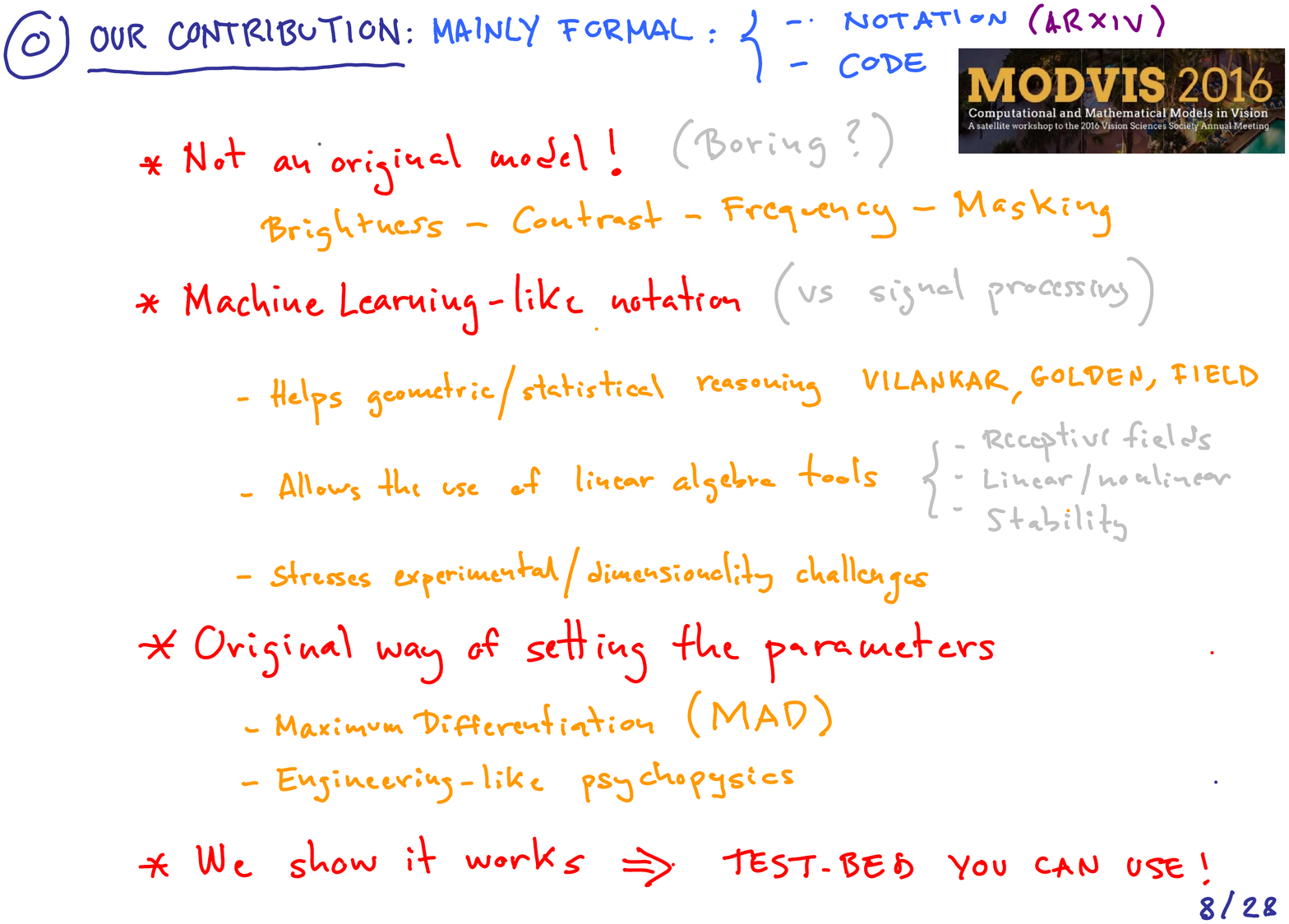} \\ [2cm]
    \hspace{-1.5cm} \includegraphics[width=14cm]{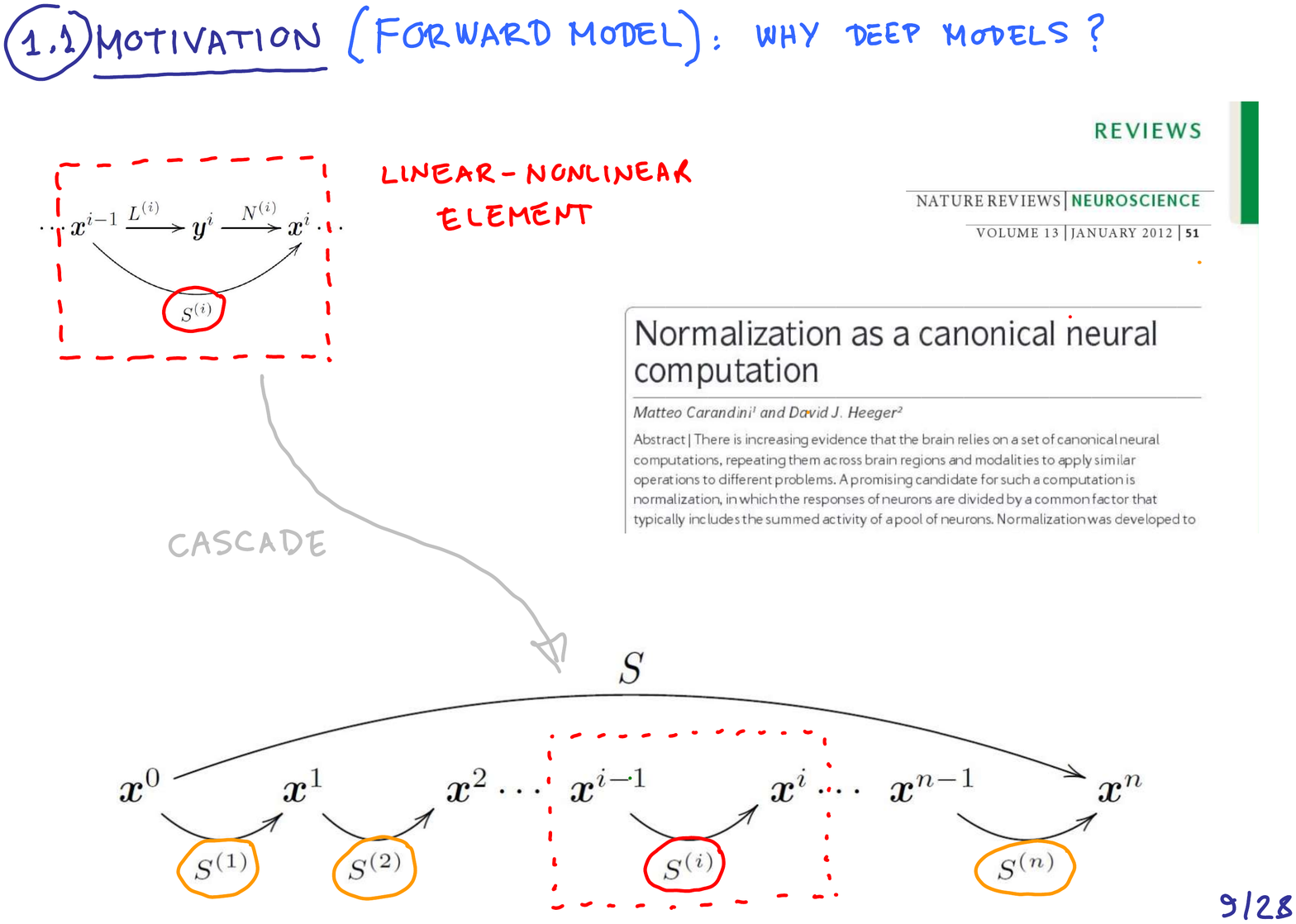}
    \end{tabular}
\end{figure}

\begin{figure}[h]
	\centering
    \small
    \setlength{\tabcolsep}{2pt}
    \vspace{-2cm}
    \begin{tabular}{c}
    \hspace{-1.5cm} \includegraphics[width=14cm]{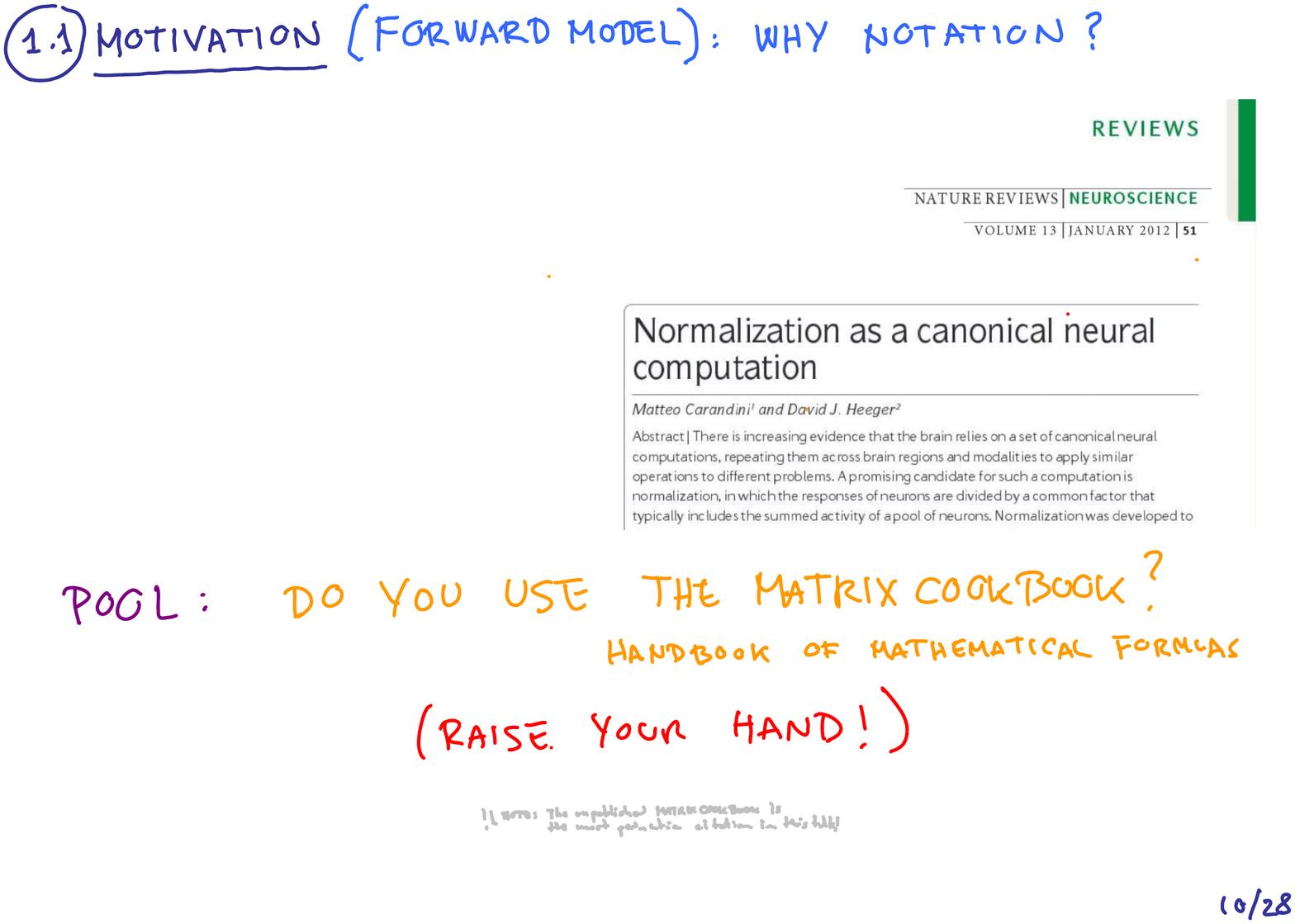} \\ [2cm]
    \hspace{-1.5cm} \includegraphics[width=14cm]{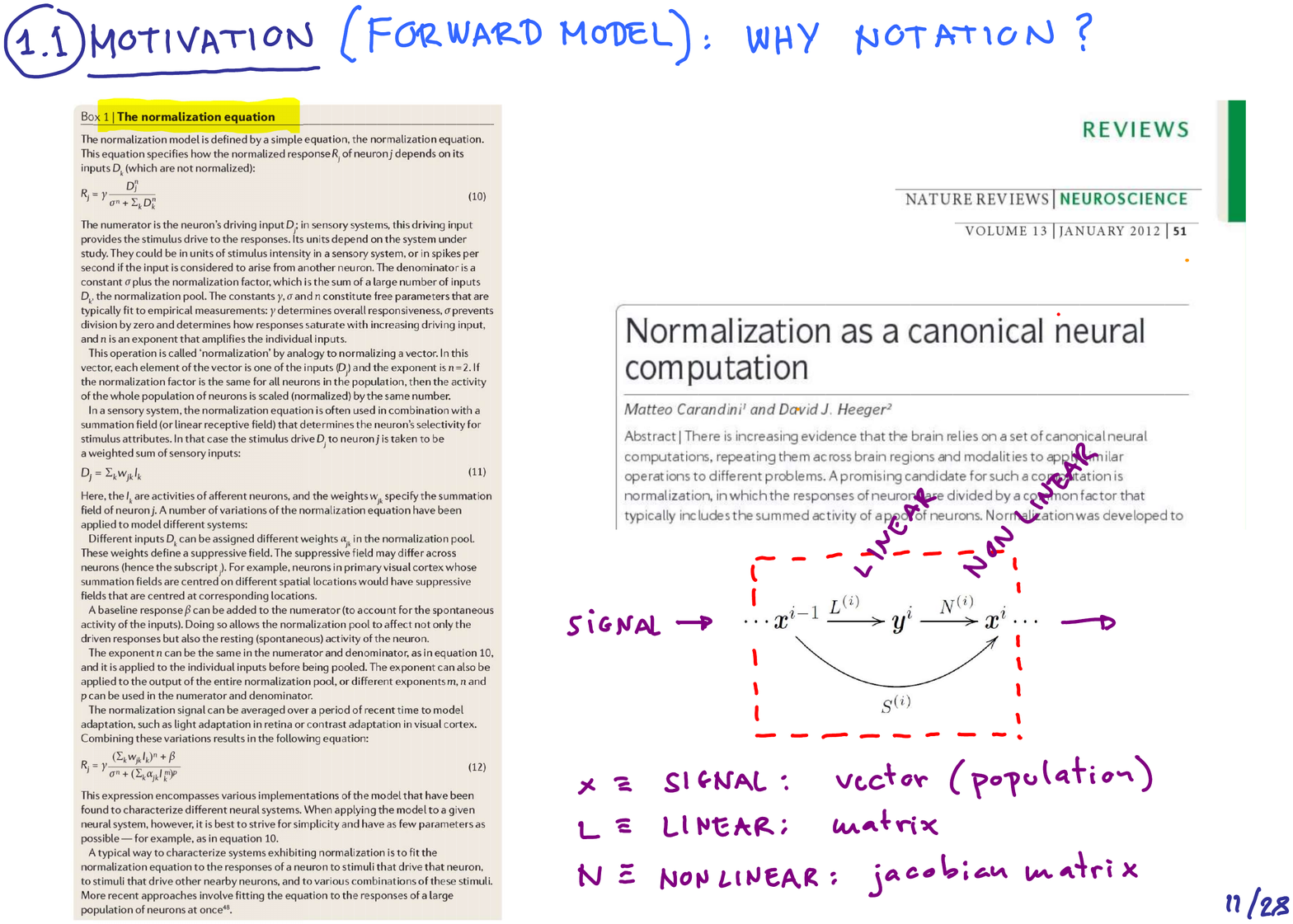}
    \end{tabular}
\end{figure}

\begin{figure}[h]
	\centering
    \small
    \setlength{\tabcolsep}{2pt}
    \vspace{-2cm}
    \begin{tabular}{c}
    \hspace{-1.5cm} \includegraphics[width=14cm]{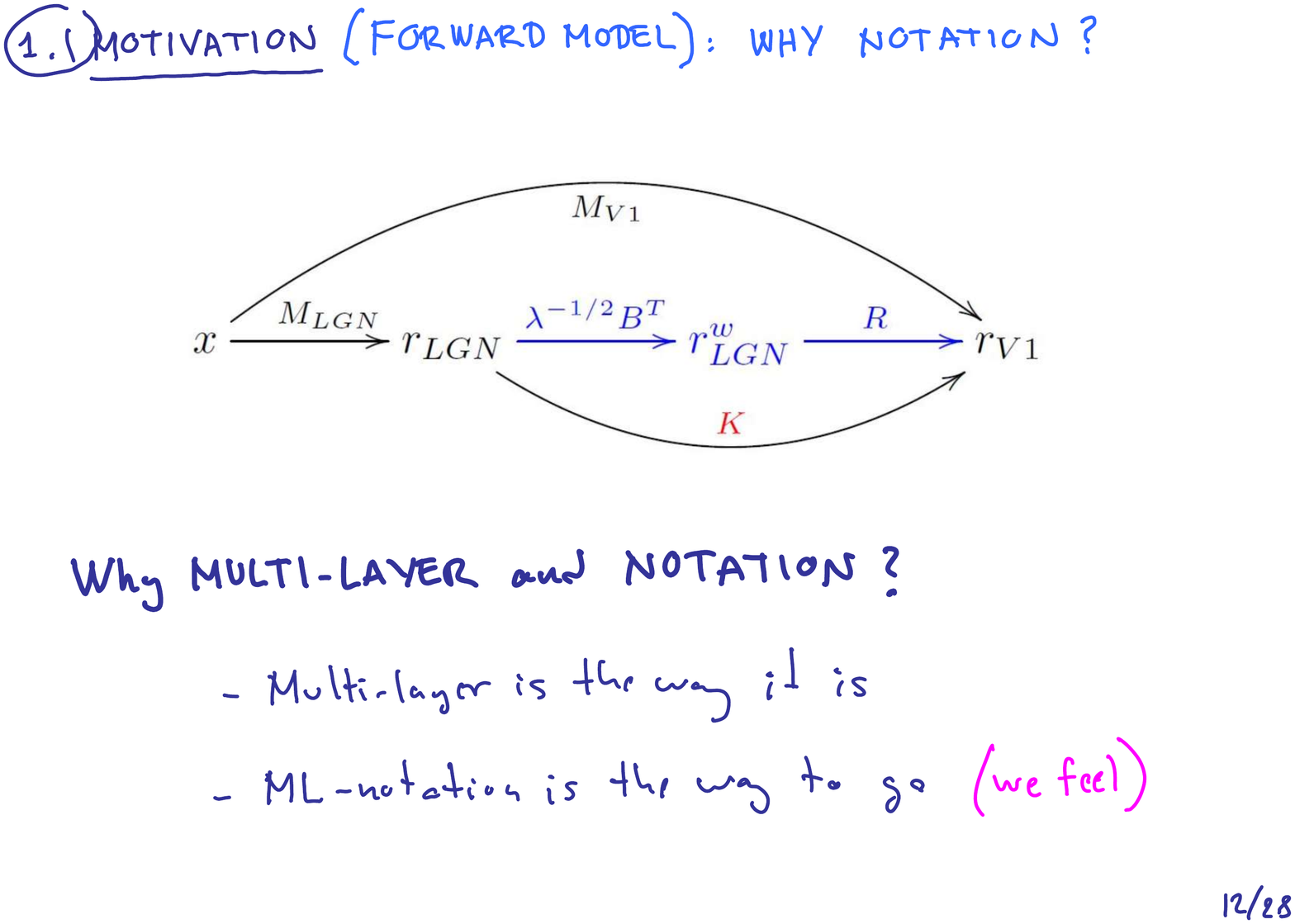} \\ [2cm]
    \hspace{-1.5cm} \includegraphics[width=14cm]{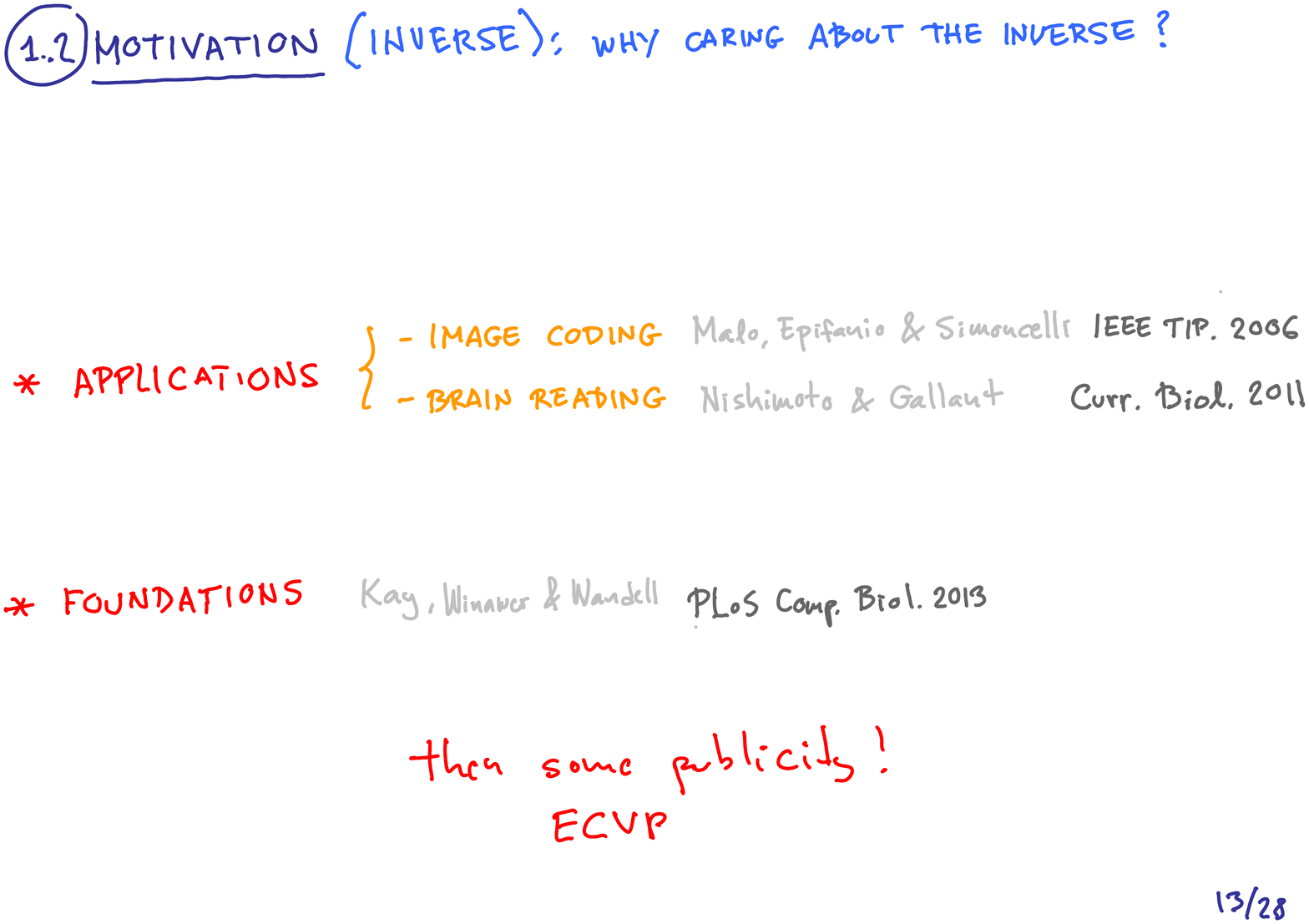}
    \end{tabular}
\end{figure}

\begin{figure}[h]
	\centering
    \small
    \setlength{\tabcolsep}{2pt}
    \vspace{-2cm}
    \begin{tabular}{c}
    \hspace{-1.5cm} \includegraphics[width=14cm]{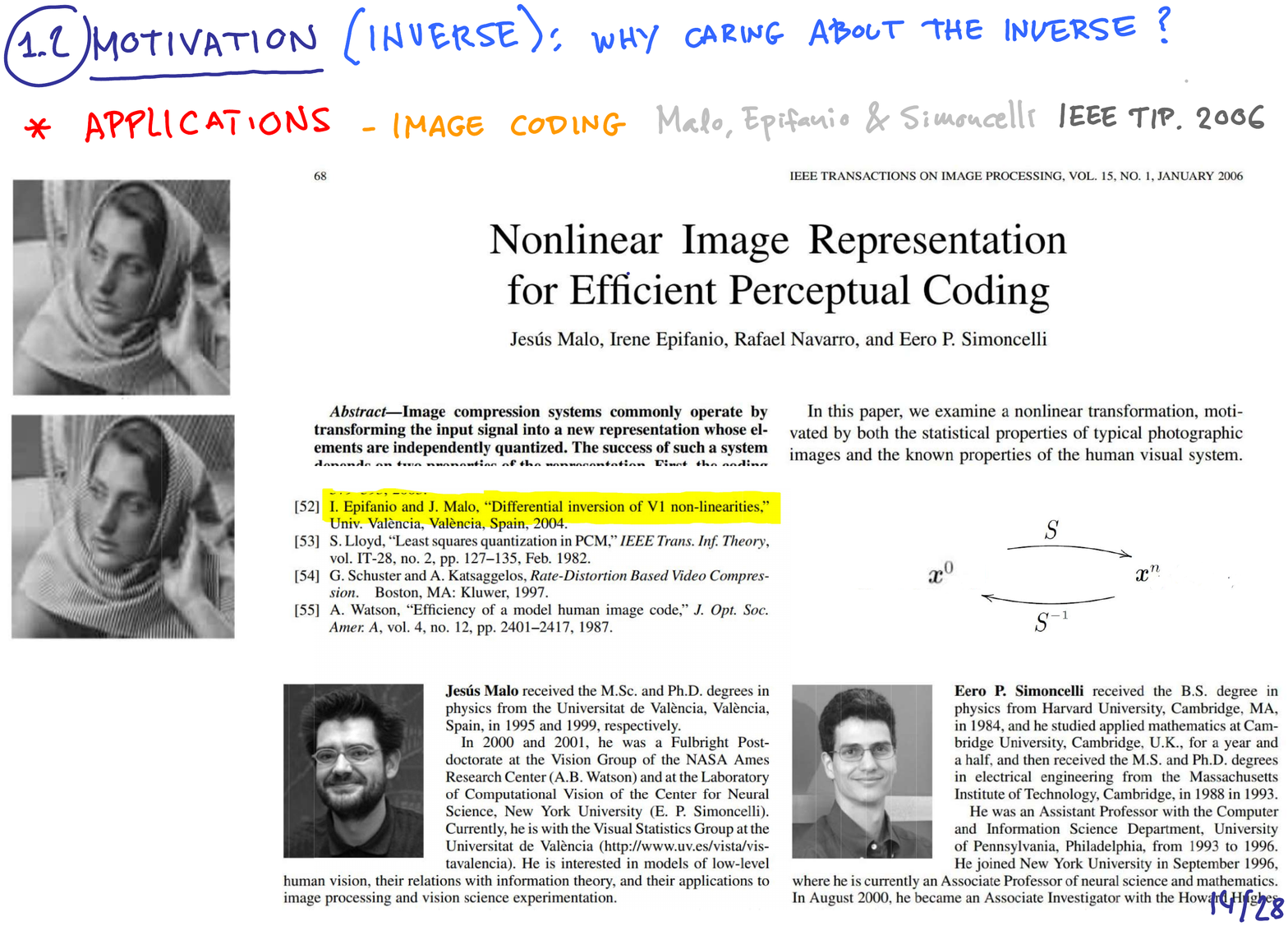} \\ [2cm]
    \hspace{-1.5cm} \includegraphics[width=14cm]{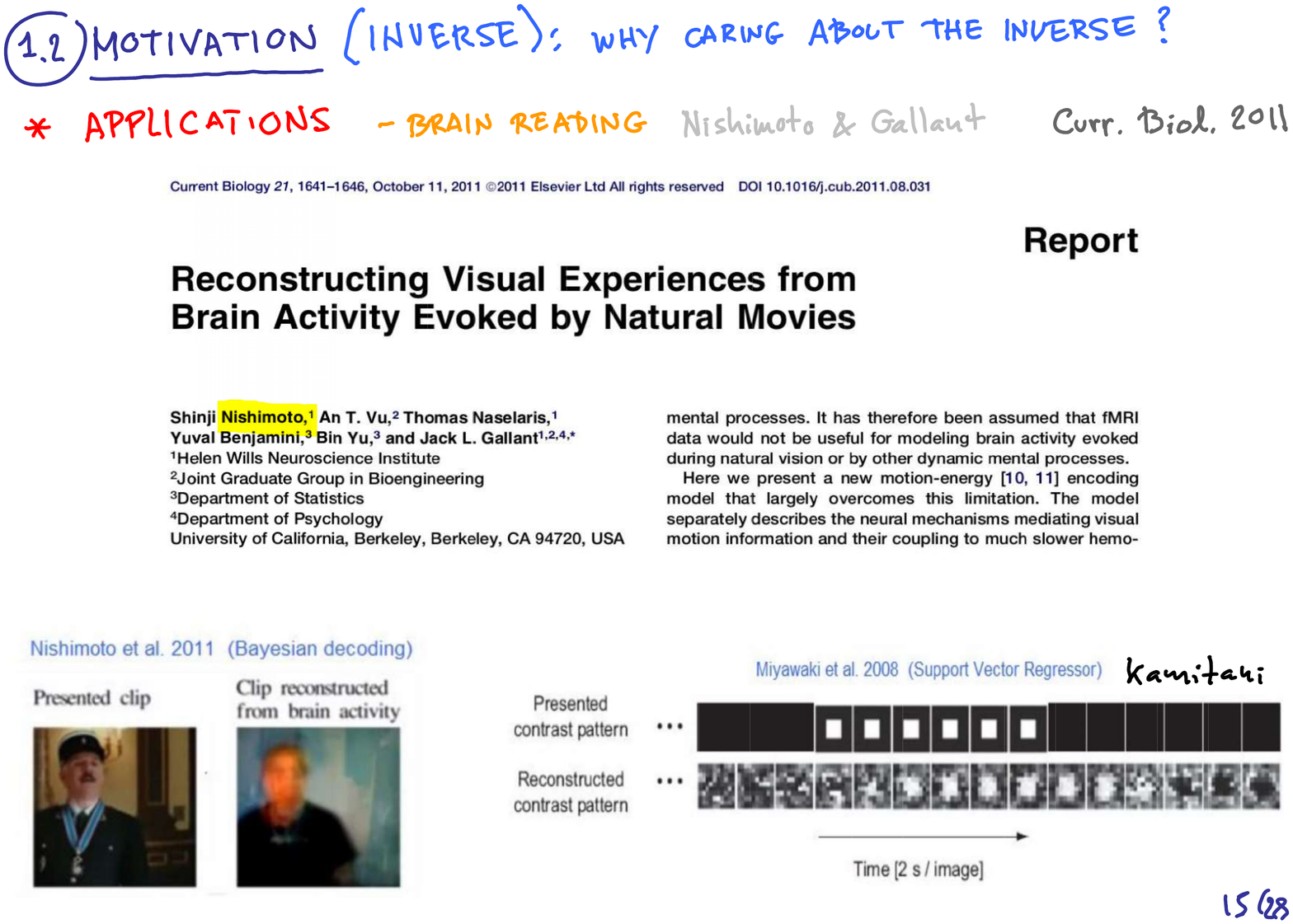}
    \end{tabular}
\end{figure}

\begin{figure}[h]
	\centering
    \small
    \setlength{\tabcolsep}{2pt}
    \vspace{-2cm}
    \begin{tabular}{c}
    \hspace{-1.5cm} \includegraphics[width=14cm]{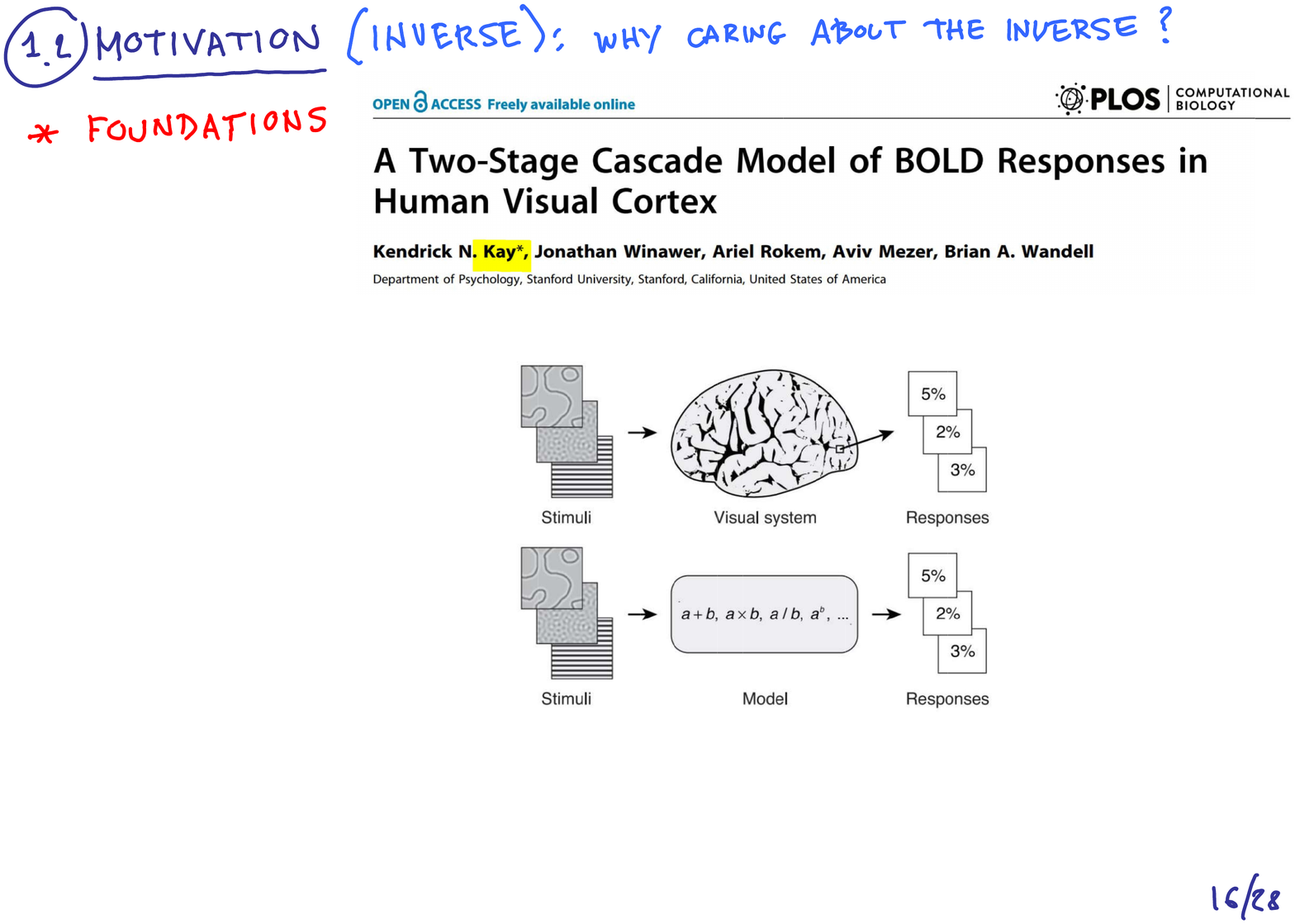} \\ [2cm]
    \hspace{-1.5cm} \includegraphics[width=14cm]{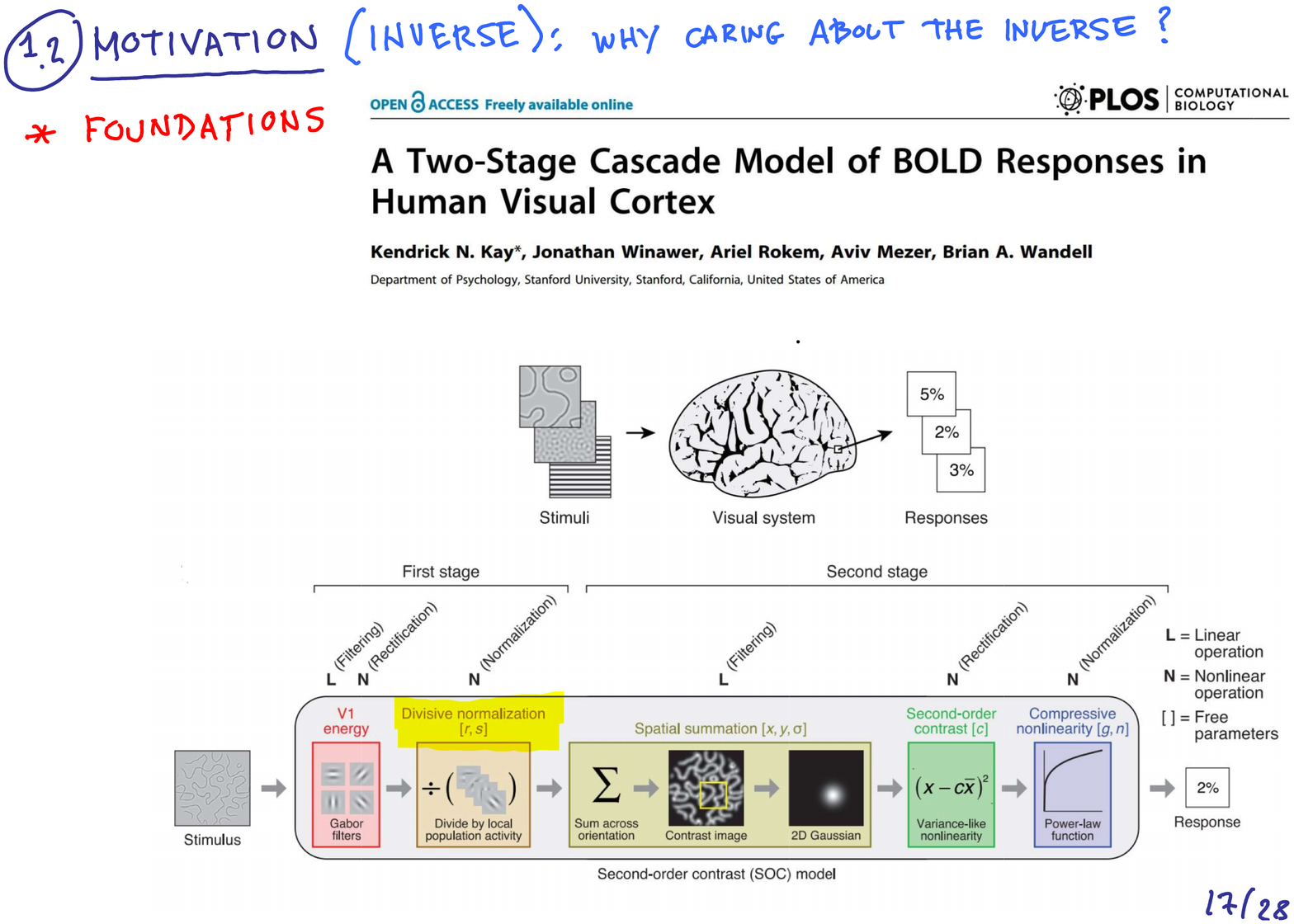}
    \end{tabular}
\end{figure}

\begin{figure}[h]
	\centering
    \small
    \setlength{\tabcolsep}{2pt}
    \vspace{-2cm}
    \begin{tabular}{c}
    \hspace{-1.5cm} \includegraphics[width=14cm]{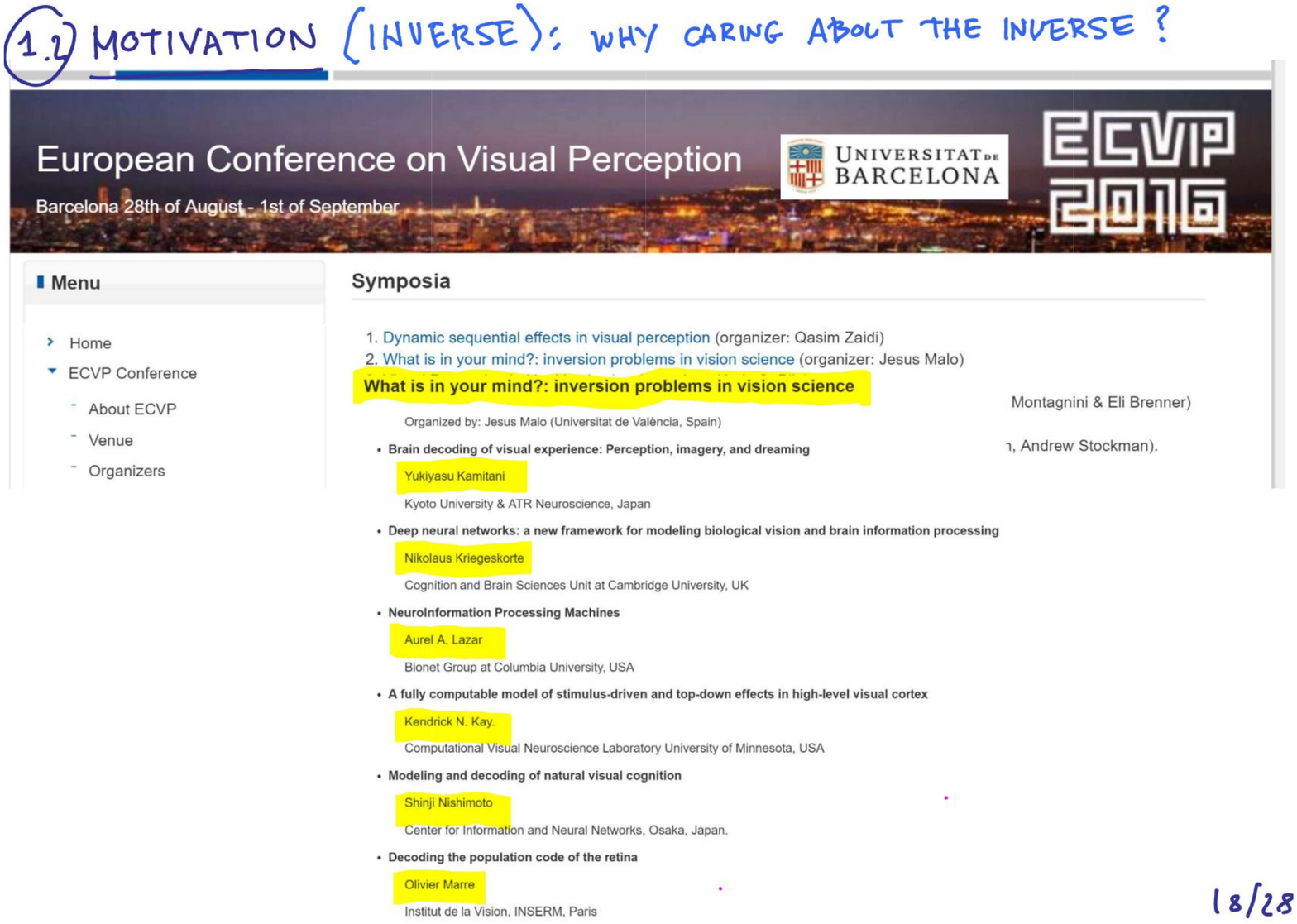} \\ [2cm]
    \hspace{-1.5cm} \includegraphics[width=14cm]{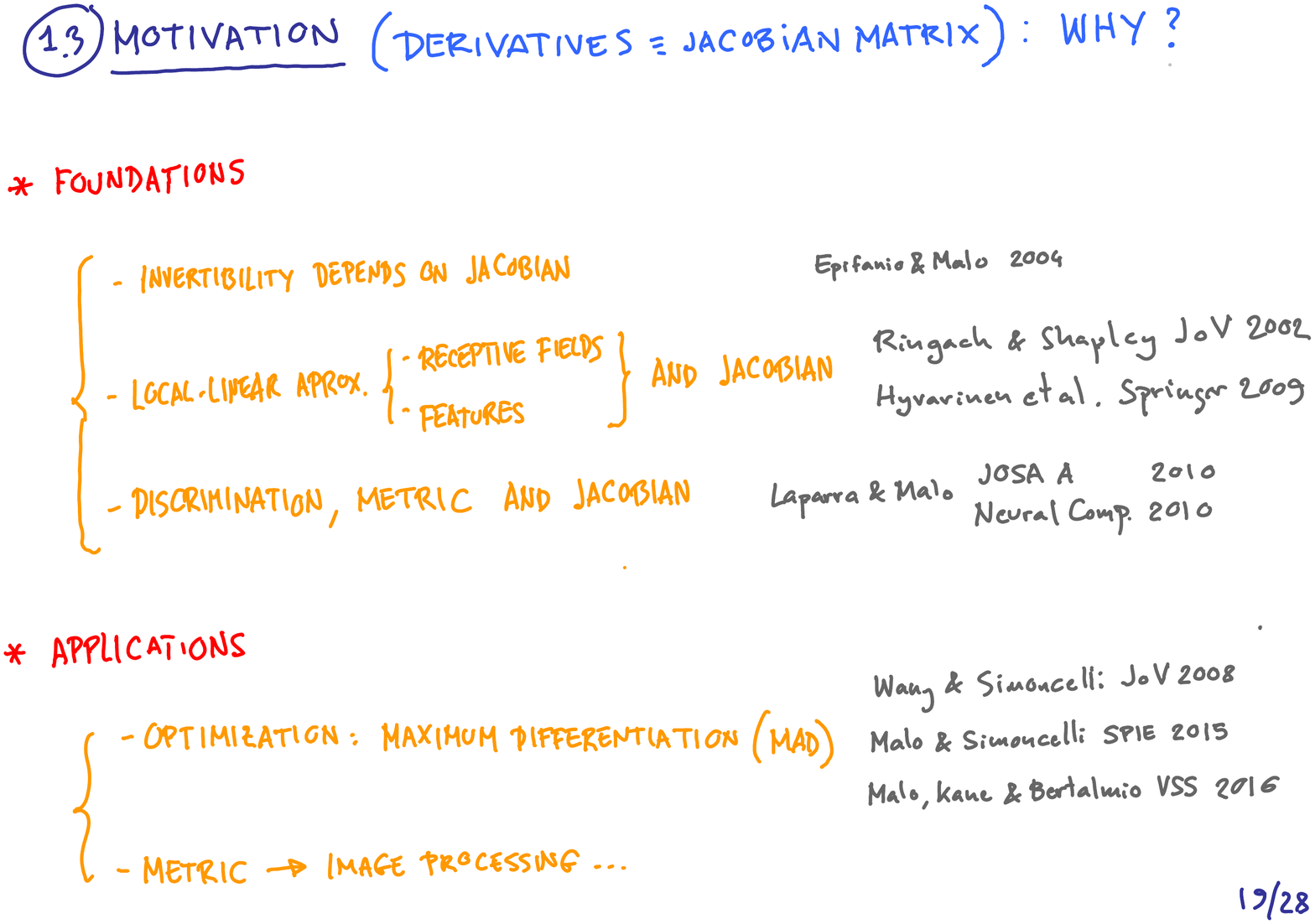}
    \end{tabular}
\end{figure}

\begin{figure}[h]
	\centering
    \small
    \setlength{\tabcolsep}{2pt}
    \vspace{-2cm}
    \begin{tabular}{c}
    \hspace{-1.5cm} \includegraphics[width=14cm]{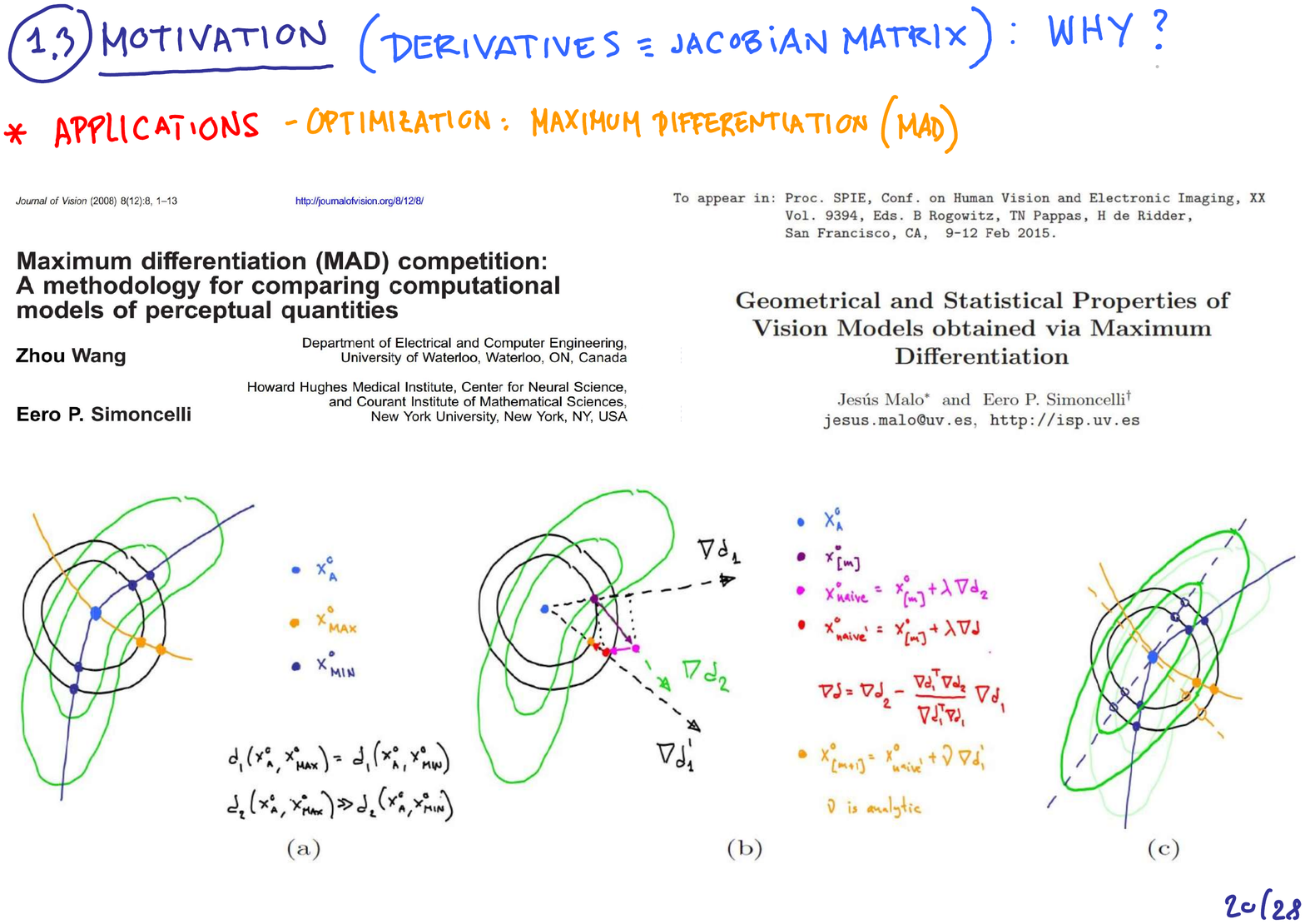} \\ [2cm]
    \hspace{-1.5cm} \includegraphics[width=14cm]{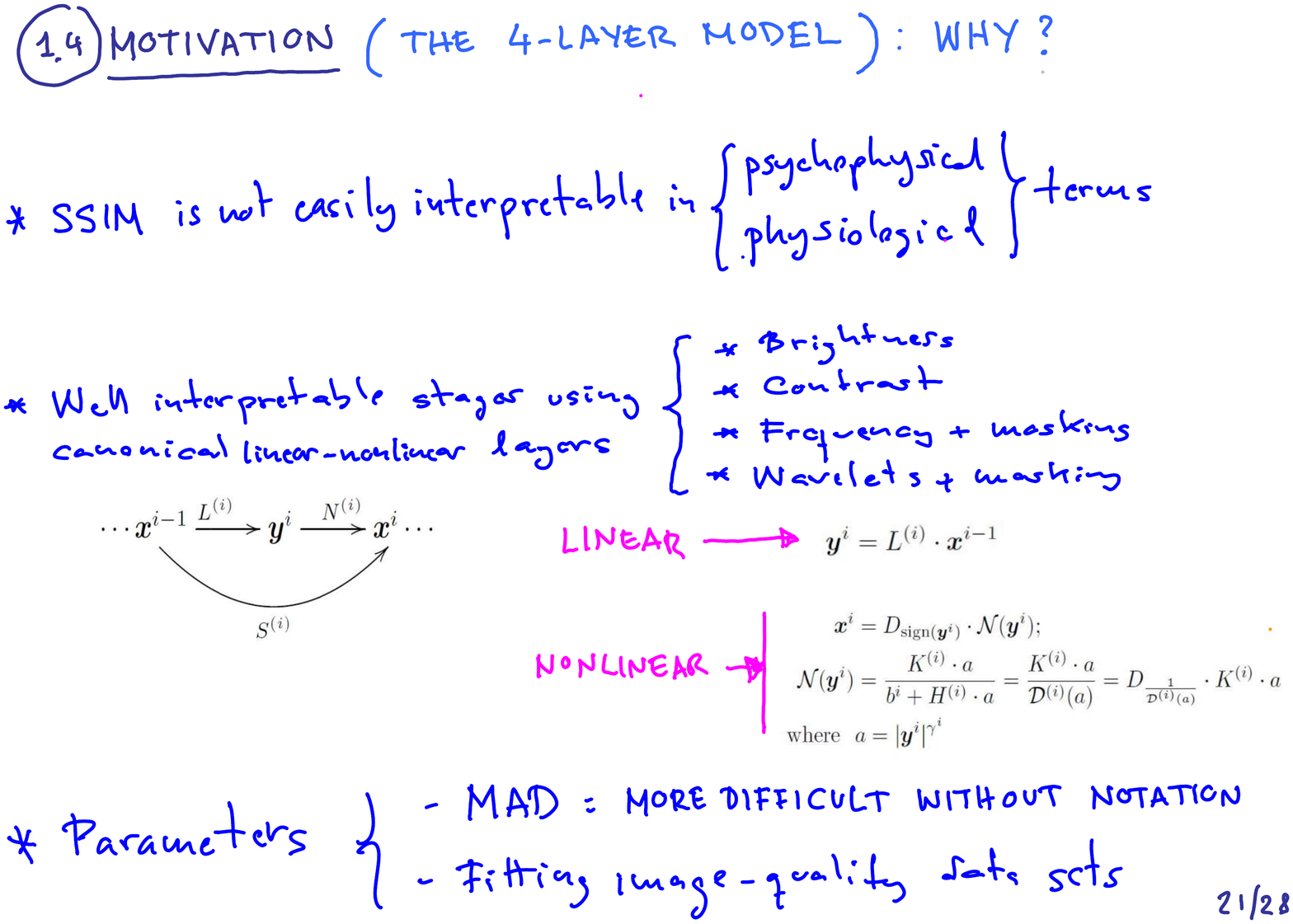}
    \end{tabular}
\end{figure}

\begin{figure}[h]
	\centering
    \small
    \setlength{\tabcolsep}{2pt}
    \vspace{-2cm}
    \begin{tabular}{c}
    \hspace{-1.5cm} \includegraphics[width=14cm]{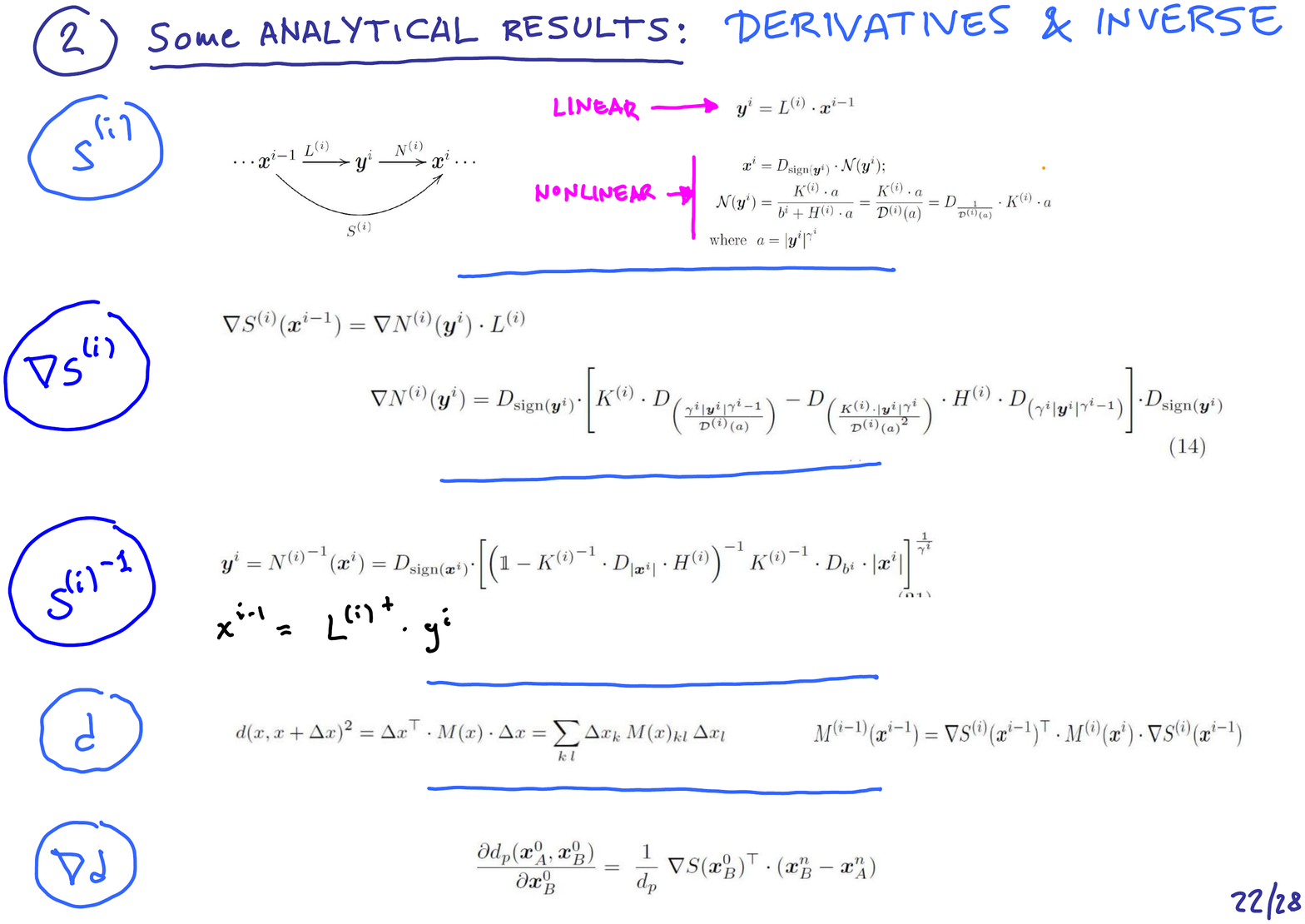} \\ [2cm]
    \hspace{-1.5cm} \includegraphics[width=14cm]{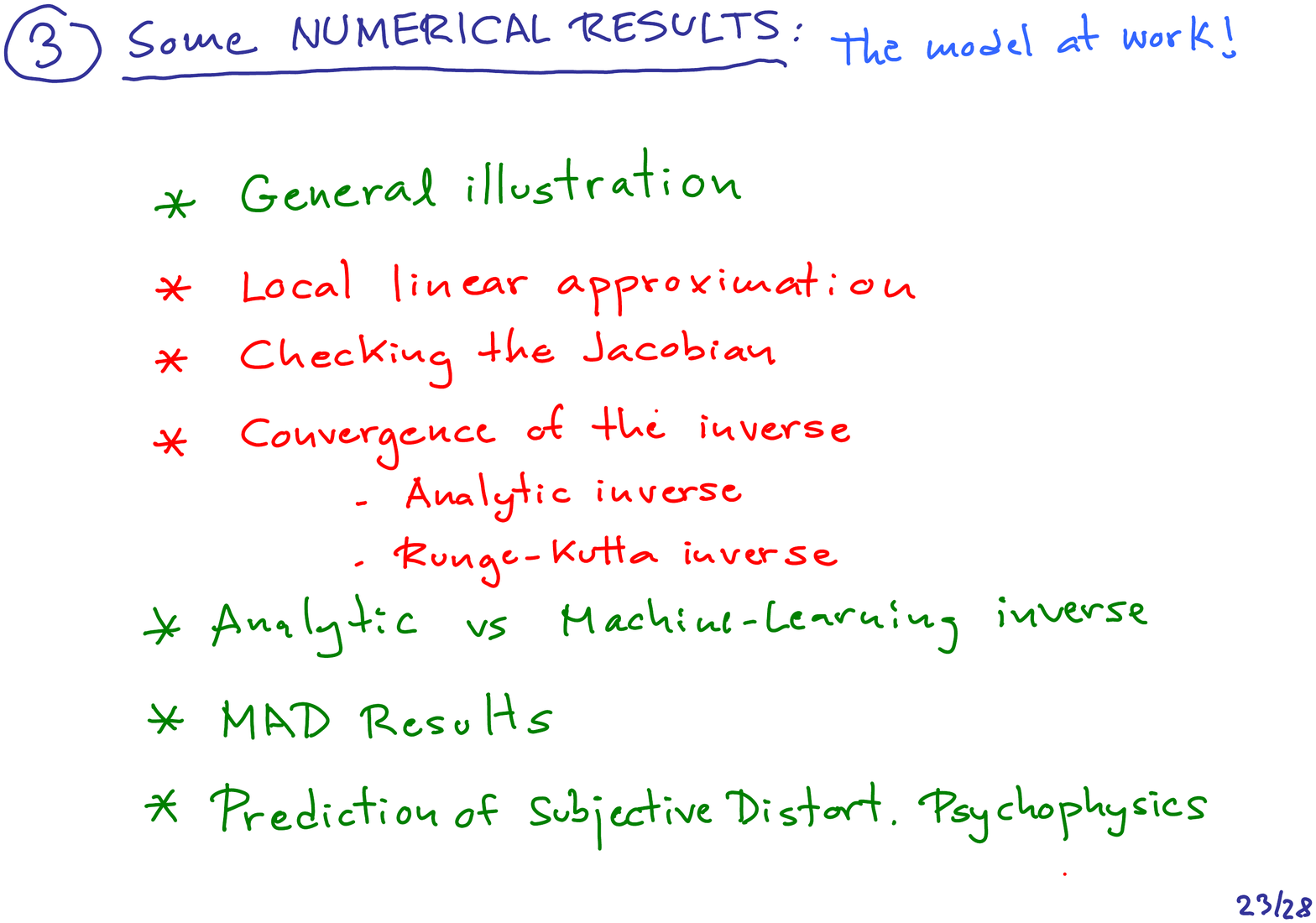}
    \end{tabular}
\end{figure}

\begin{figure}[h]
	\centering
    \small
    \setlength{\tabcolsep}{2pt}
    \vspace{-2cm}
    \begin{tabular}{c}
    \hspace{-1.5cm} \includegraphics[width=14cm]{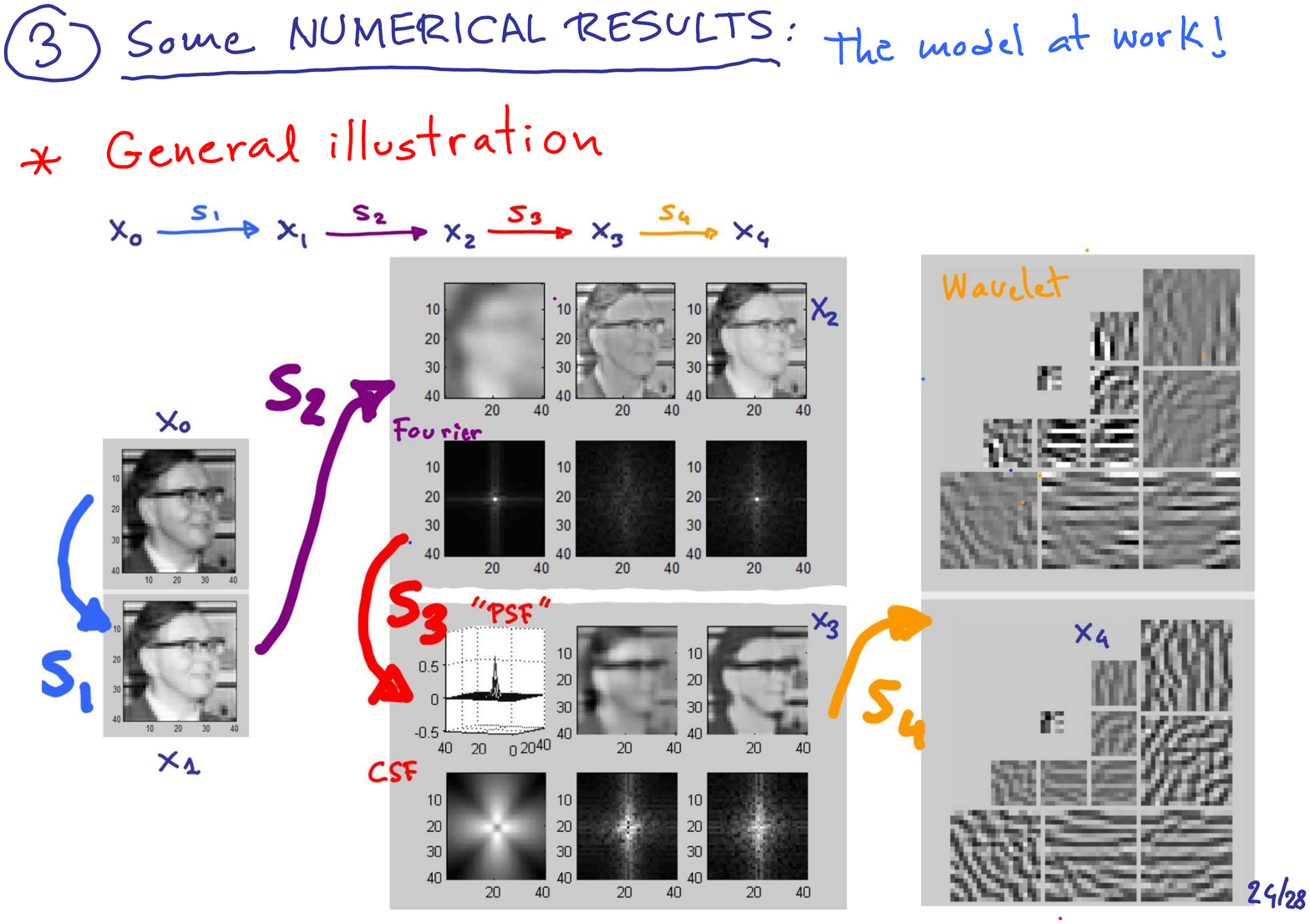} \\ [2cm]
    \hspace{-1.5cm} \includegraphics[width=14cm]{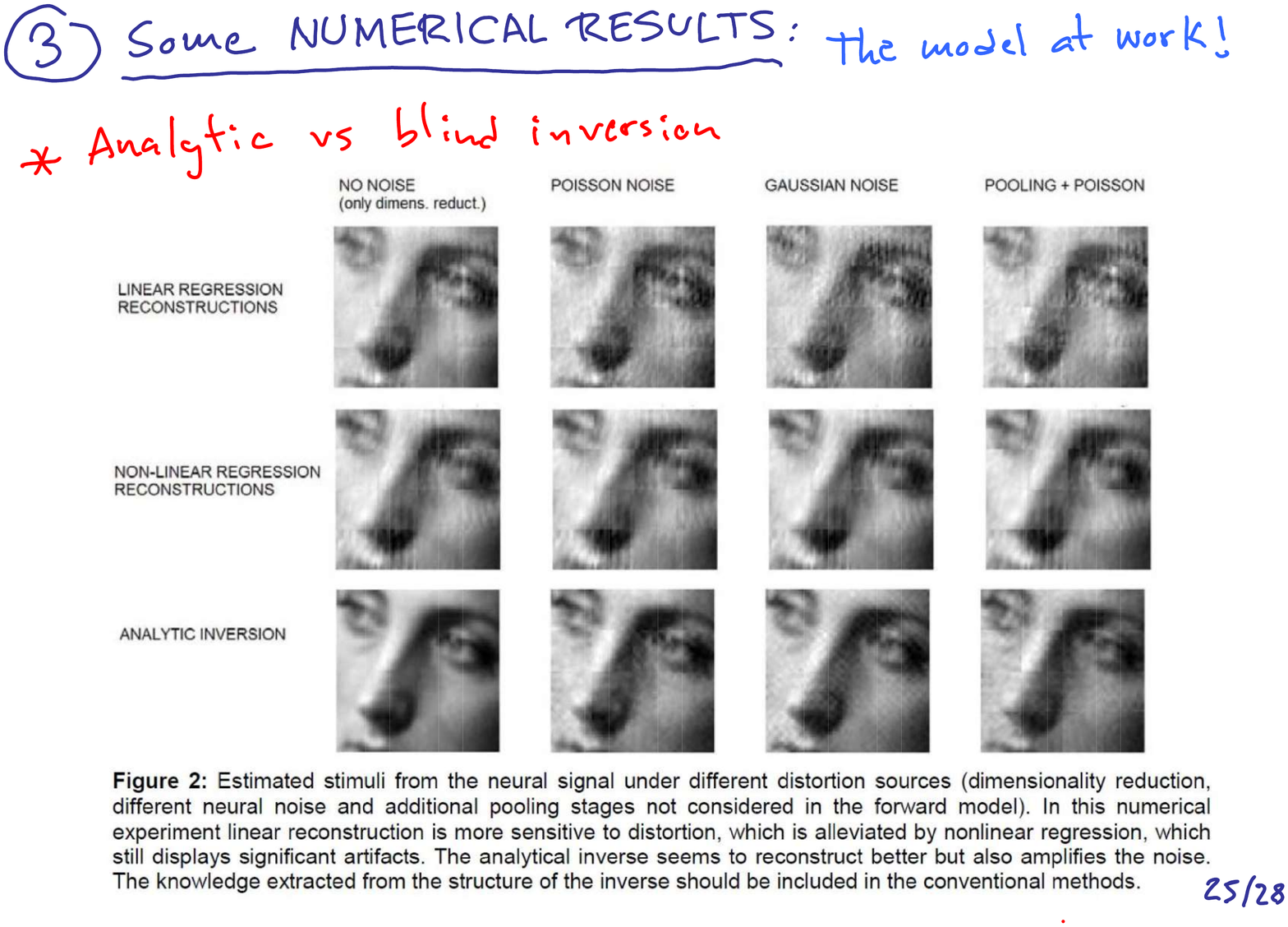}
    \end{tabular}
\end{figure}

\begin{figure}[h]
	\centering
    \small
    \setlength{\tabcolsep}{2pt}
    \vspace{-2cm}
    \begin{tabular}{c}
    \hspace{-1.5cm} \includegraphics[width=14cm]{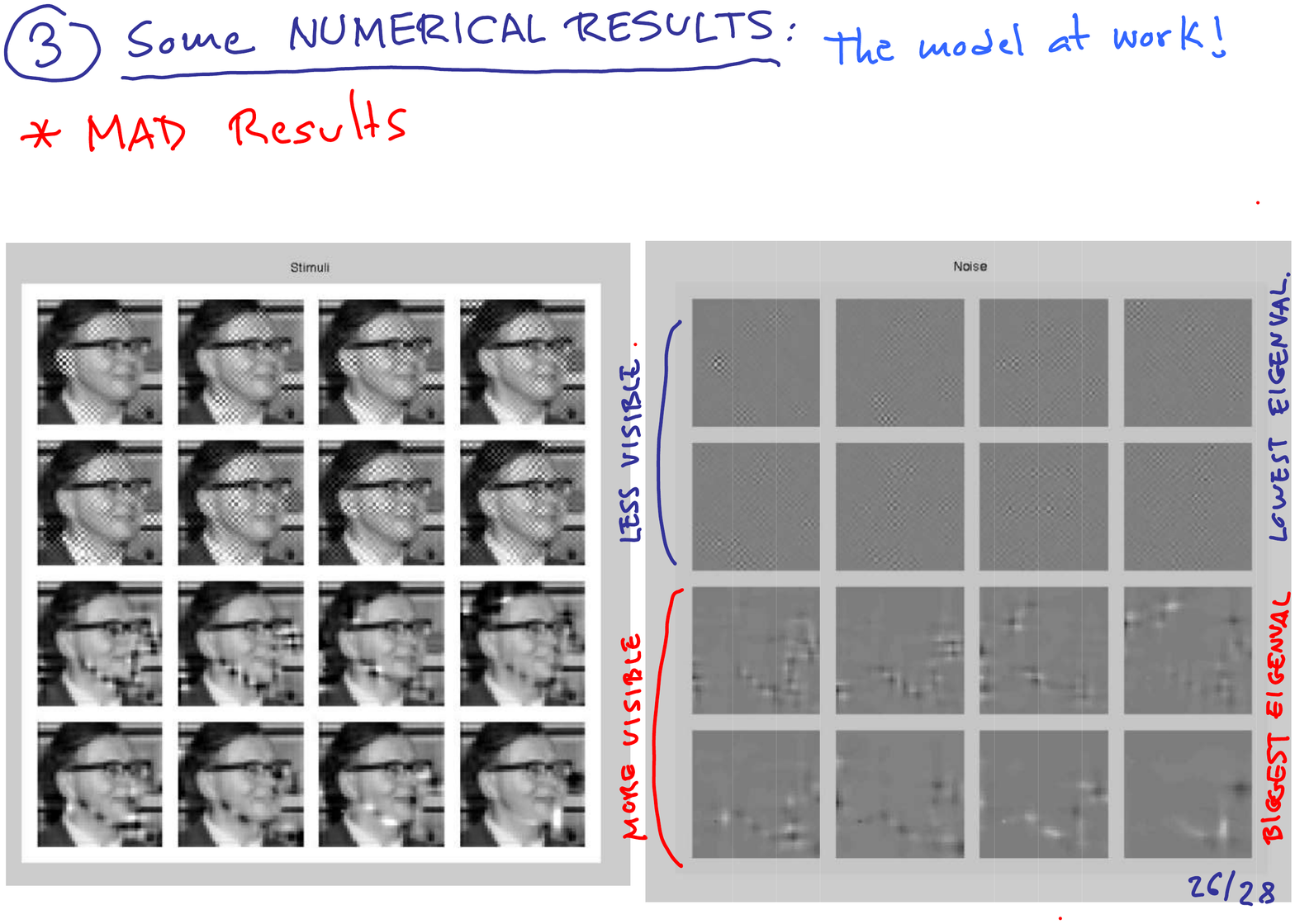} \\ [2cm]
    \hspace{-1.5cm} \includegraphics[width=14cm]{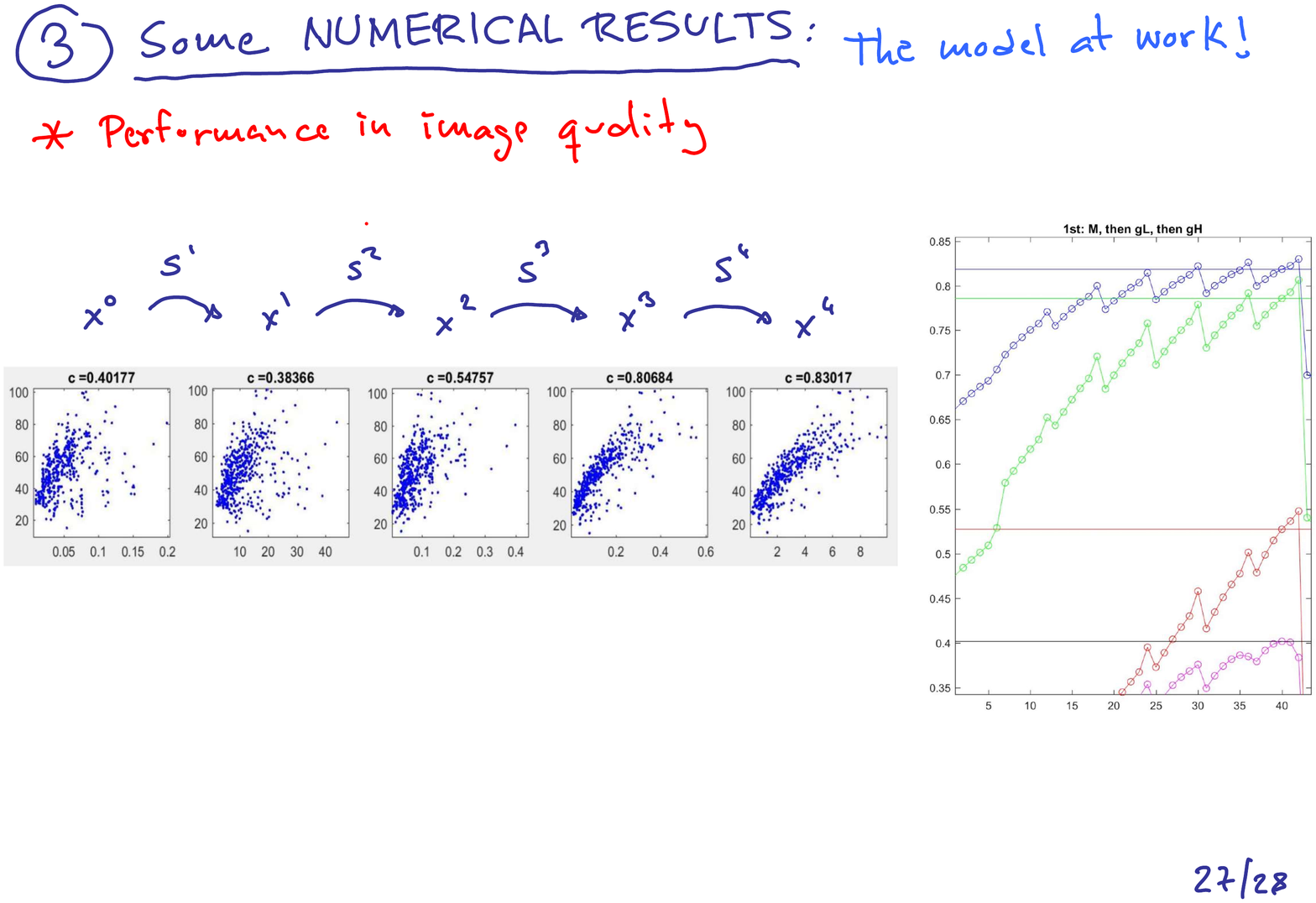}
    \end{tabular}
\end{figure}

\begin{figure}[h]
	\centering
    \small
    \setlength{\tabcolsep}{2pt}
    \vspace{-2cm}
    \begin{tabular}{c}
    \hspace{-1.5cm} \includegraphics[width=14cm]{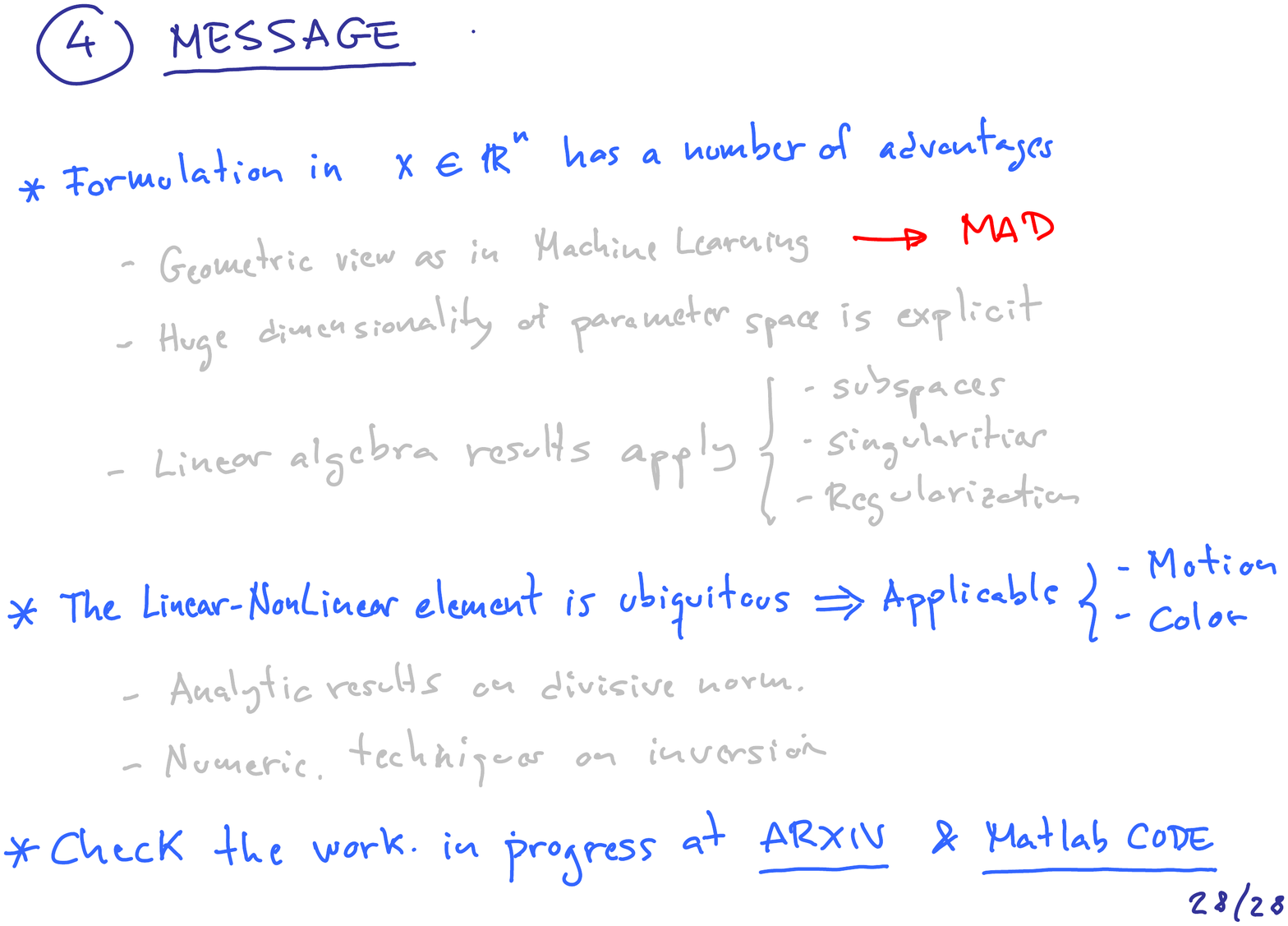} \\ [2cm]
    \hspace{-1.5cm} \includegraphics[width=14cm]{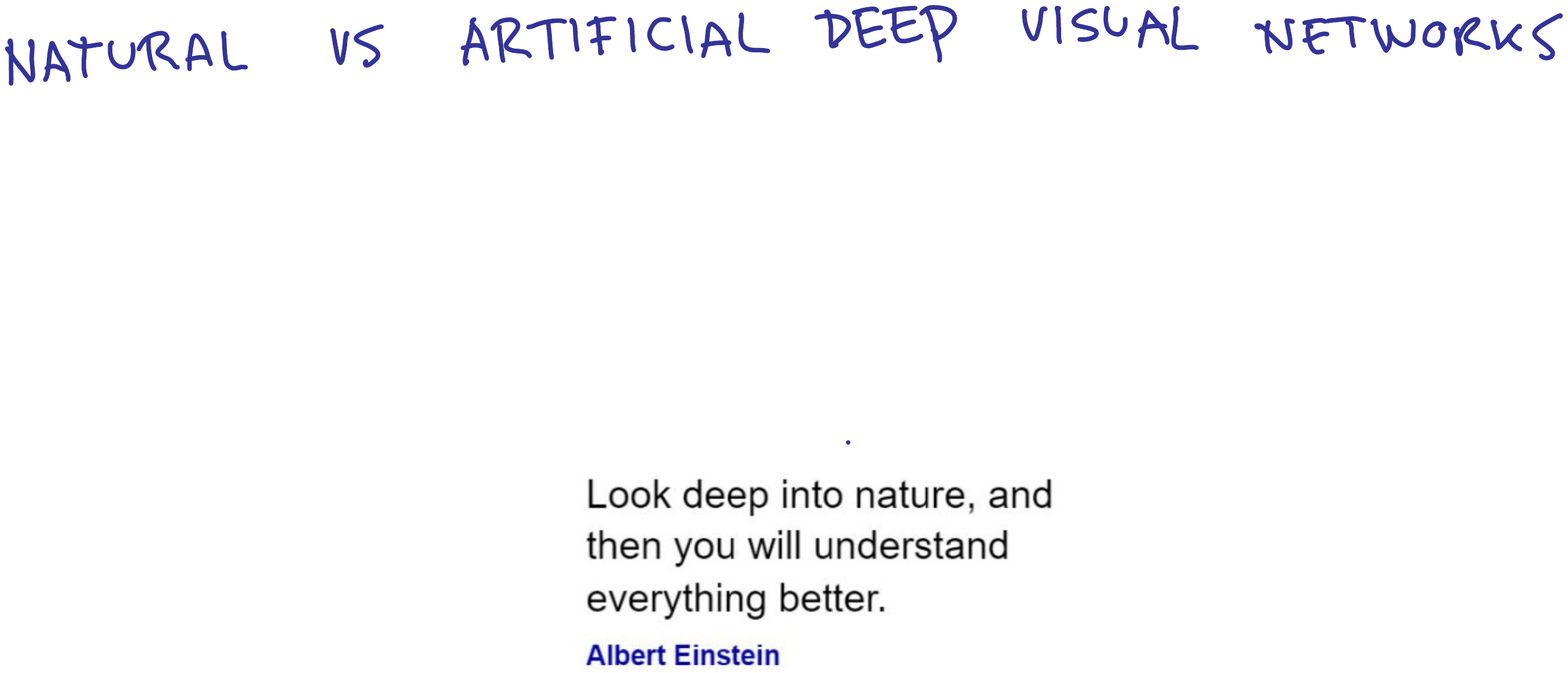}
    \end{tabular}
\end{figure}

\begin{figure}[h]
	\centering
    \small
    \setlength{\tabcolsep}{2pt}
    \vspace{-2cm}
    \begin{tabular}{c}
    \hspace{-1.5cm} \includegraphics[width=14cm]{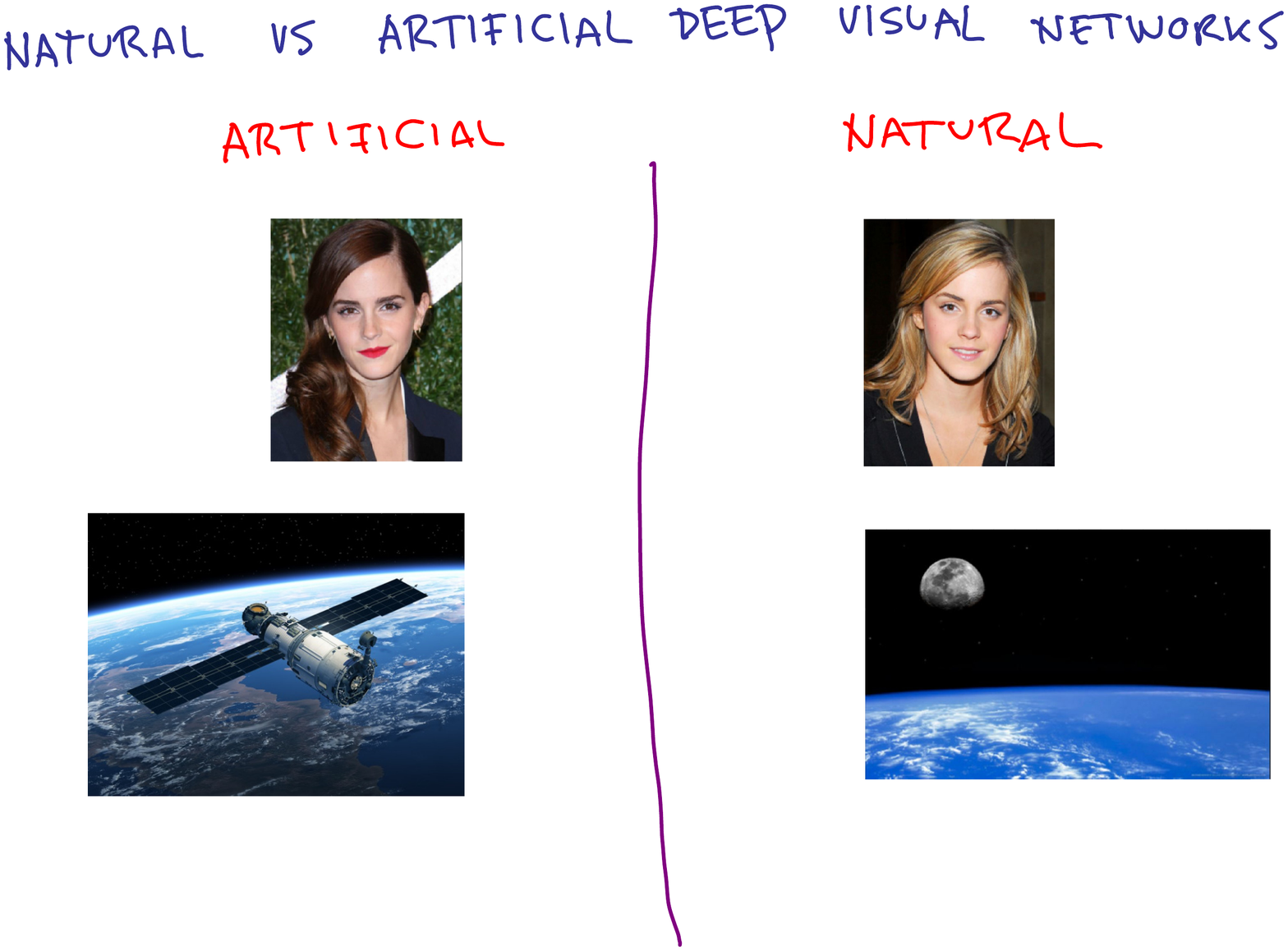} \\ [2cm]
    \hspace{-1.5cm} \includegraphics[width=14cm]{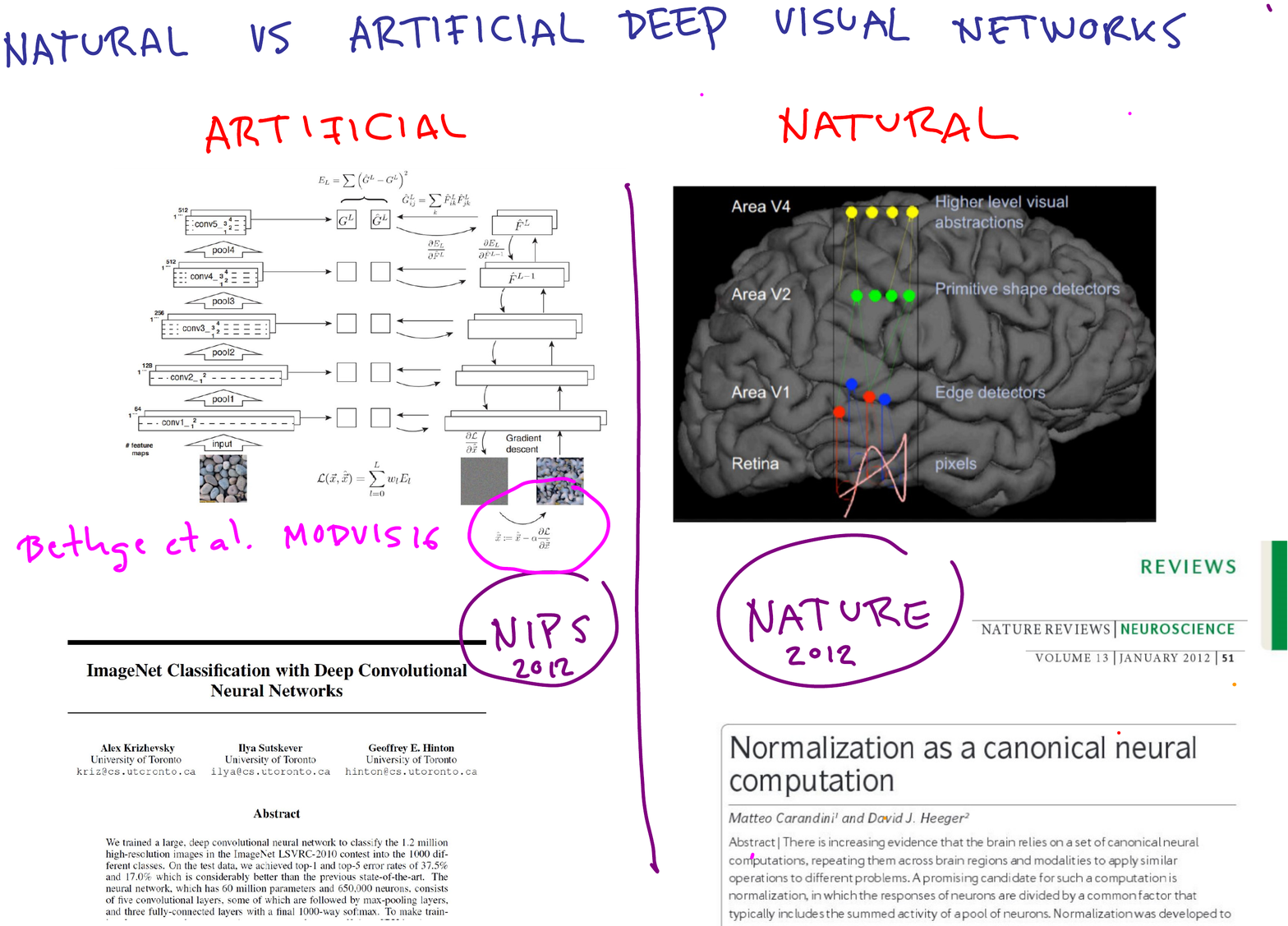}
    \end{tabular}
\end{figure}

\begin{figure}[h]
	\centering
    \small
    \setlength{\tabcolsep}{2pt}
    \vspace{-3cm}
    \begin{tabular}{c}
    \hspace{-1.5cm} \includegraphics[width=14cm]{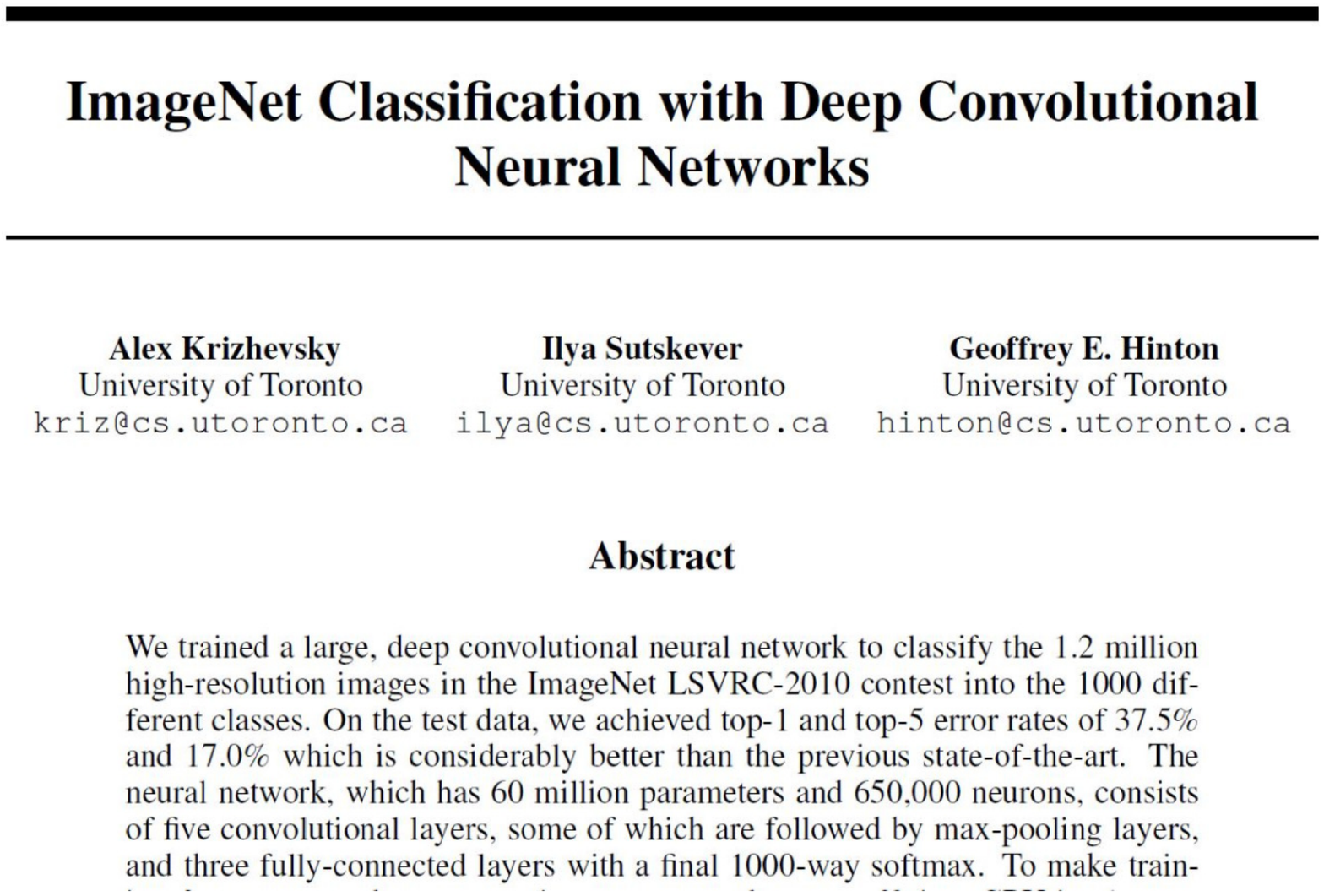} \\ [1cm]
    \hspace{-1.5cm} \includegraphics[width=14cm]{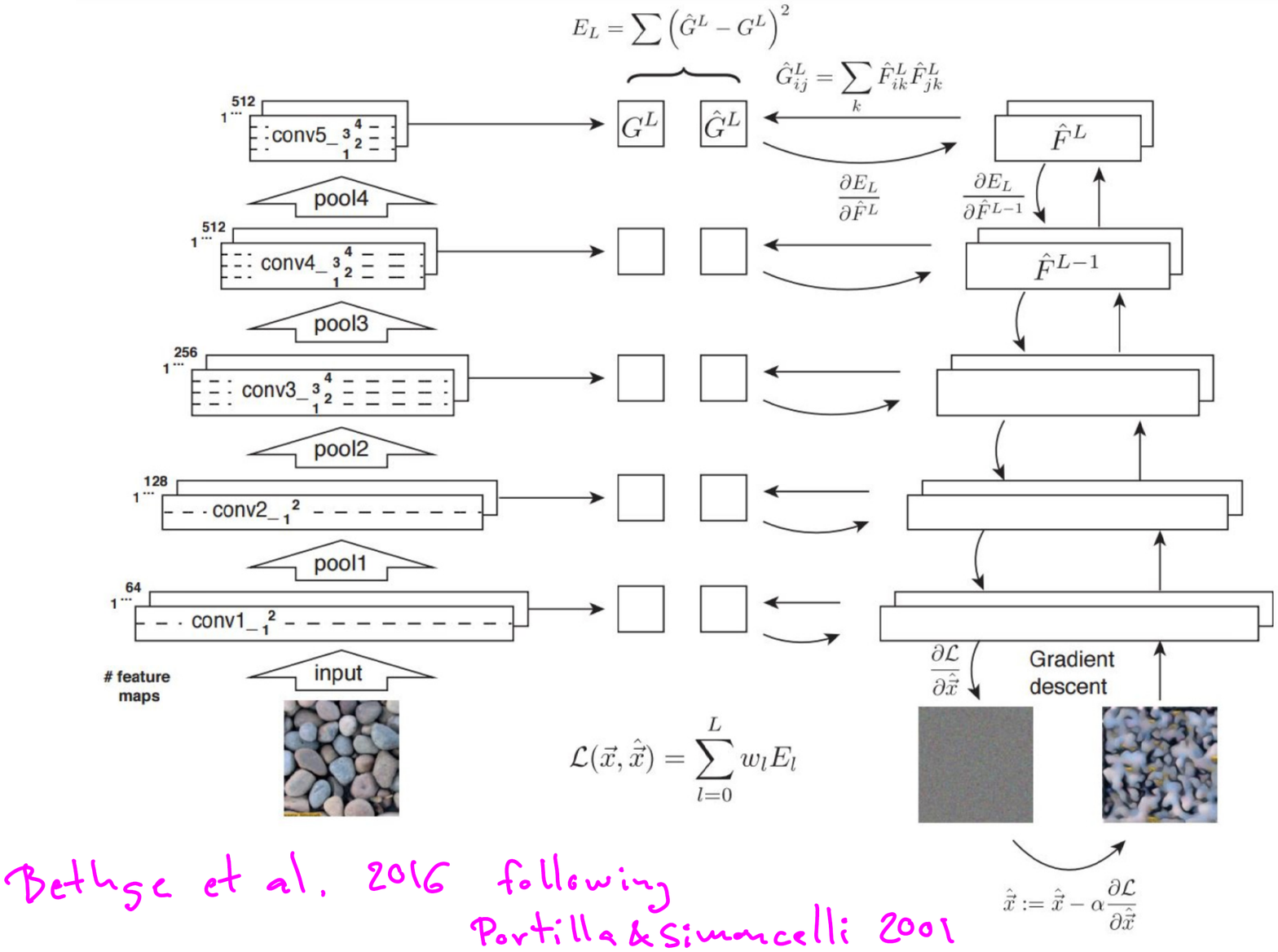}
    \end{tabular}
	\caption{Note on artificial vs natural network slides. In the recap talk at IPL on the relation of our approach
    with recent developments on Deep-Networks, I cited several works from Bethge's Lab on texture synthesis and
    artistic style transference because they use equations to compute the images which are similar to the gradient
    descent equations in MAD. Another reason supporting the interest of the notation introduced in our work.}\label{slide1}
\end{figure}

\newpage
\bibliographystyle{unsrt}
\bibliography{JesusMaloRef,references}

\end{document}